%% file: main.tex
\documentclass[12pt, draftclsnofoot, onecolumn]{IEEEtran}
\usepackage{cite}
\usepackage{setspace}
\usepackage{graphicx}
\usepackage{array}
\usepackage{float}
\usepackage{tikz}
\usepackage{tikz-3dplot}
\usetikzlibrary{shapes.geometric}
\usetikzlibrary{decorations.pathreplacing}
\usetikzlibrary{positioning}
\usetikzlibrary{patterns}
\usepackage[siunitx]{circuitikz} 
\usepackage{pgfplots}
\usepackage{pstool}  
\usepackage{mathtools}
\usepackage[linesnumbered,algoruled,boxed,lined]{algorithm2e}
\usepackage[absolute]{textpos}
\usepackage{threeparttable}
\usepackage[strict]{changepage}
\usepackage{amssymb}
\usepackage{comment}
\usepackage{bbm}
\usepackage{hhline}
\usepackage{lipsum} 
\usepackage{enumitem}
\usepackage{soul}
\usepackage{xcolor}

\usepackage{upquote}

\ifCLASSINFOpdf
\else
\fi
\usepackage{amsmath}
\ifCLASSOPTIONcompsoc
  \usepackage[caption=false,font=normalsize,labelfont=sf,textfont=sf]{subfig}
\else
  \usepackage[caption=false,font=footnotesize]{subfig}
\fi
\pgfplotsset{compat=1.14} 

\begin{document}
%
\title{A Quick Primer on Machine Learning in Wireless Communications}
%
\author{Faris B.~Mismar%
\thanks{The author is an adjunct associate professor of electrical and computer engineering at The University of Texas at Dallas (email: fbm090020@utdallas.edu).}
}%

%
%

\markboth{Revision date: 5/15/2025}%
{Mismar: Quick Primer on Machine Learning in Wireless Communications}

%




\maketitle

\begin{abstract}
This is our final issue of the quick primer on the use of Python to build a wireless communications prototype. This prototype simulates multiple-input and multiple-output (MIMO) systems for a single orthogonal frequency division multiplexing (OFDM) symbol.  Further, it shows several artificial intelligence (AI) and machine learning (ML) use cases and introduces the deepwireless library for code implementation.  The intent of this primer is to empower the reader with the means to efficiently create reproducible simulations related to AI and ML in wireless communications on inexpensive computing devices.  This primer has sprung from a draft aligned with the syllabus of a graduate course (EESC~7v86)---which we created to be first taught in Fall 2022---and has since evolved to where it stands today.
\end{abstract}


%
\IEEEpeerreviewmaketitle

\section{Introduction} 
%
%
%
%



The research domain of wireless communications and machine learning has been supplied with an abundance of publications, yielding numerous source codes implementing countless assumptions and configurations.  These source codes---especially when proprietary or released under restrictive license agreements---can make the idea of reproducibility extremely challenging.

Because of this, and keeping simplicity in mind, we share this quick primer on machine learning in wireless communications in addition to the source code related to it.  We call the essential library that implements this primer ``deepwireless.''  The objective of this primer and the deepwireless library is to fast-track the reader into the ability to build their simulations efficiently using inexpensive computing environments and open-source tools---especially ones with machine learning (ML) and artificial intelligence (AI) applications.  Because the wireless communications system of our choice supports multiple-input and multiple-output (MIMO) systems and orthogonal frequency division multiplexing (OFDM), this source code can be used for the prototyping of AI-based air interfaces (AI-AI) for 4G LTE and 5G air interfaces alike.  We hope that AI-AI would empower next-generation wireless networks beyond 5G and thus extend the longevity of these libraries and related source code.  The deepwireless library and the related source code are written in Python and are available online. Specifically, the deepwireless library is available on GitHub \cite{github} or can be installed using pip and a well-documented source code implementing all the scenarios shown in this primer using deepwireless is available on GitHub.  It is no surprise that we have chosen Python to implement deepwireless as Python has become the quintessential programming language for building AI and ML applications due to its simplicity and abundance of supporting libraries. Moreover, it is free to use.

 


\begin{table}[!t]
\begin{adjustwidth}{-.1in}{0cm}
\centering
\setlength\doublerulesep{0.5pt}
\caption{Abbreviations}
\vspace*{-1.5em}
\label{tab:abbrev}
\fontsize{7.5}{8}\selectfont
\begin{tabular}{ lm{13em}|lm{18em}|lm{16em} } 
\hhline{======}
AI & Artificial Intelligence & IDE & Integrated Development Environment & RF & Radio Frequency \\
BER & Bit Error Rate & KDE & Kernel Density Estimation & RLC & Radio Link Control \\
BLER & Block Error Rate & LLM & Large Language Model & RNN & Recurrent Neural Network \\
BS & Base Station & LMMSE & Linear MMSE & RRC & Radio Resource Control \\
CDL & Clustered Delay Line & LOS & Line Of Sight & RSRP & Reference Signal Received Power \\
CNN & Convolutional Neural Network & LSTM & Long Short-Term Memory & SDMA & Space-Division Multiple Access \\
CRC & Cyclic Redundancy Check & MAC & Medium Access Control & SGD & Stochastic Gradient Descent \\
CSI & Channel State Information & MDP & Markov Decision Process & SINR & Signal to Interference plus Noise Ratio \\
CoMP & Coordinated Multi-Point & MIMO & Multiple-Input and Multiple-Output & SISO & Single-Input and Single-Output \\
DFT & Discrete Fourier Transform & ML & Machine Learning & SM & Spatial Multiplexing \\
DNN & Deep Neural Network & MMSE & Minimum Mean Square Error & SNR & Signal to Noise Ratio \\
DQN & Deep $Q$-Network & NLOS & Non-LOS & SVD & Singular Value Decomposition \\
FEC & Forward Error Correction & OFDM & Orthogonal Frequency Division Multiplexing & UE & User Equipment \\
FFT & Fast Fourier Transform & PHY & PHYsical & ZF & Zero-Forcing \\
GAN & Generative Adversarial Network & RE & Resource Element \\
\hhline{======}
\end{tabular}
\end{adjustwidth}
\end{table}%

Also, considering the large number of acronyms and abbreviations in this primer, we build a table that can be referenced.  These abbreviations can be found in Table~\ref{tab:abbrev}.

\section{Protocol Stack}\label{sec:protocol_stack}

The wireless protocol stack enables communication between devices in cellular networks, facilitating seamless data transmission over the air.  The stack is divided into two planes: the user plane and the control plane.  The user plane protocol is responsible for handling the actual user data: the content transmitted between end devices (e.g., voice, video, and Internet traffic).  The control plane is responsible for signaling (e.g., to set up, tear down, manage, and maintain user sessions) in the network.

\begin{figure}[H]
\centering
\resizebox{0.45\textwidth}{!}{\input{figures/stack.tikz}}%
\caption{The control plane air interface protocol stack.}
\label{fig:stack}
\end{figure}%
The control plane stack is composed of multiple protocol layers as shown in Fig.~\ref{fig:stack}.  We discuss some of them here: %
\begin{enumerate}
    \item The physical (PHY) layer:  The PHY layer is broken down to two sublayers: low- and high-PHY.  The low-PHY sublayer has responsibilities that generally run at a fast time scale and are closer to the radio-frequency (RF) spectrum resources such as time synchronization, frequency alignment, and modulation and coding. 
 The high-PHY sublayer is for functions that operate at slower time scales such as channel equalization, beamforming, and MIMO processing.
    \item The medium access control (MAC) layer:  MAC manages transfers of data in the form of transport blocks, performs retransmission, handles random access contention, and ensures efficient use of the shared radio spectrum.  It works closely with the radio link control (RLC) layer, which is responsible for payload segmentation, error detection, retransmissions, flow control, and in-sequence delivery of protocol data units.
    \item The radio resource control (RRC) layer:  This is commonly referred to as ``layer-3'' signaling.  It plays a key role in managing RRC connection management, radio bearers, system information broadcast, paging, area updates, mobility (cell reselections and handovers across base stations, frequency bands, and technologies), and measurement reporting.
\end{enumerate}

Together, the PHY, MAC, and RLC layers enable reliable communication over the air interface by preparing and managing the data transmitted over the network.  A key set of procedures that bring together these layers to observe and improve the network performance is radio resource management (RRM).  RRM is a group of algorithms for controlling radio parameters to utilize the limited RF spectrum resources and radio network infrastructure as efficiently as possible.  Examples of RRM procedures are: power control, precoding, carrier aggregation, interference coordination, link adaptation, scheduling, and mobility management (i.e., handovers and cell reselection).

A valid question is often posed in this context:  What \textit{potential} value would AI and ML bring to next-generation wireless network RRM procedures that cannot be realized with ``legacy'' solutions?

\begin{enumerate}
    \item \textbf{Lower latency}: The predictive capability of AI and ML allows the network to prepare an action before it actually happens.  This is in contrast with the scenario where the network computes the action reactively (e.g., after a user has moved to the new location as opposed to predicting it).
    \item \textbf{More versatility}:  The data-driven approach resulting from UEs reporting their measurements enables model-free systems that can further improve static (and thus highly biased) legacy model-driven models.  For example:  How often is the noise profile over the air interface truly Gaussian?  How does shot noise, especially in large transmission bandwidth, impact the noise profile?  In such cases, can symbol detection be done more efficiently for a given channel condition?
    \item \textbf{Better performance}:  Legacy and voice-centric RRM procedures such as the configuration of measurement gaps during mobility stem from the days of the third generation wireless standards and still live in 5G NR.  These measurement gaps cause data transmission to pause for channel measurements to take place, which in a voice-centric network may not be a perceivable problem.  However, with explosive data transmission rates, such measurement gaps can cause a severe degradation in the throughput performance.  The use of data-driven intelligence can eliminate these gaps thus improving the performance.
    \item \textbf{More efficient computation}:  Several ML algorithms enable parallelization and collaborative learning making use of spare compute power available at neighboring BSs.  When an RRM procedure is based on these ML algorithms, it can---unlike legacy RRM procedures---benefit from this advantage for efficiency.
\end{enumerate}

There are also instances where AI and ML should not be applied to wireless network problems.  Here are some cases:
\begin{enumerate}
\item \textbf{Costs outweigh benefit}:  If training an AI model in a UE can achieve an incremental performance gain at a significant battery life expense, then it should not be done.
\item \textbf{Problem is well-understood}:  If linear calculations can solve the problem, then no need to introduce unnecessary complexity with AI, especially if AI cannot be explained.
\item \textbf{Data is insufficient}:  If UEs do not report often on their pertinent wireless measurements or if observability is lacking, then collected data could be underestimating ground truth.
\item \textbf{Real-time requirements}: If the wireless environment changes so rapidly that it becomes impractical to keep invalidating current models and training new ones.
\item \textbf{Inferior performance}:  This is an obvious yet an important one.  If introducing AI and ML causes the network performance to degrade, it should not be used.
\end{enumerate}

Now that we sufficiently motivated the protocol stack and the value of introducing AI and ML in wireless networks, we move to the system model next.  The system model describes a model-driven implementation, which we contrast with a data-driven implementation.  This enables us to then study several RRM procedures formulated as use cases solvable with AI and ML.

\section{System Model}\label{sec:system_model}

We consider a MIMO\footnote[1]{Also applies to the other single-multiple permutations: SISO, MISO, and SIMO systems alike.} system model with $N_t$ transmit antennas at the base station (BS) with a single OFDM symbol spanning over a number of narrowband subcarriers, the spacing of which is $\Delta f$, and $N_r$ receive antennas at the user equipment (UE).  The transmitting end has a total power of $P_\text{total}$ distributed across various transmit antennas such that the signal per antenna has a power of $P_x$ and an energy of $E_x$.  For one OFDM subcarrier, we write the system model as:
\begin{equation}\label{eq:system}
    \mathbf{y} = \mathbf{H}\mathbf{x} + \mathbf{n}
\end{equation}%
where $\mathbf{y} \in \mathbb{C}^{N_r}$ is a column vector containing the received OFDM signal from a transmitted signal column vector $\mathbf{x} \in \mathbb{C}^{N_t}$, which has an average per-antenna power of $\mathbb{E}[\Vert\mathbf{x} \Vert^2] \coloneqq P_x = E_x\Delta f$.  Note that $\mathbf{E} [\mathbf{x}\mathbf{x}^\ast] \coloneqq \mathbf{R}_\mathbf{xx} = P_x\mathbf{I}_{N_t}$.  Thus, $P_\text{total} = \text{tr}(\mathbf{R}_\mathbf{xx}) = N_t P_x = N_t E_x \Delta f$.  $\mathbf{H}\in \mathbb{C}^{N_r\times N_t}$ is a channel state information (CSI) matrix, and finally $\mathbf{n}\in \mathbb{C}^{N_r}$ is a column vector of additive noise the entries of which are independent and identically sampled from a zero-mean complex Normal distribution $\mathbf{n}\sim\mathcal{N}_\mathbb{C}(0, \sigma_n^2)$.  The noise power $\sigma_n^2$ is further defined with respect to the noise power spectral density $N_0$ multiplied by the bandwidth of an OFDM subcarrier (i.e., $\sigma_n^2 = N_0\Delta f$). The number of subcarriers should not exceed the Fast Fourier Transform (FFT) size, which is naturally the smallest power of two that exceeds the number of subcarriers (the difference is given to the guardband subcarriers for robustness against interference). The OFDM symbols are Gray-coded and are either based on $M$-PSK or $M$-QAM constellations, with $M$ being square (i.e., $M \in \{4, 16, 64, 256\}$). We define $k \coloneqq \log_2 M$ which represents the number of bits transmitted or received per constellation symbol.

The number of streams in this MIMO system $N_s$ is defined as the minimum of the number of antennas in both transmitter and receiver sides.  That is, $N_s \coloneqq \min(N_t, N_r)$. The number of data streams cannot exceed $N_s$. 

The total bandwidth $B$ is computed as the number of subcarriers $N_\text{SC}$ multiplied by the subcarrier spacing $\Delta f$ (i.e., $B \coloneqq N_\text{SC}\Delta f$).  If we define the capacity with respect to Shannon in bits per second (bps) as $C = B\log_2(1 + \gamma)$, then the spectral efficiency $\textsf{SE}$ (in bps/Hz) is defined as $\textsf{SE} \coloneqq C/B$ for a given bit rate $C$, with $\gamma$ being the post-equalization signal-to-noise-plus-interference ratio as we define later. Currently, the source code only supports \textit{one user} only; therefore, this model is suitable for a single UE served by a single BS.%

\textbf{Fading:} We denote $G$ as the large-scale fading coefficient which can be calculated based on  path loss and shadow fading.  The calculation of the path loss ($\mathsf{PL}$) depends on several factors such as the desired carrier frequency ($f_c$), antenna gains on both ends, path loss exponent ($\eta$), and distance of the UE from the BS ($d$).  Let us take the example of free space path loss (FSPL), which when excluding the the transmit and receive antenna gains has the formula
\begin{equation}
    \mathsf{PL} = \left ( \frac{4\pi d}{\lambda} \right )^2 = \left (\frac{4\pi d f_c}{c}\right )^2
\end{equation}
from which $\eta = 2$.  Also, $\lambda = c/f_c$ is the relationship between the wavelength $\lambda$, the carrier frequency, and the speed of light in free space $c$.  As the number of subcarriers become large, applying one large-scale fading value uniformly may no longer be accurate and thus we overload the path loss formula $\textsf{PL}(f_c)$ to become $\textsf{PL}(f_i)$ where $f_i = f_c + \left (i - \frac{N_\text{SC}}{2} \right ) \Delta f, i \in \{1, 2, \ldots, N_\text{SC}\}$.

Besides FSPL, there are other path loss models from industry standards such as the 3GPP urban micro (UMi), urban macro (UMa), and rural macro (RMa).  We refer the interested reader to \cite{3gpp38901} for more details.

For shadow fading, we introduce a Normally distributed random variable with zero mean and unity variance that is multiplied by the shadow fading margin $\sigma_\text{SF}$ and then subtracted from the path loss (in dB).  Thus, $G_\text{dB} \coloneqq 10\log(\mathsf{PL}) - \mathcal{N}(0,\sigma_\text{SF}^2)$.  For a Rayleigh fading channel, the elements of $\mathbf{H} \coloneqq [h_{ij}]_{i,j}$ are sampled from a zero-mean complex Normal distribution $h_{ij}\sim\mathcal{N}_\mathbb{C}(0,1)$ since the squared modulus of a complex Normal distributed random variable has a Rayleigh distribution.  Thus, the power gain of the channel becomes $G\Vert\mathbf{H}\Vert_F^2$, which means that to incorporate the large-scale fading to $\mathbf{H}$, we simply multiply all the elements of $\mathbf{H}$ by $\sqrt{G}$.  For Ricean fading (with presence of a line-of-sight component between the transmitter and the receiver), the elements of $\mathbf{H} \coloneqq [h_{ij}]_{i,j}, h_{ij}\sim \mathcal{N}_\mathbb{C}(\sqrt{\frac{K}{(1+K)}}, \frac{1}{K+1})$, where $K$ is the Ricean $K$-factor (when $K=0$ we are back to a Rayleigh fading channel).

Alternatively, we introduce the industry standards clustered delay line (CDL) channel models \cite{3gpp38901} for low complexity simulations since CDLs are statistical representations of the channel that a transmitted signal follows to a receiver for various scenarios: line of sight (LOS), non-line of sight (NLOS), and a combination of both.

\textbf{Eigenmodes:} The channel eigenmodes are the eigenvalues $\lambda_i$ of $\mathbf{HH}^\ast$, each of which is a representation of a parallel spatial data stream enabled by the channel, for what is known as the spatial multiplexing (SM).  The value of each eigenmode is the power gain of that stream due to the channel.  It is easy to prove that the power gain of the channel is simply the sum of the eigenmodes (i.e., $\Vert\mathbf{H}\Vert_F^2 = \sum_{i=1}^{N_r} \lambda_i$).  The proof is as follows:
\begin{equation}
\begin{aligned}
    \Vert \mathbf{H}\Vert_F^2\coloneqq \text{tr}(\mathbf{HH}^\ast) = \text{tr}(\mathbf{U}\boldsymbol\Lambda \mathbf{U}^{-1}) = \text{tr}(\mathbf{U}^{-1}\mathbf{U}\boldsymbol\Lambda) =  \text{tr}(\boldsymbol\Lambda) = \sum_i \lambda_i. \\
\end{aligned}
\end{equation}

In order to exploit these power gains (through means of a precoder $\mathbf{F}$ and the waterfilling algorithm which we briefly describe next), the channel needs to be known at the transmitter.  The precoding matrix $\mathbf{F}\in\mathbb{C}^{N_t\times N_s}$ is responsible for both power control and the mapping of the transmit antennas to a number of parallel streams $N_s\le N_t$.

\textbf{Precoding and combining:} The precoder can be used in order to enable several users to access the same resources at once such as in space-division multiple access (SDMA), cancel interference (i.e., zero-forcing) across different MIMO streams, perform SM, and perform power control per antenna, where the power control matrix can be computed using the waterfilling algorithm (assuming that the channel eigenmodes are known at the transmitter).  Waterfilling optimizes the channel capacity by allocating more transmit power towards larger eigenmodes since they have better signal quality.  For the transmit diversity mode\footnote[2]{Beamforming and transmit diversity are two modes in MISO systems.  The difference is in the precoding and thus usage:  Diversity sends multiple copies of the same signal without having to know the CSI (suitable for areas with low SNR), while beamforming adjusts the transmit phase on these signals to find an optimal signal to interference plus noise ratio (SINR) and thus needs to know the CSI (suitable for high-SNR environments).}, the Alamouti scheme precoding can be used (MIMO streams are spatially correlated and the same information is sent across multiple antennas to improve robustness).  Without precoding, the MIMO system becomes a SISO equivalent with the channel gain equal to the largest eigenmode of the channel $\mathbf{H}$. For the SM mode, where streams are spatially uncorrelated, the capacity of the system is the sum of the capacities of the comprising single channels after decomposing the MIMO channel.  Here, the precoder $\mathbf{F}$ and the combiner $\mathbf{G} \in \mathbb{C}^{N_r\times N_r}$ are derived through the singular value decomposition (SVD) of the channel $\mathbf{H} \coloneqq \mathbf{U}\boldsymbol\Sigma\mathbf{V}^\ast$ as applied onto the system \eqref{eq:system}:
\begin{equation}\label{eq:system_sm}
\begin{aligned}
        \mathbf{y} &= \mathbf{H}\mathbf{x} + \mathbf{n} \\
        &= \mathbf{U}\boldsymbol\Sigma\mathbf{V}^\ast\mathbf{x}  + \mathbf{n}  \\
        &= \mathbf{U}\boldsymbol\Sigma\mathbf{V}^\ast\underbrace{\mathbf{V}\mathbf{s}}_{\mathbf{x} \coloneqq \mathbf{F}\mathbf{s}}  + \mathbf{n}  \\
        &= \mathbf{U}\boldsymbol\Sigma\mathbf{s}  + \mathbf{n}  \\
\underbrace{\mathbf{U}^\ast}_{\coloneqq \mathbf{G}} \mathbf{y}  &=  \mathbf{U}^\ast\mathbf{U}\boldsymbol\Sigma\mathbf{s}  + \mathbf{U}^\ast \mathbf{n}  \\
\tilde{\mathbf{y}} &=  \boldsymbol\Sigma\mathbf{s}  + \tilde{\mathbf{n}}
\end{aligned}
\end{equation}
where $\boldsymbol\Sigma$ is a diagonal matrix of the square roots of the eigenmodes of the channel $\mathbf{H}$ (or, again, the eigenvalues of $\mathbf{HH}^\ast$).  It is easy to discern that the precoder $\mathbf{F}\coloneqq \mathbf{V}$, the combiner $\mathbf{G}\coloneqq\mathbf{U}^\ast$, and the diagonalized channel state information is $\boldsymbol\Sigma$. This also implies that the transmitter and the receiver have perfect knowledge of the CSI $\mathbf{H}$, otherwise the computation of $\mathbf{U}$ and $\mathbf{V}^\ast$ would not be possible.  In words, this system works by applying SVD to the known CSI.  Then, the transmitter precodes the transmitted data by pre-multiplying it by $\mathbf{V}$ while the combiner pre-multiplies the signal with $\mathbf{U}^\ast$ at the receiver end. 

\textbf{Equalization:}  To get rid of the impact of the channel (or the diagonalized channel for SM), the equalization matrix $\mathbf{W}\in \mathbb{C}^{N_t\times N_r}$ is left-multiplied by the received signal in \eqref{eq:system} to obtain 
\begin{equation}
\begin{aligned}
    \mathbf{z} \coloneqq \mathbf{W} \mathbf{y} &= \mathbf{W} \mathbf{H}\mathbf{x} + \mathbf{W} \mathbf{n} \\
    &= \mathbf{\hat x} + \mathbf{v} \\
\end{aligned}
\end{equation}%
or, analogously, in \eqref{eq:system_sm} to obtain $\mathbf{z} \coloneqq \mathbf{W}\mathbf{\tilde y}$, which is then fed into a symbol detector to identify the originally transmitted OFDM symbols in the presence of the inseparable noise.   Equalization also amplifies the noise $\mathbf{v}\coloneqq \mathbf{Wn}$ at the receiver (or $\mathbf{W\tilde n}$ in the case of precoding).

\textbf{Beamforming:} For the special case where $N_s = N_r = 1$, the precoder $\mathbf{F}$ becomes a column vector $\mathbf{f}\in\mathbb{C}^{N_t}, \Vert\mathbf{f}\Vert^2 = 1$ and the channel becomes a row vector.  Therefore, \eqref{eq:system} becomes:%
\begin{equation}\label{eq:system_beamforming}
    y = \mathbf{h}^\ast\mathbf{f} x + n,
\end{equation}%
which is the reason a beamforming channel is often known as a ``rank-1'' channel.  A well-known example of a beamforming codebook is the discrete Fourier transform (DFT) codebook.   Let the codebook $\mathcal{F}$ contain a set of these beamforming vectors $\mathbf{f}$, also known as ``grid of beams.''  Regardless of the codebook, a power-optimal beamforming precoding vector $\mathbf{f}^\star$ which maximizes the received channel gain can thus be found through a search in $\mathcal{F}$: %
\begin{equation}
    \mathbf{f}^\star \coloneqq \underset{\mathbf{f}\in\mathcal{F}}{\arg\max} \, \vert \mathbf{h}^\ast \mathbf{f} \vert ^2,
\end{equation}
which is how beamforming improves the received signal power.  

The power-optimal beamforming precoder can also be computed from the channel state information if it is known at the transmitter.  In this case, there is no need to search in a codebook.  The optimal beamforming precoder is thus:
\begin{equation}
    \mathbf{f}^\star = \underset{\mathbf{f}:\Vert\mathbf{f}\Vert^2 = 1}{\arg\max}\; \vert \mathbf{h}^\ast\mathbf{f}\vert^2 = \frac{\mathbf{h}}{\Vert \mathbf{h}\Vert},
\end{equation}
since the quantity enclosed in the modulus operator is an inner product of the channel and the precoder vectors, which is maximized when they are both equal.  This optimal beamforming precoder aligns the transmitted signal with the channel to maximize the received signal power with a gain of $\Vert \mathbf{h}\Vert^2$.  Note that there is no need for a combiner in single-user beamforming, since the received signal is already a scalar quantity.

\textbf{Cell-Free MIMO:} Cell-free MIMO, better known by its former name coordinated multi-point (CoMP) joint transmission, exploits the phenomenon when signals transmitted from two (or more) spatially separate cells, they are likely to be strongly decorrelated and MIMO can thus take advantage of them. In this case, the large-scale fading coefficient $G$ is redesigned so that different elements in $\mathbf{H}$ are multiplied with different location-dependent fading coefficients based on which transmitting cell these elements represent.

\textbf{Statistics:} 
\subsubsection{Transmit power}
The base station transmit power $P_\text{total}$ is divided across all OFDM subcarriers and transmit antennas.  Therefore, for a MIMO system with $N_t$ transmit antennas, the average power of an OFDM subcarrier before power control is $P_x = P_\text{total}/N_t$.



Industry standards use the term resource element (RE) as the smallest element in an OFDM time-frequency grid corresponding to one subcarrier-symbol.  In the \textit{special} case of a single OFDM symbol, the transmit power per RE is equal to the transmit power per antenna over one OFDM symbol, which is equal to $P_x / (N_tN_\text{SC})$.  In 4G LTE, certain REs are known as the ``reference symbols'' (or ``reference signals'' in 5G NR) and are used to estimate the channel and measure the cell coverage.   If the channel gain is $\Vert\mathbf{H}\Vert_F^2$ (or $G\Vert\mathbf{H}\Vert_F^2$ with the large-scale fading as motivated earlier), then the averaged power of these reference symbols (or signals) as measured at the receiver---known as the reference signal received power (RSRP)---is $G\Vert\mathbf{H}\Vert_F^2P_x/((N_\text{symb} \coloneqq 1) N_\text{SC})$.  The path loss is the attenuation the signal undergoes as it propagates through the channel as a result of the channel impact (i.e., large-scale fading, shadowing, etc.) as shown earlier in this section.  It is thus defined as $\textsf{PL} \coloneqq G\Vert\mathbf{H}\Vert_F^2$ and is often reported in dB scale (i.e., $10\log_{10}(\cdot)$).

\subsubsection{Signal to noise ratio}
To derive the signal to noise ratio (SNR) at the receiver, we use the system model \eqref{eq:system} and compute the autocorrelation matrix of the received signal:
\begin{equation}
\begin{aligned}
    \mathbf{R_{yy}} \coloneqq \mathbb{E}[\mathbf{yy}^\ast] &= \mathbb{E}[(\mathbf{Hx+n})(\mathbf{Hx+n})^\ast] \\
    &= \mathbb{E}[\mathbf{Hx}\mathbf{x}^\ast\mathbf{H}^\ast] + \mathbf{R_{nn}} \\
    &= P_x \mathbf{I}_{N_r} \mathbb{E} [\mathbf{H}\mathbf{H}^\ast] + \sigma_n^2\mathbf{I}_{N_r}  \\ 
    &\stackrel{(a)}{=} \underbrace{P_x \Vert\mathbf{H}\Vert_F^2}_{\text{signal}} + \underbrace{\sigma_n^2}_\text{noise},\\ 
\end{aligned}
\end{equation}
where $(a)$ is a scalar quantity derived from the normalized traces of $\mathbf{R}_\textbf{yy}$ for the ease of reading.  If the SNR at the transmitter is equal to $\rho\coloneqq P_x/\sigma_n^2$, and the channel gain is $\Vert\mathbf{H}\Vert_F^2$, then it follows that the SNR at the receiver before equalization is equal to $\gamma = \rho \Vert\mathbf{H}\Vert_F^2 / N_f$,  where $N_f$ is the noise figure.  Formally, the noise figure is the ratio of the input SNR to the output SNR at the receiver.

In the presence of an equalizer, the SNR at the receiver is different for different receive antennas due to the equalization matrix $\mathbf{W}$, the elements of which enhance (i.e., amplify) the noise independently as derived earlier in this section.   As a result, we write the  post-equalization received SNR at the $j$-th antenna as $\gamma_j$.  Details are in Section~\ref{sec:model-driven}.

The transmit SNR can also be written as $\rho = E_x/N_0$, where $E_x$ is the energy of the transmitted signal and $N_0$ is the noise power spectral density.   Given that one symbol is made of $k = \log_2 M$ bits at the transmitter, the bit energy to noise power spectral density ratio at the transmitter is $E_b/N_0 = \rho / k = E_x / (kN_0)$. Note that $E_b/N_0$ is the per-bit SNR and that both quantities are also often represented in dB scale.

At the receiver, the bit energy to noise power spectral density ratio ($E_b/N_0$) post-equalization is $\gamma / (kr) $ where $\gamma$ is the receive SNR and $r$ is the code rate.  The maximum code rate is defined as $r_\text{max} \coloneqq \textsf{SE} / k \le 1$.  Generally, $r \le r_\text{max}$, especially as $\rho$ decreases.

Also if the interference is present at the receiver, it is added to the noise as shown in Section~\ref{sec:model-driven}.  In this case, we have a received SINR measured per receive antenna. The total noise plus interference power is $\sigma_n^2 + p_\text{interference} \sigma_i^2$ as motivated in Section~\ref{sec:model-driven}.



\section{Model-Driven Implementation}\label{sec:model-driven}

\begin{figure}[!t]
\centering
\resizebox{1.05\textwidth}{!}{\input{figures/overall.tikz}}%
\caption{A block diagram of the overall model of the wireless system (thicker boxes are optional blocks).}
\label{fig:overall}
\end{figure}

We start with the implementation of various functions of wireless network through models and statistics as shown in Fig.~\ref{fig:overall}.  We call this ``model-driven'' as opposed to ``data-driven'' which is discussed later in Sections~\ref{sec:unsupervised_ml}, \ref{sec:supervised_ml}, and \ref{sec:reinf_learning}.
A $b$-bit quantizer is denoted as $Q_b(\cdot)$ and is optional.  We use $\mathbf{W} \in \mathbb{C}^{N_r\times N_t}$ as a channel equalizer, which removes the effect of the channel onto the received data.  However, equalizers require an estimate of the channel state information, which the receiver computes at its end.  This estimate is denoted as $\mathbf{\hat H}$ and has the same dimensions as $\mathbf{H}$. Channel estimation is enabled through ``pilots'' which are a sequence known to both the transmitting and receiving ends of the wireless system.  Pilot symbols and cyclic prefix (CP) symbols (seen in both 4G LTE and 5G) are not the same---pilot symbols are used to estimate the channel and CP serves two purposes: (1) reduces inter-carrier interference and (2) allows the use of circular convolution which translates to a multiplication operation in the frequency domain.

\textbf{Constellation:}  We follow the notation of \cite{proakis} in generating the symbols for both QPSK and $M$-QAM as baseband symbols.  For QPSK, the inphase ($I$) and quadrature ($Q$) branches take the values $\pm 1$ and for $M$-QAM they take values in the integer interval $[-\sqrt{M} + 1, \sqrt{M}]$.  These constellation symbols are normalized so the power per symbol is equal to unity and are Gray-coded as stated earlier.  Let the constellation be a set $\mathcal{C}$.  This constellation has a cardinality $\vert\mathcal{C}\vert = M$.

\textbf{Noise:}  At the receiver, the noise signal is a complex zero-mean Normal random variable with a variance (or noise power) of $\sigma^2 \coloneqq k_\text{B}T\Delta fN_f$, where $k_\text{B}$ is the Boltzmann constant, $T$ is the temperature in Kelvins and $N_f$ is the noise figure at the receiver, as mentioned earlier.

\textbf{Interference:} Besides the channel effect that brings interference through the off-diagonal elements on $\mathbf{H}$ (i.e., the coupling between different transmit-receive antenna pairs), another method to model interference across all subcarriers is a binomial random variable with a probability of a receive antenna interfered at being $p_\text{interference}$.  If we represent the interference signal as an independent and identically sampled from a zero-mean complex Normal random variable with $\sigma_i^2 = P_\text{interference}$ in a similar to the noise, then the additive interference per OFDM subcarrier is a column vector with the dimension $\mathbf{i}\in\mathbb{C}^{N_r}$ that contains either zeros or an interference signal.  Hence, the average interference power is equal to $p_\text{interference}\sigma_i^2$ across all OFDM subcarriers.

\textbf{Codeword Construction:}  We generate a random sequence of bits as a payload.  The maximum size of this payload in bits is $L_\text{codeword}^{(\text{max})} = kN_\text{SC}N_t$ (or $kN_\text{SC}N_s$ in the case of SM) transmitted over a duration of one OFDM symbol.  The cyclic redundancy check (CRC) polynomial generator function $\mathcal{P}(\mathbf{x})$ operates on this payload and adds an overhead equal to the CRC length in bits.  There is only one CRC sent per transmission, regardless of the channel transmission rank and it is positioned at the end of the transmission block (i.e., the codeword).  The receiver extracts the CRC and attempts to compare it with a fresh CRC computed on the received block for a match.  If either the payload or the CRC does not occupy an integer number of symbols, padding is necessary.  Hence, a codeword is a payload, a variable padding length, and a CRC.

\textbf{Padding:}  If the CRC length in bits is equal to $L_\text{CRC}$, then the CRC length (in bits) with padding included is $k\lceil L_\text{CRC}/k \rceil$.  Moreover, if the payload size is $L_\text{payload}$, then a similar formula can be derived for the payload.  However, what remains is the variable padding length.  To compute this padding length, we solve a non-convex optimization problem:

\begin{equation}
    \begin{aligned}
        \text{minimize}\colon & \qquad L_\text{padding} \\
        \text{s.t.}\colon & \qquad L_\text{codeword} \coloneqq L_\text{payload} + L_\text{padding} + L_\text{CRC} \ge 0,\\
        & \qquad L_\text{codeword} \equiv 0 \ (\text{mod}\ N_t), \\
        & \qquad L_\text{codeword} \equiv 0 \ (\text{mod}\ N_\text{SC}), \\
        & \qquad L_\text{codeword} \le L_\text{codeword}^{(\text{max})}. \\
    \end{aligned}
\end{equation}

Other techniques such as forward error correction (FEC) and bit interleaving can also be used in constructing the payload, making it more robust against channel impairments. Common examples of FEC are turbo codes and low-density parity check (LDPC) codes---none of which is implemented in the source code today.

\textbf{Quantization:} Mapping continuous and infinite values that $\mathbf{y}$ takes to a smaller set of discrete finite values makes it easier to group symbols that ``appear'' similar and treat them like so---greatly reducing computational overheads (e.g., in machine learning).  However quantization is irreversible, non-linear, and causes a degradation to the quality of these symbols known as the ``distortion''.  Therefore, employing a quantizer comes with its drawbacks and designs may avoid it.  We use a truncation quantizer and define a $b$-bit quantization function as $Q_b(\cdot)$, with $b$ being the quantization resolution.  This quantizer thus has $2^b$ quantization levels.

\textbf{Channel Estimation:} Let the pilot symbols have a length $n_\text{pilot}$ OFDM symbols. We denote the fat matrix of known pilot symbols $\mathbf{X_P}\in \mathbb{C}^{N_t\times n_\text{pilot}}$.  Note that this matrix cannot be tall because the number of pilots must at least be equal to the number of transmit antennas ($n_\text{pilot} \ge N_t$).  Let the received pilot symbol values (i.e., channel response) be stored in another matrix $\mathbf{Y_P}\in \mathbb{C}^{N_r\times n_\text{pilot}}$.  Then we can compute the least-squares (LS) estimate of the channel as follows:%
\begin{equation}
    \mathbf{\hat H}_\text{LS} = \underset{\mathbf{H}}{\arg\min}\, \Vert \mathbf{Y_P} - \mathbf{H}\mathbf{X_P}\Vert ^2,
\end{equation}
which---besides the traditional way of writing the norm as an inner product---can be solved as follows:
\begin{equation}\label{eq:H_hat}
\begin{aligned}
    \mathbf{Y_P} &= \mathbf{\hat H}\mathbf{X_P}  \\
    \mathbf{Y_P}\mathbf{X_P}^\ast &= \mathbf{\hat H}\mathbf{X_P}\mathbf{X_P}^\ast  \\
    \mathbf{Y_P}\mathbf{X_P}^\ast(\mathbf{X_P}\mathbf{X_P}^\ast)^{-1} &= \mathbf{\hat H} \\
    \mathbf{\hat H} &= \mathbf{Y_P}\mathbf{X_P}^\ast(\mathbf{X_P}\mathbf{X_P}^\ast)^{-1}.
\end{aligned}
\end{equation}
There is a little abuse of notation in the derivation above since the pilot symbols and the channel response are matrixes.

\textbf{Pilot Design:} Since the matrix $\mathbf{X_P}$ is not square, it should be carefully designed so that $\mathbf{X_P}\mathbf{X_P}^\ast$ is invertible.  An example of carefully designed codes is Zadoff-Chu codes in both 4G LTE and 5G.  However, for simplicity we exploit the idea of semi-unitary matrixes such that $\mathbf{X_P}\mathbf{X_P}^\ast = \mathbf{I}_{N_t}$ and therefore we do not need to worry about the invertibility of $\mathbf{X_P}\mathbf{X_P}^\ast$ any further.   At a high level, the construction of a unitary matrix (i.e., $\mathbf{Q}^{-1} = \mathbf{Q}^\ast$) is possible through either DFT, a QR decomposition of a matrix the elements of which are sampled from $\mathcal{N}_\mathbb{C}(0, 1)$, or a combinatorial rearrangement of orthonormal bases.


Consequently, the LS channel estimate $\mathbf{\hat H}$ from \eqref{eq:H_hat} becomes:
\begin{equation}
    \mathbf{\hat H}_\text{LS} = \mathbf{Y_P}\mathbf{X_P}^\ast
\end{equation}%

The linear MMSE estimation (LMMSE) minimizes the mean squared error in the channel estimate and is thus written as
\begin{equation}
    \mathbf{\hat H}_\text{LMMSE} \coloneqq \underset{\mathbf{\hat H}\colon \mathbf{\hat H} = \mathbf{Ay}}{\arg\min}\;\mathbb{E} [\Vert \mathbf{H} - \mathbf{\hat H} \Vert_F^2]
\end{equation}%
where $\mathbf{\hat H} = \mathbf{Ay}$ is a \textit{linear} transform (hence the qualifier ``linear'').  Solving for the Frobenius norm as the trace of an outer product and finding the gradient of the mean squared error with respect to $\mathbf{A}$ then setting it to $\mathbf{0}_{N_r\times N_t}$, we obtain the LMMSE for a given transmit SNR and the channel covariance matrix $\mathbf{R_{HH}}$:
\begin{equation}
    \mathbf{\hat H}_\text{LMMSE} = \mathbf{R_{HH}} \left ( \mathbf{R_{HH}} + \frac{1}{\rho}\mathbf{I}_{N_r} \right ) ^{-1} \mathbf{\hat H}_\text{LS}.
\end{equation}%
Note that this channel covariance (or precisely, correlation) matrix can be found from the time-based averaging of the quantity $\mathbf{H}\mathbf{H}^\ast$).  Due to LMMSE having prior knowledge of the channel statistics as evident from the channel covariance matrix above, it outperforms LS estimation.

\textbf{Channel Equalization:} As mentioned earlier, channel equalization intends to remove the impact that the channel made onto the transmitted symbols.  Two commonly used types of channel equalizations are (1) zero-forcing (ZF) and (2) minimum mean square error (MMSE) equalization.  Let us call the equalizer $\mathbf{W}$ which is computed from a given channel (either perfectly known or estimated).  Further, let us continue to use the SM type of MIMO and thus apply the equalizer to the diagonalized channel $\boldsymbol\Sigma$ system as shown in \eqref{eq:system_sm}.  In this case:
\begin{equation}\label{eq:equalization}
\begin{aligned}
    \mathbf{\tilde z} \coloneqq \mathbf{W} \mathbf{\tilde  y} &= \mathbf{W} \boldsymbol\Sigma\mathbf{s} + \mathbf{W} \mathbf{\tilde n} \\
    &= \mathbf{\hat x} + \mathbf{\tilde v}. \\
\end{aligned}
\end{equation}%

If we denote an estimated channel---regardless whether precoded or not---as $\mathbf{\hat H}$ 
then the ZF equalizer of this estimated channel is given by:
\begin{equation}\label{eq:zfequalization}
\mathbf{\hat x} = \underset{\mathbf{x}}{\arg\min}\;\Vert \mathbf{y} - \mathbf{\hat H}\mathbf{x}\Vert^2 = \mathbf{H}^\dag\mathbf{y},
\end{equation}%
from which we deduce from \eqref{eq:equalization} that $\mathbf{W}_\text{ZF} = \mathbf{H}^\dag$.  Three immediate cases emerge from the pseudo-inverse: First, if the channel $\mathbf{\hat H}$ is square and invertible (i.e., full rank), then $\mathbf{W}_\text{ZF} = \mathbf{H}^{-1}$. Second, if the channel is wide (or $N_t > N_r$) and has full row rank, then $\mathbf{H}^\dag = \mathbf{H}^\ast(\mathbf{HH^\ast})^{-1}$.  Finally, if the channel is tall (or $N_t < N_r$) and has full column rank, then $\mathbf{H}^\dag=(\mathbf{H^\ast H})^{-1}\mathbf{H}^\ast$.  This pseudo-inverse (i.e., in \eqref{eq:zfequalization}) becomes a concern if the channel matrix is ill-conditioned (e.g., close to singular)\footnote[3]{Matrix inversion---even for well-conditioned matrixes---is a computationally expensive operation, especially with the increasing deployment of massive MIMO in later 5G releases which creates a larger dimension $\mathbf{H}$.  This justifies the need for graphics processing units.}.

For ZF equalization, keeping in mind that $\mathbf{R_{xx}} = P_x\mathbf{I}_{N_t}$ and $\mathbf{R_{nn}}= \sigma_n^2\mathbf{I}_{N_r}$, the post-equalization SNR on the $j$-th decoded stream $\gamma_j^\text{ZF}$ is
\begin{equation}
    \gamma_j^\text{ZF} = \rho \frac{1}{[\mathbf{WW}^\ast]}_{j,j} = \rho\frac{1}{[(\mathbf{\hat H}^\ast\mathbf{\hat H})^{-1}]_{j,j}}.
\end{equation}%
The construct $[\mathbf{A}]_{j,j}$ means the $j$-th diagonal element of the matrix $\mathbf{A}$.  ZF equalization completely eliminates the interference due to the coupling between different transmit-receive antenna pairs, hence the name.  Due to the concerns in the pseudo-inverse formula as outlined earlier, an alternative equalization technique is considered.

The MMSE equalizer of the estimated channel minimizes the residual error term: %
\begin{equation}\label{eq:mmse-equalizer}
    \mathbf{W}_\text{MMSE} = \underset{\mathbf{W}}{\arg\min}\; \mathbb{E} [\Vert \mathbf{W}\mathbf{y} - \mathbf{x}\Vert^2 ],
\end{equation}%
which when introducing the trace operator, computing the gradient, and setting it to $\mathbf{0}_{N_t \times N_r}$, we obtain:
\begin{equation}
\begin{aligned}
    \mathbf{W}_\text{MMSE} &= \mathbf{R}_{\mathbf{xy}} \mathbf{R}_\mathbf{yy}^{-1} \\
    &=\mathbf{\hat H}^\ast(\mathbf{\hat H}\mathbf{\hat H}^\ast + \frac{1}{\rho}\mathbf{I}_{N_t})^{-1}.
\end{aligned}
\end{equation}

Similarly, the average post-equalization SNR per antenna on the $j$-th decoded stream due to the MMSE equalization equals
\begin{equation}
    \gamma_j^\text{MMSE} = \rho \frac{1}{[\mathbf{WW}^\ast]}_{j,j},
\end{equation}%
which can also be derived in terms of the channel estimate $\mathbf{\hat H}$ (for the eager reader).

At low SNRs, ZF further amplifies the noise and thus MMSE offers as a useful advantage at low SNR.  However, at high transmit SNRs (i.e., $1/\rho\to 0$), the MMSE functions just as well as a ZF equalizer since it approaches a matrix inverse.  Otherwise, the MMSE equalizer does not lead to complete elimination of the interference \cite{molisch}.  

Another equalization technique uses the maximum likelihood receiver to decode the received vector $\mathbf{y}$.  It does so by exhaustively searching for a vector of symbols (needing up to $\vert\mathcal{C}\vert^{N_t}$ searches) that \textit{maximizes} the likelihood of the noise random variable (i.e., Gaussian), or:
\begin{equation}
    \mathbf{\hat x} = \underset{\mathbf{x}}{\arg\min}\, \Vert \mathbf{y} - \mathbf{\hat H}\mathbf{x} \Vert ^2.
\end{equation}%
This equalization technique is SNR-optimal in the sense of no losses in the per-antenna SNR.


\textbf{Symbol Detection:} The last step is to convert the received symbols to the symbols belonging to the constellation.  Since $\mathbf{v}$ is also Gaussian, we can use a maximum likelihood detector (or $k$-means clustering as detailed later).  In this case, every symbol in $\mathbf{z}$, which we call $z\in\mathbb{C}$, can be found based on the Euclidean distance from the nearest symbol in the constellation $\{s_m\}_{m=0}^M$.  Thus we obtain a column vector $\mathbf{\hat m}^\star$ from $z$, which is comprised of the entries $\hat m^\star$ that fulfill the following relationship:%
\begin{equation}\label{eq:ml_detection}
\hat m^\star = \underset{m}{\arg \min}\, \vert z - s_m \vert, \qquad m \in \{0, 1, \ldots, M-1 \}.
\end{equation}
Furthermore, since every $m$ corresponds to one $I$/$Q$ symbol, we can also obtain the column vector $\mathbf{\hat x}$ the entries of which are the detected symbols.  Because every symbol represents a string of bits, the received codeword (including the padding and the CRC) can be recovered from $\mathbf{\hat x}$ for further analysis.

So far we have covered model-driven applications of Python in wireless communications.  Namely: construction of symbols, creation of a MIMO payload with CRC, creation of a channel, creation of additive noise, estimation of the channel using pilots, channel equalization, and symbol detection.  Next, we focus on a data-driven approach where the value of machine learning and artificial intelligence is demonstrated.  Three areas of interest are: unsupervised learning, supervised learning (including deep learning), and reinforcement learning. %

\section{Unsupervised Machine Learning Use Cases}\label{sec:unsupervised_ml}

\subsection{Clustering}
The use of $k$-means clustering\footnote[4]{The $k$ and the $\mathcal{C}$ in $k$-means here should not be confused with the $k$ and the $\mathcal{C}$ defined in Section~\ref{sec:system_model} which represent the number of bits per symbol in a constellation $\mathcal{C}$ of size $M$.} to perform symbol detection is the quintessential application of unsupervised learning in wireless communications.  In the code implementation of this case, a two-dimensional plane representing the $I$/$Q$ components of the complex-valued symbols is formed.  The $k$-means centroids $\mathcal{C}$ are initialized (and fixed) to these $M$ constellation symbols (this is a major difference from the common $k$-means algorithm).  Then a two-dimensional column vector $\mathbf{z}$ is constructed from the received symbols $z \coloneqq z_I + jz_Q$ as follows $\mathbf{z}^\top \coloneqq [z_I, z_Q]$.  The centroids in $\mathcal{C}$ are constructed in a similar fashion.  Next, the symbols in presence of additive noise are grouped in a way that minimizes the Euclidean distance to a centroid:%
\begin{equation}\label{eq:kmeans}
m^\star = \underset{\mathbf{c}\in\mathcal{C}\colon\vert\mathcal{C}\vert = M}{\arg \min}\, \Vert \mathbf{z} - \mathbf{c} \Vert.
\end{equation}
which is essentially similar to the result from \eqref{eq:ml_detection}.

\subsection{Kernel Density Estimation}
The kernel density estimation (KDE) is a nonparametric estimate of a given density of a random variable through a group of kernels and smoothing parameters (or bandwidths).  Let us take the conditional density $f(\mathbf{y}\mid \mathbf{H}, \mathbf{x})$.  This density is equal to $f_\mathbf{n}(\mathbf{y}-\mathbf{Hx}) = \mathcal{N}_\mathbb{C}((\mathbf{y}-\mathbf{Hx}), \sigma^2_n \mathbf{I}_{N_r})$.  However, with the introduction of the quantizer, the density is no longer Gaussian since quantization is a non-linear operation. To estimate the density in presence of the quantizer  $f(\mathbf{y}_b\mid \mathbf{H}, \mathbf{x})$, either (1) an \textit{empirical} density can be constructed (i.e., through binning intervals, counting of samples per bin, and then dividing by the total), which does not introduce any density biases (though the number of bins can be a subjective choice) or (2) KDE can be applied which uses a kernel function (e.g., a Gaussian mixture) to generate a smooth curve.

Finding an optimal number of bins to fit an empirical density can be an arduous task and defending it later can be challenging due to the subjectivity.  Thus, resorting to a nonparameteric estimate of the density through KDEs has become more convenient\footnote[5]{Well, there are some of parameters that can be adjusted in KDEs.}. 

\textbf{Further Use Cases:}  In unsupervised learning, here are a few use cases: anomaly detection and root cause analysis through clustering \cite{9500204}, quantization using $k$-means, and wireless fingerprinting (i.e., distinguishing between devices sending similar messages) or localization using clustering algorithms \cite{9448961}.

\section{Supervised Machine Learning Use Cases}\label{sec:supervised_ml}

We define the learning task of supervised machine learning using a matrix of inputs in a design matrix format $\mathbf{X}$, a supervisory signal of a column vector $\mathbf{y}$, the predicted supervisory signal of a column vector $\mathbf{\hat y}$, and a set of hyperparameters $\boldsymbol\Theta$.  The multi-dimensional space spanned by the columns of $\mathbf{X}$ define the feature space.  The full data matrix is $\mathbf{M} \coloneqq [\mathbf{X} \mid \mathbf{y}]$.  To avoid confusion with the model-based development in Section~\ref{sec:model-driven}, any symbol used moving forward is expected to refer to this terminology unless explicitly referred to an equation from Section~\ref{sec:model-driven}.

Formally, a supervised learner minimizes a loss function $L(\mathbf{y}, \mathbf{\hat y}; \boldsymbol\Theta)$ through a search space defined by the hyperparameters.  An optimizer uses the training dataset to find the optimal $\boldsymbol\Theta^\star$ and the validation set is used as a benchmark to report on the loss function.  Formally:
\begin{equation}\label{eq:loss_function}
\underset{\boldsymbol\Theta}{\text{minimize:}}\qquad L \big (\mathbf{y}, \mathbf{\hat y} \coloneqq f_{\boldsymbol\Theta}(\mathbf{X}); \boldsymbol\Theta \big ),
\end{equation}
where $f_{\boldsymbol\Theta}(\mathbf{X})$ is a supervised learning model specific function.  Depending on the data type that $\mathbf{y}$ represents, we could have a ``regressor'' for continuous supervisory data $[\mathbf{y}]_i \in \mathbb{R}$ (an example of which would be an estimate of the channel state information $\mathbf{H}$ or a ``classifier'' for categorical supervisory data $[\mathbf{y}]_i \in \mathbb{Z}$, where the class labels have no quantitative significance (such as the beam identifier or the transmit antenna identifier).  Examples of optimizers that are used extensively today are the stochastic gradient descent (SGD) and the adaptive moments (Adam).  Details on SGD and Adam can be found in \cite{goodfellow}.

The process through which wireless systems data is obtained and training is performed is elaborate.  However, in short, there are two common approaches today: (1) \textit{learn-exploit-invalidate} where the base station obtains data from a group of users it serves for a period of time and then it trains the model for another period of time (defined by a duty cycle), after which the model is used for an exploitation period (i.e., inference) until the performance falls below a certain threshold, after which the model is invalidated and the cycle repeats and (2) \textit{periodic transfer learning} where training happens periodically at a central location (base stations either stream actual data to it) or by federated learning (where each base station only shares the deep learning weights it has computed locally and the serving base station aggregates these local updates).

There are several machine learning algorithms that can be used.  However, to make this brief and relevant, we choose two algorithms: linear regression and deep learning.  Also, we only provide detail for a single wireless communications use case each to avoid overwhelming the reader with too much information.
\subsection{Linear Regression}
\textbf{Channel Estimation:} Let us take the case of a MIMO system with two transmit antennas and two receive antennas (i.e., a $2\times 2$ channel).  In such a case, we can write \eqref{eq:system} using real-valued scalar variables as:
\begin{equation}
\begin{aligned}
        y_1 & = h_{11} x_1 + h_{12} x_2 + n_1 \\
        y_2 & = h_{21} x_1 + h_{22} x_2 + n_2 \\
\end{aligned}
\end{equation}
which means that we could learn $h_{11}, h_{12}$ using $x_1, x_2$ and $y_1$ as learning features.  Similarly we could learn $h_{21}, h_{22}$ using $x_1, x_2$ and $y_2$ as learning features.  The linear regression problem thus becomes $y_1 = \mathbf{h}_1^\top\mathbf{x} + \epsilon_1$, with $\epsilon_1$ sampled from a Gaussian distribution and $\mathbf{h}_1$ being a column vector representing the respective elements of the channel.  In other words, to learn the column vector $\mathbf{h}_1^\top \coloneqq [h_{11}, h_{12}]$, we need to construct a data matrix of measurements in the format $\mathbf{M} \in\mathbb{R}^{N_\text{SC}\times 3} \coloneqq [x_1, x_2 \mid y_1]$.  This is repeated for the imaginary part and the remaining signal elements in $\mathbf{y}$.  It can also be generalized for systems other than $2\times 2$.%

\begin{figure}[!t]
\centering
\resizebox{0.45\textwidth}{!}{\input{figures/DNN.tikz}}%
\caption{Fully connected deep neural network (depth $D=3$ and width $W=6$) of an input $N$ and an output $M$.}
\label{fig:fcdnn}
\end{figure}

\subsection{Deep Learning} 
Deep neural networks (DNN) have been used successfully to perform various wireless communications procedures.  Let our DNN have a depth $D$ and width $W$.  Since the majority of variables in wireless communications can be normalized (or scaled in the interval $[0,1]$), the non-linear activation function $f(\cdot)$ can be the sigmoid function, while other choices of an activation function such as the rectified linear unit can also be considered especially if certain variables are not scaled.  If every node in one layer is connected to every node in the next layer, the DNN is known as a fully connected DNN, as in Fig.~\ref{fig:fcdnn}.

\textbf{Symbol Detection:} Let us study the case where we would like to perform symbol detection from a constellation of size $M$ in the presence of noise.  The learning features thus are the various $I$/$Q$ symbols $\mathbf{x}$ and the symbol identifier $\{0,1,\ldots,M-1\}$.  If we separate the real $\mathbf{x}_r$ and imaginary part $\mathbf{x}_i$ of these symbols into two column vectors, we thus have a full data training matrix of $\mathbf{M}\in\mathbb{R}^{M\times 3} \coloneqq [\mathbf{X} \coloneqq [\mathbf{x}_r, \mathbf{x}_i] \mid \mathbf{y}]$.  However, for inference there are more number of symbols to label than the constellation size itself.  As a result, we have to increase the size of the training dataset.  Otherwise, the model is very likely to underfit (overfitting and underfitting are not discussed in this primer).  To make the model more robust, we introduce Normal distributed noise to $\mathbf{X}$ and repeat the same supervisory labels $\mathbf{y}$ for an augmented data matrix $\mathbf{\tilde M}\in\mathbb{R}^{qM\times 3}$, with $q$ being a small positive integer as a result of this perturbation step.  This leaves us with a data \textit{inference} matrix $\mathbf{M}\in\mathbb{R}^{N_\text{SC}N_r\times 3} \coloneqq [\text{Re}[\mathbf{y}],\text{Im}[\mathbf{y}] \mid \mathbf{m}]$ which is constructed from all the data points from all receive antennas as in Section~\ref{sec:system_model}.  If a split for training and test data is required for the augmented data matrix, it must be large enough for the random split not to ruin the adequate representation of symbol identifiers.

Given the nature of the supervisory signal (i.e., representing categories that do not have any quantitative significance), it is encoded into a binary class matrix for both the training and inference data matrixes.  Thus, a softmax activation function at the output layer of the DNN and a categorical cross-entropy loss function are required.

\textit{Loss functions:} Loss functions \eqref{eq:loss_function} are what the deep learning optimizer aims at minimizing.  Examples of loss functions are the (root) mean squared error or the mean absolute error are both used to measure how far the predicted supervisory signal $\mathbf{\hat y}$ is from the true supervisory signal $\mathbf{y}$.  Mathematically, 
$\mathsf{MSE} \coloneqq \frac{1}{M} \sum_{i=1}^M ([\mathbf{\hat y}]_i - [\mathbf{y}]_i) ^ 2$ and $\mathsf{MAE} \coloneqq \frac{1}{M} \sum_{i=1}^M \big \vert [\mathbf{\hat y}]_i - [\mathbf{y}]_i \big \vert$.  To keep track of the performance of a model as a result of optimizing the loss, performance measures, known as metrics, are reported.

\textbf{Equalization and Detection in a Rotation Channel:} Let there be a channel the task of which is to rotate complex-valued OFDM subcarriers by an angle $\theta$ and add some noise (which may not necessarily be Gaussian).  In other words, for $\nu$ transmitted OFDM subcarriers, we have a system model: $\mathbf{y} = e^{j\theta}\mathbf{I}_\nu\mathbf{x} + \mathbf{n}$.  Given that these symbols have a spatial relationship with one another, as known from their $I$/$Q$ constellation, the use of convolutional neural networks (CNNs) may help recover $\mathbf{x}$ from $\mathbf{y}$.  In other words, we train the CNN to perform the channel equalization and symbol detection together. 

\textit{CNNs:} CNNs have the ability to discern spatial relationships between learning features. 
 Because of that, if we construct the learning features such that the training data $\mathbf{M} \in\mathbb{R}^{\nu\times 4} \coloneqq [\mathbf{X}\mid \mathbf{y}] = [\text{Re}(\mathbf{y}), \text{Im}(\mathbf{y}) \mid \text{Re}(\mathbf{x}), \text{Im}(\mathbf{x})]$, then a CNN can likely exploit the spatial relationship between the real and imaginary part of both the input and output symbols through the use of convolutional layers with kernel sizes equal to the number of learning features in $\mathbf{X}$, which equals $2$.  Also, to capture any spatial relationship across symbols, we set the stride value to $1$ so the kernel moves one symbol component per convolution step.  CNNs can also be useful in problems related to multi-user beamforming strategies for massive MIMO, reflective intelligent surfaces, scheduling, and more.
 
\textit{Reproducibility:}  Unlike the feedforward deep neural networks, CNNs have issues related to reproducibility, even with the seed being fixed.  Therefore, it is a good practice to save the model weights once a satisfactory result is obtained.

\textbf{Beam Prediction in Trajectory:} If a user's trajectory follows a street or any path with many users on it, then the construction of a time series that can predict the next SINR-optimal beam from a codebook of beamforming vectors (or a grid of beams) becomes a possibility through the use of recurrent neural networks (RNNs).  RNNs can process sequential data and are thus suitable for time series.  We examine the long short-term memory (LSTM) type of RNN here.

\textit{LSTMs:}  LSTMs can remember information for an extended period larger than that of RNNs, making them suitable for our problem.  LSTMs comprise multiple elements: (1) memory cell state, which is generally common in RNNs (2) forget gate and (3) input gate. The memory cell states simply record information. The input gate is used to update the memory cell state. The forget gate can learn to reset the state of the
memory cell when the stored information is no longer needed (i.e., invalidation of the memory cell). If we denote vectors of the candidate memory cell (i.e., new information) as $\mathbf{\tilde{c}}_t$, the forget gate as $\mathbf{f}_t$, the input gate as $\mathbf{i}_t$, then we can represent the impact of the forget and input gate on the memory cell output as follows: $\mathbf{c}_t \coloneqq \mathbf{c}_{t-1} \odot \mathbf{f}_t + \mathbf{\tilde{c}}_t \odot \mathbf{i}_t$.
 
\textit{Feature engineering:} Feature engineering for time series can be in general a bit tricky compared to regression or classification that is not time-based.  Let our full data matrix be $\mathbf{M} \coloneqq [\mathbf{X}\mid\mathbf{y}]$ which captures historical data of $N$ learning features over $T$ time steps, each time step has $M$ records.  Thus, $\mathbf{M}\in\mathbb{R}^{MT\times N}$.  Specifically, for beam prediction $\mathbf{X}$ captures the time stamp, the user identifier, the received power, and the received SINR.  Optionally, it can also include the longitude and latitude coordinates of each user.  These dictate the dimension of the learning features $N$.  Also, $\mathbf{y}$ is the target variable that captures the beam identifier corresponding to the highest SINR at that \textit{time} for that user.  With a sufficiently explained $\mathbf{M}$, we quickly go over a few feature engineering techniques suitable for time series to improve the predictability:
\begin{enumerate}
    \item Shifting and Lagging: A timeshift is applied block-wise with past and future shift
values on $\mathbf{M}$. That is, the dataset becomes $[\mathbf{M}_{t-k} \mid \ldots \mid \mathbf{M} \mid \ldots \mid \mathbf{M}_{t+\ell}]$, with $k$ and $\ell$ being the lag and lead shifts respectively.  The lead time $\ell$ defines the prediction horizon which is how many time steps in the future is the LSTM expected to predict.  This shifting and lagging operation leads to an additional $k + \ell$ columns in the dataset.  The undefined values as a result of these operations can be filled with the last known value to avoid reducing row count.
    \item Differencing: The differencing operation reduces trend in the time series and makes the series more stationary.  Let us define the $j$-th order difference as a column vector $\boldsymbol\Delta_{t-j} := \mathbf{x}_t - \mathbf{x}_{t-j}, j\in \{1,2,3,\ldots, N\}$ and apply it column-wise on the learning features $[\boldsymbol\Delta_{t-k} \mid \ldots \mid \mathbf{0} \mid \ldots \mid\boldsymbol\Delta_{t+\ell}]$.
\end{enumerate}
Thus an engineered data matrix $\mathbf{\tilde M}$ is constructed as %
\begin{equation}
    \mathbf{\tilde M} \coloneqq [\boldsymbol\Delta_{t-k} \mid \mathbf{M}_{t-k} \mid \ldots \mid \mathbf{0} \mid \mathbf{M}_t \mid \ldots \mid\boldsymbol\Delta_{t+\ell} \mid \mathbf{M}_{t+\ell}],
\end{equation} %
which is a matrix in $\mathbb{R}^{MT\times (2N(k+\ell) + 1)}$. This matrix is then partitioned into $\mathbf{\tilde M} = [\mathbf{X} \mid \mathbf{y}_\ell]$, with $\mathbf{y}_\ell$ being the beam identifier at a single future lookahead value\footnote[6]{Actually, forecasting multiple time steps is also possible either as a single shot or through an autoregressive approach where one prediction is made at a time and then fed back to the model.} equal to the prediction horizon of interest (also known as the ``offset'') $\ell$.  These partitions are further converted into tensors having the shape (timesteps, record, and features).

There are more feature engineering techniques that can further be applied to time series.  For example, besides the scaling part (which we have mentioned earlier) and applying $\log(\cdot)$ to positive-valued features, the application of FFT onto a learning feature $x(t)$ extract frequencies that have high information content (e.g., as measured by the power spectral density $\vert X(f)\vert ^2$) can be often useful.  In particular, in transforming the learning feature to a sinusoid $\varphi_k \colon x(t) \mapsto \cos(2\pi f_k x(t))$, FFT can help determine which frequencies $f$ should be used in constructing this transformation.  There are some nuances with using FFTs: First, if the periodicity of this signal $x(t)$ is $T_s$, then the frequency of this signal is $1/T_s = F_s$.  Also, if the signal is aperiodic, then a window function can be applied first.  Second, if the FFT has $N_\text{FFT}$ points, then $\Delta f \coloneqq F_s / N_\text{FFT}$.  Third, the computation of $f_k = k\Delta f, k \in \{0,1,\ldots, N_\text{FFT} - 1\}$ which corresponds to the frequencies in the interval $[-F_s/2, F_s/2]$.  Finally, plotting the power spectral density versus frequency provides an indication which frequencies $f_k$ have the highest information content and thus should be used in the transformation $\varphi_k$ (there could be several suitable $k$'s). 

\textit{Training-test split:}  Different from random splitting, a time series split of the engineered data matrix $\mathbf{\tilde M}$ has two types of constraints it has to fulfill: (1) no randomness hence the sequence (order of data within column vectors) is undisturbed and (2) the split must abide by the time boundaries (e.g., if the measurements are over a radio frame, then splits can only happen at the boundary of whole radio frames).  Therefore, for a training split ratio $r_\text{training}$, the first  $\lfloor \lfloor r_\text{training} MT \rfloor / T \rfloor T$ rows belong to the training set while the remainder belongs to the test dataset.  Notice that we cannot set $r_\text{training}$ arbitrarily as it has a time boundary it is adjusted to.

\textit{Reproducibility:} Similar to CNNs, LSTMs also have problems with reproducibility.

\textbf{Channel State Information Compression:}  Since different OFDM subcarriers experience correlated fading despite the different frequencies \cite{10706622,8938771}, applying parts of the CSI towards other OFDM subcarriers while removing redundant other parts or even exploiting the channel sparsity (i.e., effectively ``compressing'' the channel) is beneficial in reducing the control plane overhead, especially for users that are in high SNR environments.  The compression ratio $0 \le \kappa < 1$ is the ratio that determines the latent layer dimension and is agreed upon by both the UE and BS: A smaller latent dimension enables transmission of less bits.  The UE compresses the channel using the encoder to a dimension $D\coloneqq N_rN_t\lceil(1 - \kappa)N_\text{SC}\rceil$ while the BS decompresses it and reconstructs the CSI---with loss---using the decoder.

\textit{Autoencoders:} Deep autoencoders are a ``self-supervised'' learning technique that uses deep neural networks to efficiently learn how to compress and encode data.  They then learn to reconstruct the data back from the compressed encoded representation to another that is close to the original input (hence self-supervised).  Autoencoder compression is a lossy compression technique since the reconstructed data is not equal to the original.  Autoencoders are comprised of three different layers: an encoder, a latent layer, and a decoder.

\textbf{Further Use Cases:}  In supervised learning, other types of learners not covered in this primer release can be explored such as ensemble learners.  Additionally, more use wireless cases can be explored such as predicting inter-band handover success \cite{8403587, bandswitching} which has today become 3GPP Rel-16 conditional handovers, signal denoising using autoencoders, improving the conditions in which joint transmission can be triggered in CoMP \cite{8665922} or cell-free MIMO, and predicting an optimal transmit beam per users in a trajectory using LSTMs \cite{10122528}.


\section{Reinforcement Learning Use Cases}\label{sec:reinf_learning}

\begin{figure}[!t]
\centering
\resizebox{0.55\textwidth}{!}{\input{figures/reinf.tikz}}%
\caption{Actions, states, and rewards: reinforcement learning.}
\label{fig:reinf_learning}
\end{figure}

In the case where training data is not in a data matrix format (or simply unavailable), a policy (e.g., of a game) can be used to train an agent to ``win'' by learning a reward-maximizing policy.  A policy $\pi(\cdot\mid \cdot)$ maps the state-action space to a probability from the vantage point of the agent.  Formally:
\begin{equation}
    \pi(a \in \mathcal{A} \mid s \in \mathcal{S}) \coloneqq \mathbb{P}[A=a\mid S=s] \colon \mathcal{S} \times \mathcal{A} \rightarrow [0, 1].
\end{equation}

In reinforcement learning, an agent (an algorithm) interacts with the environment (a wireless network) through an action $a\in\mathcal{A}$ at a time step $t$ towards the environment for the environment to return an instantaneous reward $R\in\mathbb{R}$ in return and move into a new observation state $s\in\mathcal{S}$ as a result of this action.  This interaction is outlined in Fig.~\ref{fig:reinf_learning}.  To formulate a problem as a reinforcement learning problem, a Markov decision process (MDP) is defined.  MDPs are a tuple of five elements: actions, states, rewards, probability of state transitions as observed by the environment, and a reward discounting factor.  MDPs assume the Markov property:  The future state $s^\prime$ depends only on the current state $s$ and action $a$ and not on the sequence of past states (i.e., memoryless).  Therefore, the probability of state transitions is $\mathbb{P}[S^\prime = s^\prime\mid S=s, A=a]$.

Since the true policy $\pi(\cdot\mid\cdot)$ is not always revealed or known, an ``off-policy'' algorithm known as $Q$-learning \cite{sutton2018reinforcement} is used to implicitly learn the true policy or find the optimal action-value function for the policy $\pi$ written as $Q^\star_\pi(s,a)$.  This can be done through several ways, but the $\epsilon$-greedy method is popular.  In this approach, a choice between exploration (random search) and exploitation (reuse of learning) is made with a probability $\epsilon$, which is decayed every new learning episode as the agent learns more about the environment, making exploration less likely.  During every episode, the $Q$-learner improves its ``experience'' as measured by the action-value function $Q_\pi(s,a)$.   The ultimate goal is for $Q$-learning to learn an optimal representation of the policy.  That is, find $Q_\pi^\star(s, a)$ which represents an optimal strategy to choose an action $a^\prime$ that maximizes the expected future discounted reward given the current state $s$ and action $a$.  Mathematically:
\begin{equation}\label{eq:bellman}
    Q^\star_\pi(s,a) = \mathbb{E}_{s^\prime} \big [R + \gamma \max_{a^\prime} \, Q^\star_\pi(s^\prime, a^\prime) \,\big\vert\, s, a \big],
\end{equation}%
where $R$ is the present reward and $0 < \gamma < 1$ is the reward discounting factor.  This is known as the Bellman equation.

To solve this equation and compute this expectation, there are two approaches: One approach does this iteratively and hopes that as $i\to+\infty$ that $Q_i(s,a) = Q^\star(s,a)$.  Another approach uses a deep neural network as a universal function approximator and trains it so that $Q(s,a;\boldsymbol \Theta) = Q_\pi^\star(s,a)$.  This is the essence of $Q$-learning with deep neural networks, also known as the deep $Q$-network (DQN).  The difference between the action-value function at time $t$ and that of the optimal action-value function is known as the ``regret.''  In other words, the regret at the $i$-th time instant is $Q^\star(s,a) - Q_i(s,a)$ and as $i\to +\infty$, regret should ideally approach zero.

Thus, there are two flavors of $Q$-learning used today: tabular and deep.  We will focus our development only on DQNs.  The reason is because for the tabular $Q$-learning to function as desired, a discretization of the observation space into bins is necessary so one state represents vectors to bin indexes.  This prevents a good isolation between the agent and environment classes, since the agent has to be \textit{explicitly} aware of the environment bounds.  Thanks to neural networks, the vector of states can be used to train (and later infer) the ``best'' action without explicitly learning about these environment bounds.

While optimizing the loss function, neural networks can be stuck in local minima, oscillate, or even diverge.  This is why the idea of sampling experience from a replay buffer that stores the tuple $(s_t, a_t, R_t, s_{t+1})$ obtained at time $t$ can help since the behavior distribution is averaged over many of its previous states hence smoothing out learning and avoiding these oscillations or divergence during the gradient descent \cite{mnih2013playingatarideepreinforcement}.

In constructing an environment that represents the wireless network, we use Python's \verb|gym| library which allows us to divide the observation space into discrete or continuous states, necessary for $Q$-learning to be able to learn.  For this purpose, we build a power control algorithm using reinforcement learning.  The states $\mathcal{S}$ are discretizations of the environment observations which include the current received SINR, the received SINR after power control, and the power control command.  Power control commands can increase or decrease the power as dictated by the codebook $\Gamma \coloneqq \{-3, -1, 1, 3\}$ dB.  We define a target SINR which the power control aspires to achieve.  The instantaneous reward $R$ is computed in terms of the difference in SINR after the power control command was implemented.  The set of actions $\mathcal{A}$ has the following elements:
\begin{itemize}
    \item $a_0$ advances to the next value in the power control codebook, 
    \item $a_1$ retracts to the previous value in $\Gamma$, and 
    \item $a_2$ does nothing.
\end{itemize}

\textbf{Further Use Cases:}  Other problems that can be solved with reinforcement learning are joint power control and interference coordination \cite{8938771}, link adaptation improvement, resource scheduling, and optimal predictive beam allocation.



\section{Performance Measures}\label{sec:performance}
Two groups of performance measures in the source code are observed and defined below.  For the radio performance measures, the symbols refer to the model-based development in Section~\ref{sec:model-driven}.%
\subsection{Radio Performance Measures}
\begin{enumerate}
\item \textbf{Error vector:} Defined as the difference of the vectorized channel estimate from the true vectorized channel.  Therefore, $\mathbf{e} \coloneqq \mathbf{vec}(\mathbf{H}) - \mathbf{vec}(\mathbf{\hat H})$.
\item \textbf{Estimation mean squared error:} Defined as the square of the Euclidean norm divided by the number of elements in the channel matrix.  Thus $\mathsf{MSE} \coloneqq \Vert \mathbf{e}\Vert^2 / (N_rN_t)$.
\item \textbf{Information bit rate ($C$):} This is the payload size (in bits) divided by the transmit time interval (in seconds) for units of bits/s.  It is further related to the spectral efficiency $\mathsf{SE}$ as $C = B \cdot \textsf{SE} = \Delta f N_\text{SC} r\log_2 M$.
\item \textbf{Sum rate capacity:} This is a generalization of the information bit rate across multiple streams.  For example, in SM, we write the sum rate as $C = \sum_{j = 1}^{N_s} B_j\log_2(1 + \gamma_j)$.
\item \textbf{Bit error rate (BER):} Defined as the number of bits received incorrectly relative to the total number of bits transmitted across all antennas.
\item \textbf{Block error rate (BLER):} Defined at the receiver as the number of transmissions the CRC of which is incorrect divided by the total number of transmissions (Hence, $\mathsf{BLER} \coloneqq \mathbbm{P}[\mathcal{P}(\mathbf{x}) \oplus \mathcal{P}(\mathbf{\hat x}) \neq 0]$, with $\mathcal{P}(\cdot)$ being the CRC polynomial generator function).  Note that if this polynomial were set to all-ones, the $\mathsf{BLER}$ would become equal to $1 - (1 - \mathsf{BER})^{L_\text{codeword}}$ since the bit errors are independent and identically distributed.
\end{enumerate}

\subsection{Machine Learning Performance Measures}
These measures are meant for supervised learning only since the ground truth is known.  Let us call the ground truth $\mathbf{y}$ and the predicted value based on the machine learning prediction $\mathbf{\hat y}$.
\begin{itemize}
    \item \textbf{Accuracy:} Defined as the complement of the mean classification error: $\frac{1}{M}\sum_{i=1}^M \mathbbm{1}\big [ [\mathbf{y}]_i \neq [\mathbf{\hat y}]_i] \big ]$.  Note that accuracy can become meaningless if one category has rare occurrence since the classifier points to the majority class \cite{bishop}.
    \item \textbf{Precision and recall:} Used when the classification is imbalanced and the classes are not distributed equally or close enough.  The reader is invited to look up their definition in other references (e.g., \cite{goodfellow}).
\end{itemize} %

While these measures are easier to explain in a binary classification, it is more difficult---though still relevant---for multi-class classification.  There are other measures that can be used, such as the F1 score and the receiver operating characteristic area under the curve, which we invite the reader to look up.

For regression, the root MSE, MSE, and MAE quantities are all used as performance measures (besides being \textit{de jure} loss functions).   The \textit{normalized} MSE (NMSE), defined as the MSE divided by the variance of the ground truth $\sigma_y^2$, has gained popularity in recent publications despite being suitable for linear relationships.  In addition, the coefficient of determination ($R^2$) can also be used, though it is meaningful in linear regression. Otherwise for non-linear relations, imposing a high bias model (in our case a linear relationship) can cause this quantity to be negative.  Definition of bias and the bias-variance trade-off can be found in several references (e.g., \cite{bishop}).  There are suitable measures for non-linear relationships such as the maximal information coefficient, which the reader is also invited to look up.

\section{Using the Code}\label{sec:using_code}
\begin{table}[!t]
\begin{adjustwidth}{-0.5in}{0cm}
\centering
\setlength\doublerulesep{0.5pt}
\caption{System Parameters}
\vspace*{-1.5em}
\label{tab:parameters}
\footnotesize
\begin{tabular}{p{0.3\linewidth}p{0.42\linewidth}m{0.28\linewidth} } 
\hhline{===}
Parameter & Description & Python \\
\hline
Random seed & This is the seed used by the pseudo-random number generator.  Helps with reproducibility of results. & \verb|seed = 42| \\
Number of transmit antennas ($N_t$) & The number of rows in the channel $\mathbf{H}$. & \verb|N_t = 4| \\
Number of receive antennas ($N_r$) & The number of columns in the channel $\mathbf{H}$.  & \verb|N_r = 4| \\
Number of OFDM subcarriers per symbol ($N_\text{SC}$) & These subcarriers determine the transmission bandwidth.  & \verb|N_sc = 64| \\
Total transmit power [W] ($P_\text{total}$) & This is the transmit power across all antennas. & \verb|transmit_power = 1| \\
OFDM subcarrier spacing [Hz] ($\Delta f$) & Determines the OFDM transmission bandwidth and the OFDM subcarrier spacing. & \verb|Df = 15e3| \\
Center frequency [Hz] ($f_c$) & The center frequency of the transmit band.  Used for the large-scale fading and the computation of the DFT. & \verb|f_c = 1800e6| \\
Precoder ($\mathbf{F}$) & This is the channel precoder (identity, SVD, SVD and waterfilling, and DFT beamforming). & \verb|precoder = 'identity'| \\
Channel type & Specifies the channel model (CDL-A, CDL-C, CDL-E, Rayleigh, Ricean, and DeepMIMO \cite{Alkhateeb2019}). & \verb|channel_type = 'CDL-E'| \\
Path loss model & Selects a model for large scale fading. & \verb|pathloss_model = 'FSPL'|\\
Payload mode & Random or defined by an $8$-bit $32\times 32$ bitmap file. & \verb|payload_mode = 'random'| \\
Interference power [dBm] ($\sigma^2_i$) & This implements an additive Gaussian interference source at the receiver. & \verb|interference_power_dBm = -105| \\
Interference probability ($p_\text{interference}$) & Probability of the interference occurring on a subcarrier. & \verb|p_interference = 0.00| \\
Constellation ($\mathcal{C}$) & This is the modulation scheme of the OFDM subcarriers. & \verb|constellation = 'QAM'| \\
Constellation size ($M$) & Number of symbols in the constellation.  Must be a power of two. & \verb|M_constellation = 16| \\
Maximum number of transmissions & The number of full $N_t\times N_\text{SC}$ OFDM subcarriers to be transmitted (this can be converted to payload size in bits). & \verb|max_transmission = 700| \\
CRC generator $\mathcal{P}(\mathbf{x})$ & A string of bits that represent the CRC generator polynomial. & \verb|crc_polynomial = 0b1100_0100| \\
Channel compression ratio ($\kappa$) & The ratio at which the channel $\mathbf{H}$ is compressed at the transmitter before it is sent to be reconstructed at the receiver (set to zero for no compression).  & \verb|channel_compression_ratio = 0.00| \\
Number of pilot symbols ($n_\text{pilot}$) & Length of pilots sequence required for channel estimation (at least equal to $N_t$). & \verb|n_pilot = None| \\
Transmit SNR [dB] ($\rho$) & The transmitted signal SNR. & \verb|transmit_SNR_dB = [-10, -5,| \\
& & \verb|   0, 5, 10, 15, 20, 25, 30]| \\
Channel estimation algorithm for MIMO channel & This specifies the algorithm used to estimate the transmit channel (i.e., perfect, LMMSE, and LS). & \verb|MIMO_estimation = 'perfect'| \\
Channel equalization algorithm & Specifies the algorithm used to equalize the estimated channel (i.e, MMSE and ZF). & \verb|MIMO_equalization = 'MMSE'| \\
Symbol detection algorithm & Specifies the algorithm used to detect the OFDM symbols transmitted (i.e., max likelihood, k-means, ensemble, and DNN). & \verb|symbol_detection = 'ML'| \\
Independent variable & A quantity from the radio performance measures in Section~\ref{sec:performance}. & \verb|x_label = 'EbN0_dB'| \\
Dependent variable & A quantity from the radio performance measures in Section~\ref{sec:performance}. & \verb|y_label = 'BER'| \\
\hhline{===}
\end{tabular}
\end{adjustwidth}
\end{table}%

\textbf{Libraries:}  Deepwireless can be installed through pip from a terminal window (e.g., UNIX shell or Command Prompt) as follows: %

{%
\centering
\verb|pip3 install deepwireless|
\par 
}%

Python 3.13 needs to be installed on the machine that uses deepwireless.  The more advanced user can download deepwireless and make changes from the GitHub location (Git needs to be installed on the same machine): %

{%
\centering
\verb|git clone https://github.com/farismismar/deepwireless|
\par 
}%

\textbf{Scenarios:}  The source code implementing the various use cases in this primer can be downloaded from \cite{github}.  There are two source files (in addition to the libraries above):  The first file \verb|scenario_one.py| implements the parameters in Table~\ref{tab:parameters} for a model-driven implementation of a wireless system.  The reader should try and adjust these parameters to develop a basic understanding and become familiar with the code.  The second file \verb|scenario_two.py| implements all the other use cases in the data-driven implementation (i.e.,  supervised learning, unsupervised learning, autoencoders, and reinforcement learning).  These two scenario files are found under the \verb|test| folder with other auxiliary files (i.e., \texttt{autoencoder.py}, \texttt{DQNLearningAgent.py}, \texttt{QLearningAgent.py}, and \texttt{environment.py}) being reusable with minimal modifications.

\textbf{Code Execution and Output:}  The code can be run either from a terminal window (e.g., \texttt{python3 \textless file\textgreater.py}), where \texttt{\textless file\textgreater.py} is a placeholder, or an integrated development environment (IDE).  IDEs offer several advantages over terminal windows such as debugging, faster code development, and display of plots inline.  The output of the run of the model-based part (i.e., \texttt{scenario\_one.py}) is two dataframes 
\verb|df_detailed_results| and \verb|df_results|.  These files contain the radio performance measures as defined in Section~\ref{sec:performance}.  

The run of the data-driven part \verb|scenario_two.py| has several outputs depending on the use case.  The CNN learning part returns the received noisy and rotated symbols, the transmitted symbols, and the denoised data after equalization using the CNN.  The LSTM learning part returns the predicted time series along with the model score.  The reinforcement learning part returns a group of variables: \verb|Q_values|, \texttt{losses}, \verb|optimal_episode|, \verb|optimal_reward|, 
\verb|optimal_environment_progress|, and \verb|optimal_action_progress|.  The last two variables are actually based on the actions and states which are dictated by the environment and the optimization objective.  We also show output plots in Figs.~\ref{fig:output1}-\ref{fig:output9} below.

\begin{figure}[H]
    \centering
    \resizebox{0.4\textwidth}{!}{\input{figures/constellation.tikz}}%
    \caption{$16$-QAM constellation with Gray code which for any two adjacent symbols only a change of one bit is permissible.}%
    \label{fig:output1}
\end{figure}

\begin{figure}[H]
    \centering
    \subfloat[BLER and BER vs $E_b/N_0$]{\resizebox{0.4\textwidth}{!}{\input{figures/performance.tikz}}}%
    \hfil
    \subfloat[BER vs $E_b/N_0$ with machine learning]{\resizebox{0.4\textwidth}{!}{\input{figures/symbol_detection.tikz}}}%
    \caption{Bit and block error rate performance curve (left) and bit error rate performance curve for three different algorithms (\text{right}).  The curves are for a MIMO system using $16$-QAM and the parameters as shown in Table~\ref{tab:parameters}.}%
    \label{fig:output2}
\end{figure}

\begin{figure}[H]
    \centering
    \subfloat[Without quantization]{\resizebox{0.35\textwidth}{!}{\input{figures/IQ_binf.tikz}}}%
    \hfil
    \subfloat[With quantization $(b = 3)$]{\resizebox{0.35\textwidth}{!}{\input{figures/IQ_b3.tikz}}}%
    \caption{Received signal constellation plot without (left) and with quantization at $b = 3$ (right).}%
    \label{fig:output3}
\end{figure}

\begin{figure}[H]
    \centering
    \subfloat[Gray-coded QPSK constellation]{\resizebox{0.32\textwidth}{!}{\input{figures/qpsk_constellation_constellation.tikz}}}%
    \hfil
    \subfloat[Rotation channel and noise effect]{\resizebox{0.32\textwidth}{!}{\input{figures/constellation_noisy_symbols.tikz}}}%
    \hfil
    \subfloat[After CNN channel equalization]{\resizebox{0.32\textwidth}{!}{\input{figures/constellation_equalized_symbols.tikz}}}%
    \caption{CNNs equalizing a rotation channel with the presence of shot noise (using QPSK constellation).  In this case, $\rho = 30$ dB and $\theta = \pi/24$.}%
    \label{fig:output4}
\end{figure}

\begin{figure}[H]
    \centering
    \resizebox{0.45\textwidth}{!}{\input{figures/pdf_empirical_sinr.tikz}}%
    \caption{The empirical and KDE-based PDF of the received SINR at $\rho = 30$ dB.}%
    \label{fig:output5}
\end{figure}%

\begin{figure}[H]
    \centering
    \subfloat[LSTM learning performance]{\resizebox{0.32\textwidth}{!}{\input{figures/history_keras_lstm.tikz}}}%
    \hfil
    \subfloat[CNN learning performance]{\resizebox{0.32\textwidth}{!}{\input{figures/history_keras_CNN_equalization_learning.tikz}}}%
    \hfil
    \subfloat[Autoencoder learning performance]{\resizebox{0.3\textwidth}{!}{\input{figures/history_keras_autoencoder_compression_0.25.tikz}}}%
    \caption{Deep learning performance: LSTM learning performance in proactive beam prediction (left), CNN learning performance in rotation channel equalization, and autoencoders learning performance for channel compression (right).}%
    \label{fig:output6}
\end{figure}%

\begin{figure}[H]
    \centering
    \subfloat[Original channel]{\includegraphics[width=0.33\textwidth]{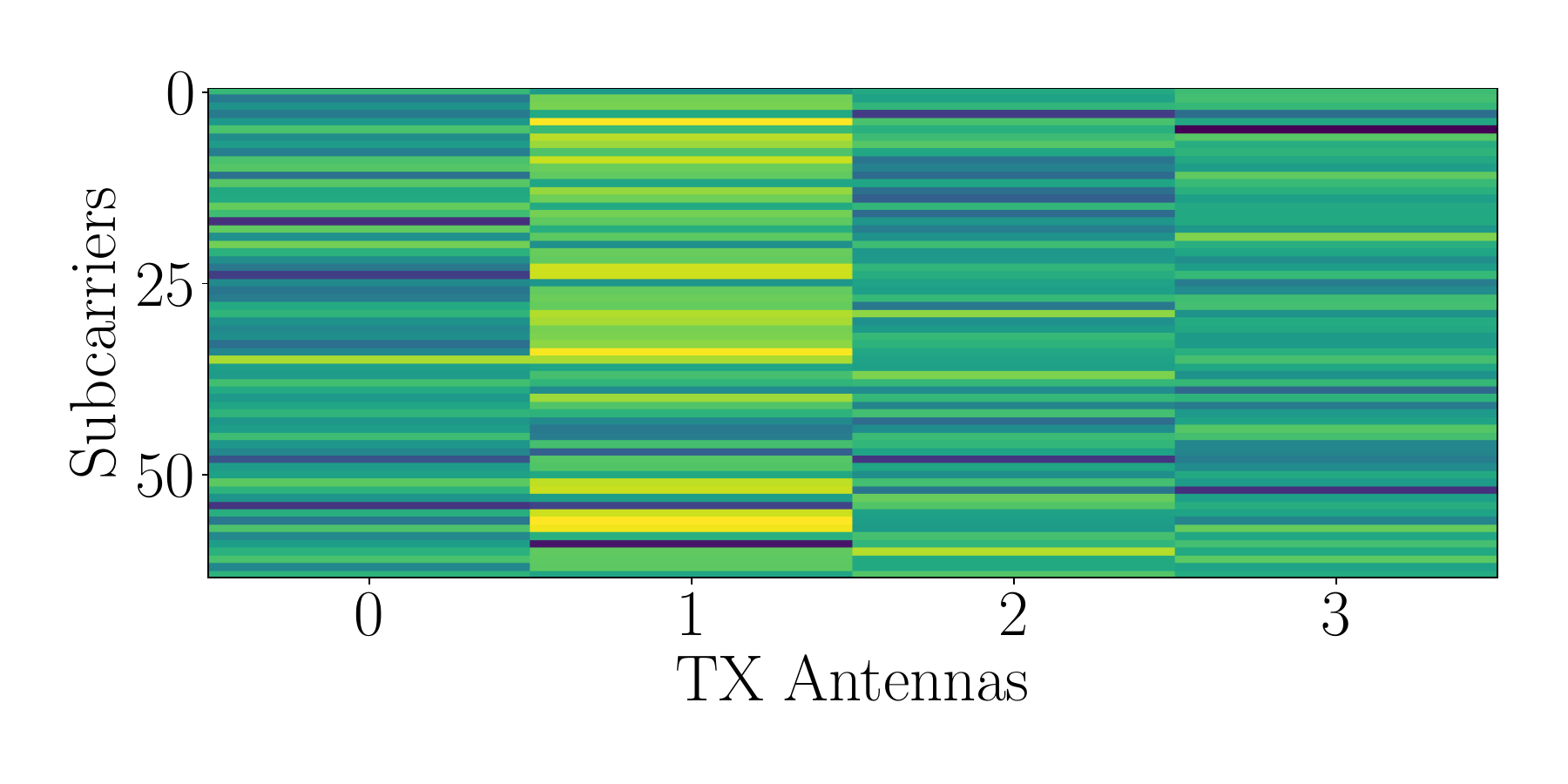}}%
    \hfil
    \subfloat[Compressed ($\kappa=0.25$) channel]{\includegraphics[width=0.33\textwidth]{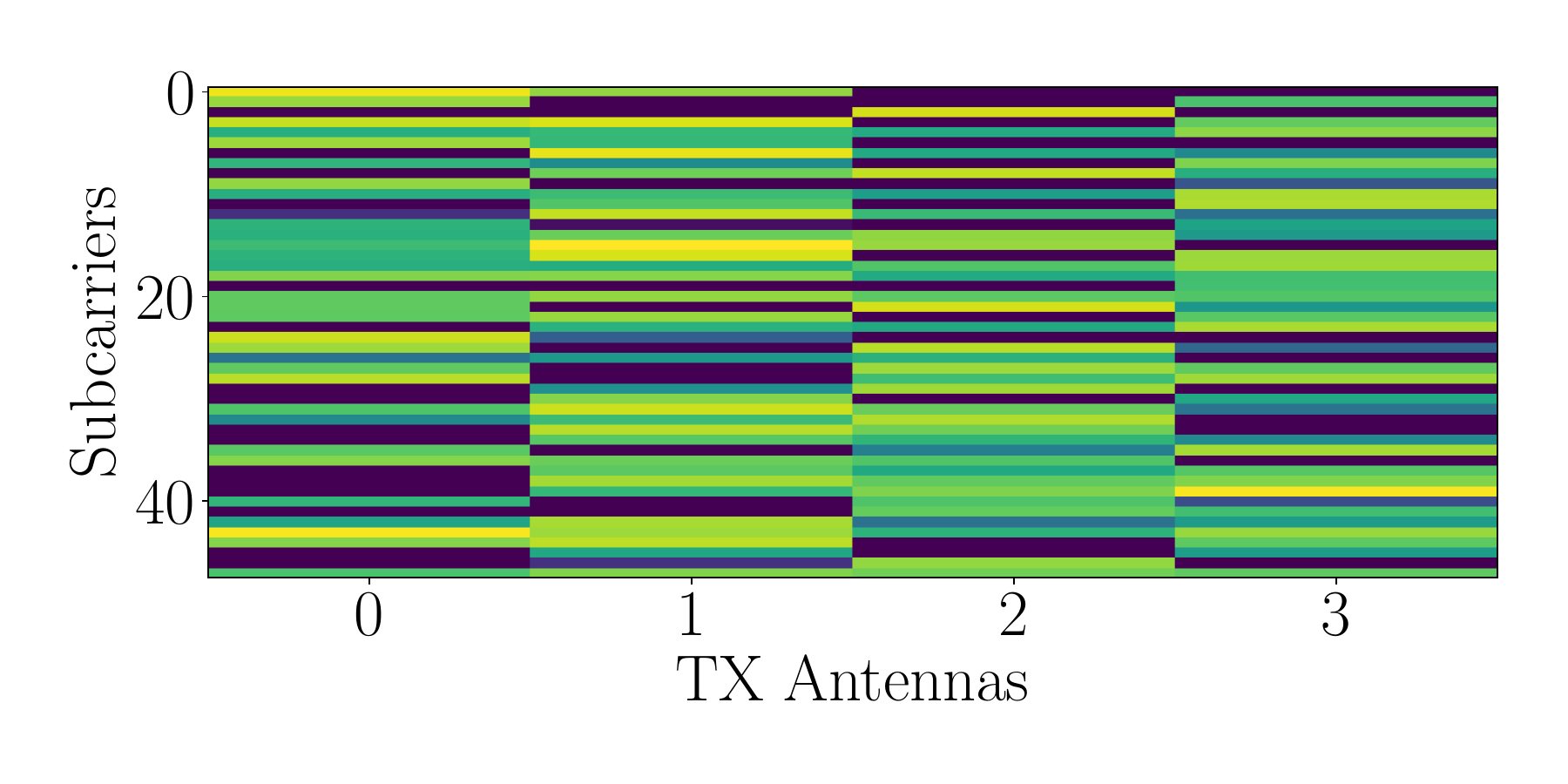}}%
    \hfil
    \subfloat[Reconstructed channel]{\includegraphics[width=0.33\textwidth]{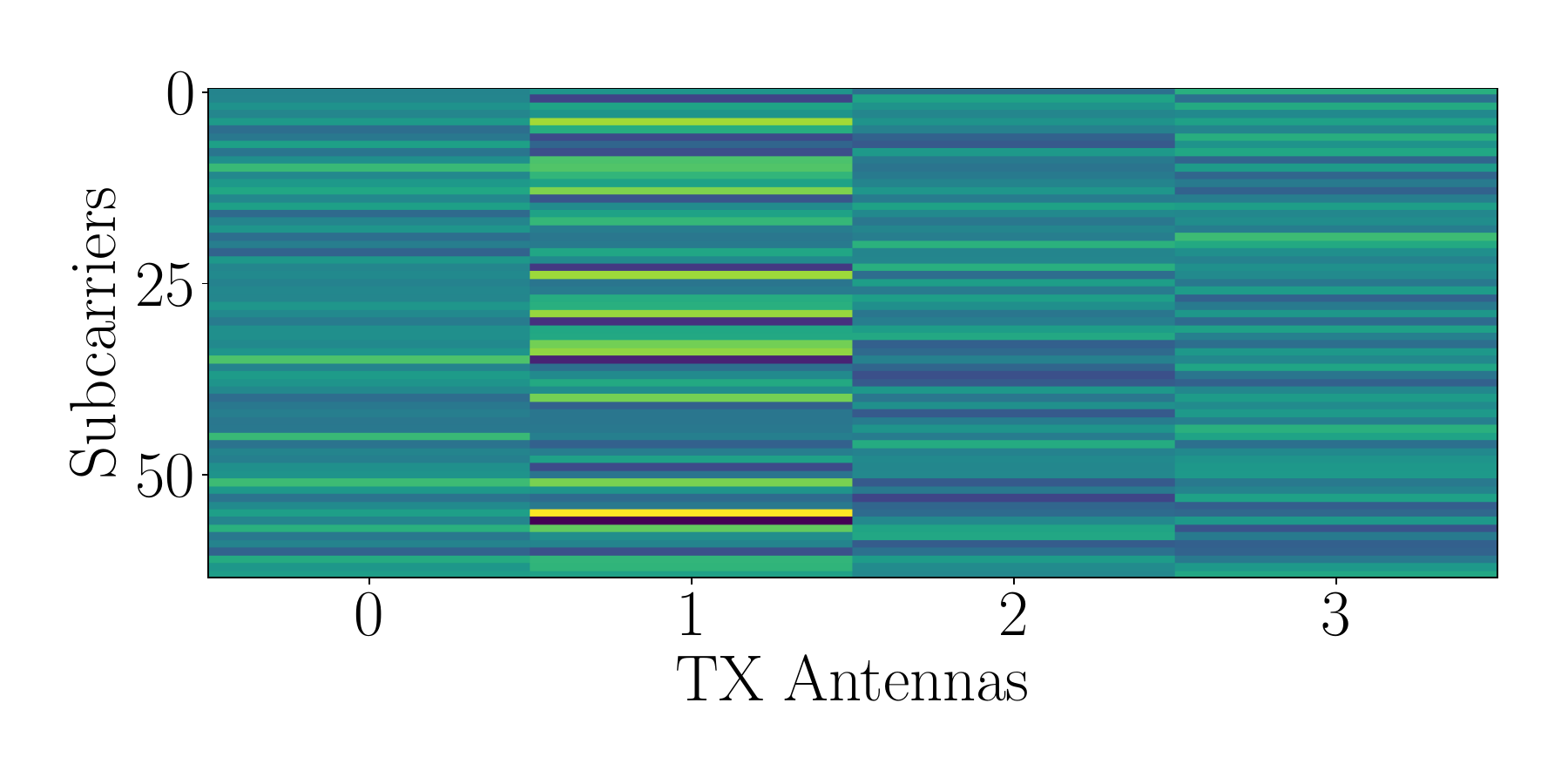}}%
    \caption{Channel compression of a CDL-E channel ($4\times 4$ MIMO and 64 OFDM subcarriers) and its reconstruction using deep autoencoders.}
    \label{fig:output7}
\end{figure}

\begin{figure}[H]
    \centering
    \subfloat[Tabular $Q$-learning]{\resizebox{0.35\textwidth}{!}{\input{figures/Qlearning_perf_tabular.tikz}}}%
    \hfil
    \subfloat[Deep $Q$-learning using deep $Q$ network]{\resizebox{0.35\textwidth}{!}{\input{figures/Qlearning_perf_dqn.tikz}}}%
    \caption{$Q$-learning using tabular (left) and deep learning (right).  The experience as measured by the value of $Q_\pi(s,a)$ goes up with more training episodes.}%
    \label{fig:output8}
\end{figure}

\begin{figure}[H]
    \centering
    \subfloat[Environment (tabular)]{\resizebox{0.2\textwidth}{!}{\input{figures/environment_SINR_dB_tabular.tikz}}}%
    \hfil
    \subfloat[Environment (deep)]{\resizebox{0.2\textwidth}{!}{\input{figures/environment_SINR_dB_dqn.tikz}}}%
    \hfil
    \subfloat[Actions (tabular)]{\resizebox{0.2\textwidth}{!}{\input{figures/actions_tabular.tikz}}}%
    \hfil
    \subfloat[Actions (deep)]{\resizebox{0.2\textwidth}{!}{\input{figures/actions_dqn.tikz}}}%
    \caption{The result of the interaction between the environment and the tabular and deep $Q$-learning algorithms.}%
    \label{fig:output9}
\end{figure}

\section{Further Topics}\label{sec:further}

Congratulations on making it this far into this primer!  We hope that it has kindled in you the desire to walk this path in your learning or research.  For this purpose, we leave a few more topics for you to consider as you probe further.

\begin{itemize}
    \item Generative Adversarial Networks (GANs):  GANs can also be applied in the wireless space.  Pilot-free channel estimation out of band is a good example.  Using measurements of an existing band (e.g., a heatmap) or the CSI, and a few measurements of an out of band spectrum, can a GAN generate the entire heatmap for the measurements (or the CSI) of this out of band carrier?
    \item Large Language Models (LLMs): While mostly text-based, LLMs have demonstrated an ability to provide insights that can be used towards intelligent automation with human supervision.  For example, generation of corrective actions (e.g., configuration scripts) for the network operation center to use as a solution for a network fault event.
\end{itemize}

These two topics fall into what is collectively known today ``Generative AI'' (as opposed to discriminative AI).  In the space of wireless communications, here are a few more topics to consider as next steps:
\begin{itemize}
    \item Multiple OFDM symbols:  Currently a single OFDM symbol is used in the simulator.  Realistic systems have more OFDM symbols per time slot.  For example, both 4G LTE and NR have $7$ OFDM symbols per timeslot ($0.5$ ms).
    \item Forward error correction can further improve the system performance due to coding gain. LDPC outperforms Turbo codes at high code rates, while Turbo codes outperform LDPC when the code rate is low.
    \item Multiple concurrent users to enable SDMA or multi-user MIMO (potentially with mobility and Doppler effect addressed).
    \item Link adaptation:  Optimize coding schemes and modulation to improve the BLER curve performance beyond current standards.
\end{itemize}


\section{Conclusion}\label{sec:conclusion}

In this primer, we summarized the various concepts required to build a wireless communication prototype that supports a single OFDM symbol MIMO system.  We showed how the prototype can be implemented in Python using the deepwireless library and showed a few use cases of machine learning (supervised, self-supervised, and unsupervised) and reinforcement learning across few layers in the air interface protocol stack.  As this is our final issue, we hope that both clarity and flow have made reading this primer beneficial to the reader.  Feedback is always welcome.
\newpage
\bibliographystyle{IEEEtran}
\bibliography{main.bib}


%








\end{document}

%% file: figures/stack.tikz
\begin{tikzpicture}[node distance=0cm,outer sep=0pt,inner sep=2pt]

\newcommand{\controlplane}{{"PHY","MAC","RLC",".\ .\ .","RRC"}}

\foreach \x in {1,...,5} {
        \node[draw, fill=gray!0,minimum width=2.5cm, minimum height=1cm] (m-\x) at (\x*3, -2.5) {\pgfmathparse{\controlplane[\x-1]}\pgfmathresult};
};

\node[draw, below of=m-3, yshift=2.5em, fill=gray!30,minimum width=14.5cm, minimum height=0.5cm] {RRM};

\end{tikzpicture}%

%% file: figures/overall.tikz
\begin{tikzpicture}[style=thick, node distance=3cm, scale=3, >=latex]

    \node [coordinate, name=input] {};
    \node [rectangle, draw, right of=input, line width=0.75mm, node distance=1.5cm,
            text width=3em, text centered, minimum height=2em] (precoder) {$\mathbf{F}$};
    
    \node [rectangle, draw, right of=precoder, minimum width=6em, node distance=2.5cm,
            text width=4em, text centered, minimum height=2em] (channel) {Channel Effect};
    \draw [draw,->] (input) -- node[left, xshift=-0.75em] {$\mathbf{x}$} (precoder);
    \draw [draw,->] (precoder) -- node {} (channel);

    \node [rectangle, draw, below of=channel, node distance=1.5cm, line width=0.75mm, text width=5em, text centered, minimum height=2em] (estimator) {Estimation}; 
    \node [coordinate, left=of estimator.165, xshift=4.5em] (pilot_in) {};
    \draw [draw,->] (pilot_in) -- node [left, xshift=-1.9em] {$\mathbf{X}_\mathbf{p}$} (estimator.165);
    \node [coordinate, left=of estimator.195, xshift=4.5em] (pilot_out) {};
    \draw [draw,->] (pilot_out) -- node [left, xshift=-1.9em] {$\mathbf{Y}_\mathbf{p}$} (estimator.195);

    \draw [draw,->] (channel) -- node {} (estimator);

    \node [circle, draw, right of=channel, node distance=2.5cm,
    text width=1em, text centered, minimum height=1em] (adder) {$+$};
    \draw [draw,->] (channel) -- node {} (adder);

    \node [circle, draw, right of=adder, node distance=1.5cm, line width=0.75mm,
        text width=1em, text centered, minimum height=1em] (adder2) {$+$};
    \draw [draw,->] (adder) -- node {} (adder2);

    \node [coordinate, above of=adder, node distance=2cm, name=noiseinput] {};
    \draw [draw,->] (noiseinput) -- node[above, yshift=2.3em] {$\mathbf{n}$} (adder);

    \node [rectangle, draw, line width=0.75mm, right of=adder2, node distance=2cm, 
        text width=2.5em, text centered, minimum height=2em] (quantizer) {$Q_b(\cdot)$};
    
    \draw [draw,->] (adder2) -- node[above] {$\mathbf{y}$} (quantizer);

    \node [coordinate, above of=adder2, node distance=2cm, name=interf] {};
    \node at (interf) [above] {$\mathbf{i}$}; 
    
    \tikzstyle{myswitch}=[closing switch,line width=0.3] 
    \draw (interf) to[myswitch] node[right, yshift=1.5em, xshift=0.45em] {$p_\text{interference}$} (adder2) {};
         
    \node [rectangle, draw, line width=0.75mm, right of=quantizer, node distance=3cm, 
        text width=2.5em, text centered, minimum height=2em] (combiner) {$\mathbf{G}$};
    
    \draw [draw,->] (quantizer) -- node[above] {$\mathbf{y}_b$} (combiner);

    \node [rectangle, draw, right of=combiner, 
        text width=2.5em, text centered, minimum height=2em] (equalizer) {$\mathbf{W}$};

    \draw [draw,->] (combiner) -- node[above] {$\mathbf{\tilde y}$} (equalizer);
    
    \node [rectangle, draw, right of=equalizer, node distance=6cm,
        text width=5em, text centered, minimum height=2em] (detection) {Symbol\\Detection};

    \draw [draw,->] (equalizer) -- node[above] {$\mathbf{z}\coloneqq\mathbf{W}\mathbf{\tilde y}  = \mathbf{\hat x} + \mathbf{\tilde v}$} (detection);
    
    \node [coordinate, right of=detection, name=output] {};
    \draw [draw,->] (detection) -- node[above] {$\mathbf{\hat x}$} (output);

    \draw [->] (estimator.0) -- ++ (0,0) -| node[above, pos=0.27] {$\mathbf{\hat H}$} (equalizer);
    

\end{tikzpicture}%

%% file: figures/DNN.tikz
\begin{tikzpicture}[thick, scale=0.5, node distance=2em, >=latex, cnode/.style={draw=black,fill=#1,minimum width=0.3cm,circle}]

    \foreach \x in {1,...,4}
    {
      \pgfmathparse{\x == 3 ? "\vdots" : "x_\x"}
       \pgfmathparse{\x== 4? "x_{N}" : "\pgfmathresult"}
        \node[cnode=gray!10,label=180:${\pgfmathresult}$] (x-\x) at (0,{-2*\x}) {};
    }
    
    \foreach \x in {1,...,6}
    {
        \node[cnode=gray!30] (h1-\x) at (3,{-2*\x+2}) {}; %
        \node[cnode=gray!30] (h2-\x) at (6,{-2*\x+2}) {};
        \node[cnode=gray!30] (h3-\x) at (9,{-2*\x+2}) {};
    }

    \foreach \x in {1,...,3}
    {
        \pgfmathparse{\x == 2 ? "\vdots" : "y_\x"}
        \pgfmathparse{\x== 3? "y_{M}" : "\pgfmathresult"}
        \node[cnode=gray!60,label=0:${\pgfmathresult}$] (y-\x) at (12,{-2*\x-1}) {}; %
    }

    
    \foreach \x in {1,...,4}
    {   
        \foreach \y in {1,...,6}
        {  
            \draw (x-\x) -- (h1-\y); 
        }
    }
  
    \foreach \x in {1,...,6}
    {   
        \foreach \y in {1,...,6}
        {  
            \draw (h1-\x) -- (h2-\y);
            \draw (h2-\x) -- (h3-\y);
        }
    }

    \foreach \x in {1,...,6}
    {   
        \foreach \y in {1,...,3}
        {  
            \draw (h3-\x) -- (y-\y);
        }
    }

    
\end{tikzpicture}

%% file: figures/reinf.tikz
\begin{tikzpicture}[style=thick, node distance=3cm, scale=1, >=latex]
    \node [rectangle, draw, text width=12em, text centered, rounded corners, minimum height=2em] (Agent) {Algorithm (Agent)};
    \node [rectangle, draw, 
    text width=8em, text centered, rounded corners, minimum height=1em, node distance=1.5cm, below of=Agent] (Environment) {Wireless Network (Environment)};
    \path [draw, -latex] (Agent.0) --++ (1em,0em) |- node [text width=5em,near start,yshift=-0.1cm,xshift=3em]{$a_t$} (Environment.0);
    \path [draw, -latex] (Environment.185) --++ (-5em,0em) |- node [text width=5em, xshift=0.2cm,yshift=0cm] {$s_{t+1}$} (Agent.175);
    \path [draw, -latex] (Environment.175) --++ (-4em,0em) |- node [text width=5em, xshift=1.1cm,yshift=-0.25cm] {$R_{t+1}$} (Agent.185);
    
    \draw [draw, -latex] (Environment.175) --++ (-2em,0em) node[above,xshift=1em] (r_t) {$R_t$};
    \draw [draw, -latex] (Environment.185) --++ (-2em,0em) node[below,xshift=1em] (s_t) {$s_t$};
    
    \draw[dashed, line width=0.5] (s_t) ++ (-1em,-0.1em) -- ++(0,2.1em);

\end{tikzpicture}

%% file: figures/constellation.tikz
\begin{tikzpicture}

\begin{axis}[
tick align=outside,
tick pos=left,
x grid style={white!69.0196078431373!black},
xlabel={$I$},
xmajorgrids,
xmin=-3.5, xmax=3.95,
xtick style={color=black},
xtick={-4,-3,...,4},
y grid style={white!69.0196078431373!black},
ylabel={$Q$},
ymajorgrids,
ymin=-3.5, ymax=3.5,
ytick style={color=black},
ytick={-4,-3,...,4},
]
\addplot [draw=black, fill=black, mark=*, only marks]
table{%
x  y
-3 -3
-3 -1
-3 1
-3 3
-1 3
-1 1
-1 -1
-1 -3
1 -3
1 -1
1 1
1 3
3 3
3 1
3 -1
3 -3
};
\draw (axis cs:-3,-3.1) node[
  scale=0.7,
  anchor=base west,
  text=black,
  rotate=0.0
]{0000};
\draw (axis cs:-3,-1.1) node[
  scale=0.7,
  anchor=base west,
  text=black,
  rotate=0.0
]{0001};
\draw (axis cs:-3,1.05) node[
  scale=0.7,
  anchor=base west,
  text=black,
  rotate=0.0
]{0011};
\draw (axis cs:-3,3.05) node[
  scale=0.7,
  anchor=base west,
  text=black,
  rotate=0.0
]{0010};
\draw (axis cs:-1,3.05) node[
  scale=0.7,
  anchor=base west,
  text=black,
  rotate=0.0
]{0110};
\draw (axis cs:-1,1.05) node[
  scale=0.7,
  anchor=base west,
  text=black,
  rotate=0.0
]{0111};
\draw (axis cs:-1,-1.1) node[
  scale=0.7,
  anchor=base west,
  text=black,
  rotate=0.0
]{0101};
\draw (axis cs:-1,-3.1) node[
  scale=0.7,
  anchor=base west,
  text=black,
  rotate=0.0
]{0100};
\draw (axis cs:1,-3.1) node[
  scale=0.7,
  anchor=base west,
  text=black,
  rotate=0.0
]{1100};
\draw (axis cs:1,-1.1) node[
  scale=0.7,
  anchor=base west,
  text=black,
  rotate=0.0
]{1101};
\draw (axis cs:1,1.05) node[
  scale=0.7,
  anchor=base west,
  text=black,
  rotate=0.0
]{1111};
\draw (axis cs:1,3.05) node[
  scale=0.7,
  anchor=base west,
  text=black,
  rotate=0.0
]{1110};
\draw (axis cs:3,3.05) node[
  scale=0.7,
  anchor=base west,
  text=black,
  rotate=0.0
]{1010};
\draw (axis cs:3,1.05) node[
  scale=0.7,
  anchor=base west,
  text=black,
  rotate=0.0
]{1011};
\draw (axis cs:3,-1.1) node[
  scale=0.7,
  anchor=base west,
  text=black,
  rotate=0.0
]{1001};
\draw (axis cs:3,-3.1) node[
  scale=0.7,
  anchor=base west,
  text=black,
  rotate=0.0
]{1000};
\end{axis}

\end{tikzpicture}

%% file: figures/performance.tikz
\begin{tikzpicture}

\begin{semilogyaxis}[
width=5in,
height=4in,
tick align=outside,
tick pos=left,
x grid style={white!69.0196078431373!black},
xlabel={Transmit $E_b/N_0$ [dB]},
xmajorgrids,
xmin=-7, xmax=25,
xtick style={color=black},
y grid style={white!69.0196078431373!black},
ylabel={Error Rate},
ymajorgrids,
ymin=1e-4, ymax=1,
scale=0.6,
yminorgrids,
ytick style={color=black},
legend cell align={left},
legend style={at={(0.98,0.98)}, fill opacity=0.6, draw opacity=1, text opacity=1, draw=white!80!black, nodes={scale=.8}},
]
\addplot [thick, black, opacity=0.7, mark=*, mark size=3, mark options={solid,fill=red,draw=black}]
table {%
-6.02059991327962 0.676666666666667
-1.02059991327962 0.446666666666667
3.97940008672038 0.25
8.97940008672037 0.0666666666666667
13.9794000867204 0.00666666666666667
};
\addlegendentry{BLER};

\addplot [thick, black, opacity=0.7, mark=square*, mark size=3, mark options={solid,fill=blue,draw=black}]
table {%
-6.02059991327962 0.177034120734908
-1.02059991327962 0.1099343832021
3.97940008672038 0.0555905511811024
8.97940008672037 0.0217158792650919
13.9794000867204 0.0062237532808399
18.9794000867204 0.00125328083989501
23.9794000867204 0.000121391076115486
};
\addlegendentry{BER};
\end{semilogyaxis}

\end{tikzpicture}

%% file: figures/symbol_detection.tikz
\begin{tikzpicture}

\begin{semilogyaxis}[
width=5in,
height=4in,
tick align=outside,
tick pos=left,
x grid style={white!69.0196078431373!black},
xlabel={Transmit $E_b/N_0$ [dB]},
xmajorgrids,
xmin=-7, xmax=25,
xtick style={color=black},
y grid style={white!69.0196078431373!black},
ylabel={Bit Error Rate},
ymajorgrids,
ymin=1e-4, ymax=1,
scale=0.6,
yminorgrids,
ytick style={color=black},
legend cell align={left},
legend style={at={(0.59,0.36)}, fill opacity=0.6, draw opacity=1, text opacity=1, draw=white!80!black, nodes={scale=.8}},
]
\addplot [line width=1.2, black, opacity=0.7, mark=*, mark size=2, mark options={solid,fill=red!60,draw=red}]
table {%
x y
-6.02059991327962 0.241440288713911
-1.02059991327962 0.19251312335958
3.97940008672038 0.141496062992126
8.97940008672037 0.0910433070866142
13.9794000867204 0.0481036745406824
18.9794000867204 0.0187959317585302
23.9794000867204 0.00518372703412073
};
\addlegendentry{DNN};

\addplot [line width=1.2, black, dashed, opacity=0.7, mark=o, mark phase=1, mark size=2, mark repeat=2, mark options={solid,fill=red,draw=black}]
table {%
-6.02059991327962 0.177050524934383
-1.02059991327962 0.110472440944882
3.97940008672038 0.0554494750656168
8.97940008672037 0.021240157480315
13.9794000867204 0.00633858267716535
18.9794000867204 0.00123031496062992
23.9794000867204 9.51443569553806e-05
};
\addlegendentry{Ensemble learning};

\addplot [line width=1.2, black, opacity=0.7, mark=square*, mark phase=2, mark repeat=2, mark size=2, mark options={solid,fill=blue!60,draw=blue}]
table {%
-6.02059991327962 0.177034120734908
-1.02059991327962 0.1099343832021
3.97940008672038 0.0555905511811024
8.97940008672037 0.0217158792650919
13.9794000867204 0.0062237532808399
18.9794000867204 0.00125328083989501
23.9794000867204 0.000121391076115486
};
\addlegendentry{K-means clustering};

\addplot [line width=1.2, black, opacity=0.7, mark=triangle*, mark phase=3, mark repeat=2, mark size=3, mark options={solid,fill=green!60,draw=green}]
table {%
-6.02059991327962 0.177034120734908
-1.02059991327962 0.1099343832021
3.97940008672038 0.0555905511811024
8.97940008672037 0.0217158792650919
13.9794000867204 0.0062237532808399
18.9794000867204 0.00125328083989501
23.9794000867204 0.000121391076115486
};
\addlegendentry{Maximum likelihood};

\end{semilogyaxis}

\end{tikzpicture}

%% file: figures/IQ_binf.tikz
\begin{tikzpicture}

\definecolor{color0}{rgb}{0.12156862745098,0.466666666666667,0.705882352941177}
\definecolor{color1}{rgb}{1,0.498039215686275,0.0549019607843137}
\definecolor{color2}{rgb}{0.172549019607843,0.627450980392157,0.172549019607843}
\definecolor{color3}{rgb}{0.83921568627451,0.152941176470588,0.156862745098039}

\begin{axis}[
tick align=outside,
tick pos=left,
x grid style={white!69.0196078431373!black},
xlabel={$I$},
xmajorgrids,
xmin=-8.34788922818466, xmax=8.72367643706217,
xtick style={color=black},
xtick={-10,-8, ..., 8},
y grid style={white!69.0196078431373!black},
ylabel={$Q$},
ymajorgrids,
ymin=-8.8937499463347, ymax=8.54077638718658,
ytick style={color=black},
ytick={-10,-8,-6,-4,-2,0,2,4,6,8,10},
legend columns=2,
legend cell align={left},
legend style={at={(0.52,0.98)}, fill opacity=0.6, draw opacity=1, text opacity=1, draw=white!80!black, nodes={scale=.8}},
]
\addplot [draw=color0, fill=color0, mark=*, only marks]
table{%
x  y
2.20462331292044 -0.496470357339567
-2.21553995157935 -2.36543355428095
-3.62350607689092 -3.49678477036686
0.830905900997968 0.776044711124545
-0.124521772785041 6.00963457012575
-4.73748716741297 -1.10518509546841
-1.91569894389066 0.559593128638593
-1.98712506992784 -5.53925219903892
3.54698428926185 1.5032898625328
-5.55036885225263 4.86598886495408
1.92123251809107 0.00540426789058479
-0.255432765061923 -4.39567675594098
-0.08479682285201 3.45338155358895
-0.986449285001263 4.59366280139902
0.724579540288393 1.99428923315306
-2.15823301442251 0.808543601595385
1.75372410393028 -1.63327987660223
-1.38615934356171 0.898842064225216
2.47158403225179 -0.527234308529175
-3.16839743478271 -5.03480149319587
-3.80413001143664 -4.22025138122978
0.493513626017084 -0.13382156610701
0.196151931105085 2.15719032852107
1.6237097349887 5.5890412112717
-2.19652652860925 1.80442099371716
-0.302834191171258 -0.151255901007524
-0.0896105828476219 -1.75218548820981
-1.98225358394963 -1.27761266402124
3.49988761136337 -1.23804100373631
1.54706802915874 -3.18859670261658
-0.484249602372211 3.22048182649933
-0.25062835149553 2.03241014427617
-0.0923413857637392 4.67198786781936
-1.30357674438044 1.9763442413684
-0.729990087185651 -1.24872328930615
0.76543931309409 0.152930364212156
-1.57625312481116 2.08360651380523
-0.287439538983803 0.362379369457273
0.750040562455655 4.38704893051327
-0.341030264481945 0.474766014876379
0.346664646700774 0.809200322218001
1.57877911168468 1.99995187127102
-2.52244477790595 2.06869125656274
-1.84969587694859 -0.0131635644436842
0.302750331471649 -0.273141759493056
3.63350998278705 0.283240878613489
0.0321657354480179 0.942547211012776
-0.094287448619603 1.65229674407518
-0.987894866992016 2.24166075104399
-0.64718617035214 0.217773450713567
1.65174300590924 0.831105777150654
3.10495603037242 4.9183348048598
6.0393801962164 -1.5758194829071
2.26549437446723 1.45476961216535
-2.54938457867527 1.28839270115808
1.16177746836051 2.92106315277702
-2.64272646850055 5.10669269377125
-1.08523233139296 -0.571805239563856
1.17822301360208 -1.23519362710163
-3.73669347177959 0.0963112177715972
3.78499394056736 -1.5725551254599
2.21517642924856 -0.867688202224095
-1.01370781351857 1.49841073577338
0.387990139569305 -0.220731348129304
};
\addlegendentry{TX ant. 0}
\addplot [draw=color1, fill=color1, mark=*, only marks]
table{%
x  y
-0.00286867662760087 -2.48964322648886
-0.929855368489875 -1.33502345501365
-1.97630644113768 -2.49790993652648
1.85273058799309 -1.04472292842609
0.409706614956851 -0.793003830753937
-8.18310844063925 3.19515854139131
-3.67280377356621 -5.9719415782571
-1.30942217590835 0.198288268389891
-2.3118345289357 1.84480346649844
2.65427314972989 4.65895710905381
1.72325597346 2.92327819971501
1.69889294521386 8.102263207127
-5.1005594389759 1.44739634521224
0.927200712944866 -3.19518881912268
-0.986399255385102 -1.76752011786876
0.487594304880679 2.70203796565789
-0.125334116789933 -1.7013609556758
0.66512684398109 -2.12886579060095
-3.70433520420803 -2.95475902293524
-1.24740423902932 4.016066832117
1.88990415308676 -1.23526882808219
-3.24361820406405 2.79231204053291
2.1117876903997 1.45908640878914
-0.57019393478771 0.773291628717855
0.0135594326048911 -2.70753581912582
0.384455091895622 -1.9760808809188
1.51858130294638 4.99358112116761
-4.84598682303754 -0.331639959608725
-3.15428957021678 -2.57515988183246
-3.16637327997296 0.335008378315202
-1.26694610524908 -1.16463901512408
-0.444337283582528 3.71570129778816
2.966708325103 -0.962752367316602
2.82198666024122 -2.49306388615346
5.03775811931875 -3.36786287170852
4.80001932722972 4.12674765452272
1.38117354413952 1.8270457023614
-0.119937345841491 -8.14139325800933
3.4949103660523 1.75382306093072
-1.44073254762436 -1.57269336636354
2.93065631730003 -2.58631363782535
-7.89757044881384 1.78795368867433
0.322016486055956 -0.792560170057449
-3.61013808482906 -0.364435134530671
0.699617301110102 -2.38340083245259
4.91242957001671 -1.34427082456334
0.592347691458563 -0.544361063640378
1.60802968955761 -3.49178120622699
-0.935726794217437 -1.31352884744626
0.342276700501386 -1.46113057956618
1.35964550645854 -1.28494971862898
-0.90783580195689 -0.816984175168924
-3.85758054301838 -2.86255938835597
2.04693278588207 -2.75686411284066
-6.29459681771307 2.57519225709566
0.390825207117523 2.77123486751007
-1.76149171264626 -7.28599265362655
1.41923946959259 -6.29100381489249
-3.5174157697533 -2.69090025943991
3.66866446953761 -2.36136872270833
1.72087044161822 -2.67624292800036
2.23770018304077 -1.2935511715603
-5.4790384799955 -1.23737560636615
-1.15747822893164 2.04070209093032
};
\addlegendentry{TX ant. 1}
\addplot [draw=color2, fill=color2, mark=*, only marks]
table{%
x  y
1.87976179466527 -0.0558705837946524
0.441050517011634 1.09390755299272
-1.71865632881197 0.239119176058567
0.621615221818769 -0.00030391801198415
2.48561295800306 1.17078174274668
-0.325933494210744 1.58726474521612
2.61284680369118 2.8314794339546
0.674635800226515 -0.392550203182414
0.160346349196227 -0.82560482004023
-0.580277656869511 -1.00287807871113
1.10590636051352 -0.137087332439886
1.08806732406524 -1.32978962288567
-0.294105710133705 2.48789253303539
1.49026566920941 1.62146410255765
3.85530887817719 -0.542310181287766
-1.05698314161152 1.31195500529399
-0.565505142593767 0.797376049221354
3.14665047776529 0.689472222456954
2.30040889877782 -0.0747104210929261
4.55346023877597 -1.83292407760348
-0.0711603963416276 -0.957395374439986
1.26117866982084 1.17277030424483
-1.14202200647508 -0.0625664656029203
0.40978231879735 2.9091960664525
2.23355105375577 4.29850413029495
-0.0771056731353221 -1.97881418316424
-1.01989665845042 -0.162096186751514
1.28815433390438 1.05796030933792
-2.16545041049058 0.681796791395084
2.83501306837409 -1.70812899798946
2.38846405490111 -0.219514838090865
-0.910954929684632 1.55395093064495
0.631802863323555 -1.30324480761324
0.821384744456951 -3.20489575619066
-0.239728892325159 -0.393875458551638
-1.30845885334441 1.11490646238923
0.138860025954983 0.850243674709293
-0.0703533173195319 0.867712802022786
1.39867914364384 -2.21027458944875
3.45389699585003 0.826962875290994
2.09574803667975 -1.34984920142635
-2.19546437906714 -0.994632390993003
-1.03783177808046 1.59490333308709
0.950338490849939 1.75043716914831
-0.844240061744676 -1.24776733659954
-1.07074803516281 -2.23105515572219
-0.676592826170411 0.822595184347555
0.159955501745368 0.17648317705474
-0.972889111489008 0.298656855069956
0.501237409052644 -0.761135318592148
3.00244527640501 -2.45243389389771
0.184081113418543 -0.802355741337096
0.451942872801806 1.16480280603522
-0.0446566279209491 2.3196834751072
0.708794301109272 0.595727389469049
-1.25695066387735 -2.93694959656129
0.933054000440548 -1.71122937093628
0.0943111243198613 -1.77620677565707
2.43140416785144 -0.200464728188448
-0.0957190579707484 1.18113074339819
-0.0764925050670301 -0.91659127418925
-4.20309884281022 -1.39005493599051
1.06246244688603 0.605367521787686
-0.879357948424944 -0.528532601824403
};
\addlegendentry{TX ant. 2}
\addplot [draw=color3, fill=color3, mark=*, only marks]
table{%
x  y
1.56382027404971 1.74412149850338
-0.844367143707454 -2.24109744678099
2.13216502882227 -1.37512890265789
-0.609932950039953 -3.30015370545166
1.0413113913138 -2.21739072941302
-0.893123541597849 -2.03984339946012
3.35870958619325 0.652486731552951
1.50525846687011 -2.32200887881929
3.3224165422102 4.01285588398297
0.89477471324827 1.28316364695145
-0.934920983663902 -1.56600674104328
-7.84576928118214 -6.05675626666404
-0.152254343180309 -1.2997949255522
-0.565693481017887 1.69022734935971
8.86714843173233 -4.87211440853723
-9.53526210660483 -2.03226073228758
-3.16652199893918 -0.00563708428619425
1.01844988722906 -1.27245669006789
-0.537916821899006 -4.68116255975953
-2.42841167847943 -1.35738190125963
-1.01436343277695 2.07951366144228
-0.444904599703027 -0.295488723338038
1.65907718986025 4.67815234421257
0.57957029958245 0.270089399387776
0.236064381914326 1.50561240369859
-1.65452392260277 -3.00045772593955
-0.275499159428185 -2.8924314415173
3.84273098603418 0.138031581912474
2.1602645157198 1.30875623149467
-4.40463295360483 2.21969861356784
4.34979913933397 0.383014055197411
0.896017567017724 0.110101494826575
-2.31398848247501 -1.92068082166334
0.298881669521283 1.35887922321568
-1.00846035866688 -0.408030569075918
6.19298026887943 -4.73079536338793
1.28758542451501 -3.52939126586534
3.69886967414974 -0.446763946121191
0.16787888733436 0.808796403574273
1.63104479157636 -3.48772901539096
-1.1754574098902 0.134552360872434
1.35945820938536 1.46664112070023
-5.63192292628859 0.388042852363498
-7.10825185357404 -0.513871417749062
-1.0745293517909 1.49005895975609
8.07790910948032 -0.913050712494527
-2.97495326319624 2.63366318786372
3.44510360542437 1.49243598867907
0.29063679032556 -0.259498887970816
-0.496104804427356 0.418618976989393
0.793491060544071 2.23504202082205
0.0269237689028418 5.18301329861639
-3.15089764421029 6.33534161606502
-0.128485348445387 0.253763803927132
-2.08256642745272 0.960166696149951
1.10851185994768 3.39997379946577
-0.0647833445162275 -0.773906879600135
3.35789722906215 -3.40757823676079
5.79034396509576 0.477424846520782
1.62505563949098 1.71665175613164
1.14179255398227 -5.09396613664275
1.91039173663572 -3.78826245429527
2.4944688503799 -1.25248807505755
-1.82586452917355 0.65138145714529
};
\addlegendentry{TX ant. 3}
\end{axis}

\end{tikzpicture}

%% file: figures/IQ_b3.tikz
\begin{tikzpicture}

\definecolor{color0}{rgb}{0.12156862745098,0.466666666666667,0.705882352941177}
\definecolor{color1}{rgb}{1,0.498039215686275,0.0549019607843137}
\definecolor{color2}{rgb}{0.172549019607843,0.627450980392157,0.172549019607843}
\definecolor{color3}{rgb}{0.83921568627451,0.152941176470588,0.156862745098039}

\begin{axis}[
tick align=outside,
tick pos=left,
x grid style={white!69.0196078431373!black},
xlabel={$I$},
xmajorgrids,
xmin=-8.34788922818466, xmax=8.72367643706217,
xtick style={color=black},
xtick={-10,-8,-6,-4,-2,0,2,4,6,8,10},
y grid style={white!69.0196078431373!black},
ylabel={$Q$},
ymajorgrids,
ymin=-8.0397707016735, ymax=7.56353073119669,
ytick style={color=black},
ytick={-10,-8,-6,-4,-2,0,2,4,6,8},
legend columns=2,
legend cell align={left},
legend style={at={(0.52,0.98)}, fill opacity=0.6, draw opacity=1, text opacity=1, draw=white!80!black, nodes={scale=.8}},
]
\addplot [draw=color0, fill=color0, mark=*, only marks]
table{%
x  y
1.63335611957695 -0.267632114701918
-1.725249342573 -2.66761927520253
-2.90043064840743 -2.66761927520253
-0.595393489306875 0.903575287722281
-0.595393489306875 4.98010042992283
-2.90043064840743 -0.267632114701918
-1.725249342573 -0.267632114701918
-0.595393489306875 -3.94570732201316
3.94133479797661 0.903575287722281
-4.06121047933675 4.98010042992283
1.63335611957695 -0.267632114701918
0.512048016945023 -3.94570732201316
-0.595393489306875 3.56726808198386
-1.725249342573 3.56726808198386
0.512048016945023 2.15532185650843
-1.725249342573 -0.267632114701918
1.63335611957695 -0.267632114701918
-0.595393489306875 0.903575287722281
2.79210151757381 -1.43590501750126
-2.90043064840743 -3.94570732201316
-2.90043064840743 -3.94570732201316
1.63335611957695 0.903575287722281
0.512048016945023 0.903575287722281
1.63335611957695 4.98010042992283
-1.725249342573 2.15532185650843
0.512048016945023 -0.267632114701918
-1.725249342573 -1.43590501750126
-0.595393489306875 -0.267632114701918
2.79210151757381 -1.43590501750126
1.63335611957695 -3.94570732201316
-0.595393489306875 3.56726808198386
-0.595393489306875 0.903575287722281
0.512048016945023 3.56726808198386
-1.725249342573 0.903575287722281
-0.595393489306875 -0.267632114701918
0.512048016945023 -0.267632114701918
-0.595393489306875 0.903575287722281
0.512048016945023 0.903575287722281
1.63335611957695 3.56726808198386
-0.595393489306875 -0.267632114701918
-0.595393489306875 0.903575287722281
1.63335611957695 0.903575287722281
-2.90043064840743 2.15532185650843
-1.725249342573 -0.267632114701918
1.63335611957695 -1.43590501750126
2.79210151757381 0.903575287722281
0.512048016945023 -0.267632114701918
-0.595393489306875 0.903575287722281
-0.595393489306875 2.15532185650843
-0.595393489306875 -0.267632114701918
1.63335611957695 -0.267632114701918
3.94133479797661 4.98010042992283
3.94133479797661 -1.43590501750126
1.63335611957695 2.15532185650843
-2.90043064840743 2.15532185650843
1.63335611957695 3.56726808198386
-1.725249342573 4.98010042992283
-0.595393489306875 0.903575287722281
-0.595393489306875 -1.43590501750126
-2.90043064840743 2.15532185650843
2.79210151757381 -1.43590501750126
1.63335611957695 -0.267632114701918
-1.725249342573 0.903575287722281
0.512048016945023 -0.267632114701918
};
\addlegendentry{TX ant. 0}
\addplot [draw=color1, fill=color1, mark=*, only marks]
table{%
x  y
-0.820759073681326 -2.23802489063385
0.478550622250234 -0.84384766134304
-2.21872410904248 -2.23802489063385
1.80932640382045 -0.84384766134304
0.478550622250234 0.431774152461006
-4.95318347442646 1.73929201667044
-2.21872410904248 -5.94866202043369
-0.820759073681326 -0.84384766134304
-0.820759073681326 1.73929201667044
1.80932640382045 5.10440446227077
1.80932640382045 1.73929201667044
1.80932640382045 5.10440446227077
-3.65730462756435 0.431774152461006
1.80932640382045 -2.23802489063385
-0.820759073681326 -0.84384766134304
-0.820759073681326 3.31954717562581
-0.820759073681326 -0.84384766134304
0.478550622250234 -0.84384766134304
-2.21872410904248 -2.23802489063385
-0.820759073681326 3.31954717562581
1.80932640382045 -0.84384766134304
-2.21872410904248 3.31954717562581
0.478550622250234 1.73929201667044
-0.820759073681326 -0.84384766134304
-0.820759073681326 -0.84384766134304
0.478550622250234 -0.84384766134304
1.80932640382045 5.10440446227077
-3.65730462756435 -0.84384766134304
-3.65730462756435 -2.23802489063385
-2.21872410904248 0.431774152461006
-2.21872410904248 -0.84384766134304
-0.820759073681326 1.73929201667044
1.80932640382045 -2.23802489063385
3.17058264269733 -2.23802489063385
4.44176661183061 -4.02957863994648
4.44176661183061 3.31954717562581
1.80932640382045 1.73929201667044
-0.820759073681326 -5.94866202043369
3.17058264269733 1.73929201667044
-0.820759073681326 -0.84384766134304
3.17058264269733 -2.23802489063385
-4.95318347442646 1.73929201667044
-0.820759073681326 0.431774152461006
-2.21872410904248 0.431774152461006
0.478550622250234 -0.84384766134304
3.17058264269733 0.431774152461006
0.478550622250234 -0.84384766134304
1.80932640382045 -4.02957863994648
-0.820759073681326 -0.84384766134304
-0.820759073681326 -0.84384766134304
0.478550622250234 0.431774152461006
0.478550622250234 -0.84384766134304
-3.65730462756435 -2.23802489063385
1.80932640382045 -2.23802489063385
-4.95318347442646 1.73929201667044
0.478550622250234 1.73929201667044
-0.820759073681326 -5.94866202043369
1.80932640382045 -4.02957863994648
-2.21872410904248 -2.23802489063385
3.17058264269733 -2.23802489063385
1.80932640382045 -2.23802489063385
1.80932640382045 0.431774152461006
-4.95318347442646 -0.84384766134304
-0.820759073681326 1.73929201667044
};
\addlegendentry{TX ant. 1}
\addplot [draw=color2, fill=color2, mark=*, only marks]
table{%
x  y
0.773089215857839 -0.394175838336479
0.773089215857839 -0.394175838336479
-2.01500852747959 0.300128233206444
1.73483430224612 0.996578805482312
1.73483430224612 2.54831098266703
0.773089215857839 0.996578805482312
2.72027494401951 0.996578805482312
0.773089215857839 0.300128233206444
-0.166855197873718 -1.09161475958973
-0.166855197873718 -1.09161475958973
0.773089215857839 -0.394175838336479
0.773089215857839 -1.09161475958973
-0.166855197873718 2.54831098266703
0.773089215857839 1.71313246488126
1.73483430224612 -0.394175838336479
-1.09359451664902 0.996578805482312
0.773089215857839 0.996578805482312
2.72027494401951 0.996578805482312
2.72027494401951 0.996578805482312
3.71818212552 -1.80802256829361
-0.166855197873718 -1.09161475958973
0.773089215857839 0.300128233206444
-1.09359451664902 0.300128233206444
-0.166855197873718 2.54831098266703
1.73483430224612 2.54831098266703
-0.166855197873718 -1.09161475958973
-1.09359451664902 0.300128233206444
-0.166855197873718 2.54831098266703
-0.166855197873718 0.996578805482312
1.73483430224612 -1.09161475958973
1.73483430224612 -0.394175838336479
-0.166855197873718 0.996578805482312
1.73483430224612 -1.80802256829361
0.773089215857839 -1.80802256829361
-0.166855197873718 -1.09161475958973
-1.09359451664902 0.996578805482312
-0.166855197873718 -0.394175838336479
0.773089215857839 0.996578805482312
1.73483430224612 -1.80802256829361
2.72027494401951 0.300128233206444
0.773089215857839 -1.09161475958973
-2.01500852747959 -1.09161475958973
0.773089215857839 0.996578805482312
-0.166855197873718 0.996578805482312
-1.09359451664902 -0.394175838336479
-1.09359451664902 -2.6360601696802
-1.09359451664902 -0.394175838336479
-0.166855197873718 0.300128233206444
-1.09359451664902 -0.394175838336479
0.773089215857839 -1.09161475958973
1.73483430224612 -1.09161475958973
0.773089215857839 -1.80802256829361
1.73483430224612 0.996578805482312
-0.166855197873718 0.996578805482312
0.773089215857839 0.300128233206444
-1.09359451664902 -2.6360601696802
0.773089215857839 -1.09161475958973
0.773089215857839 -1.09161475958973
1.73483430224612 0.300128233206444
-0.166855197873718 0.996578805482312
-0.166855197873718 -0.394175838336479
-2.97195981551876 -1.80802256829361
0.773089215857839 -0.394175838336479
-1.09359451664902 -1.09161475958973
};
\addlegendentry{TX ant. 2}
\addplot [draw=color3, fill=color3, mark=*, only marks]
table{%
x  y
0.850346553987639 0.554821796386528
-0.653909667101899 -0.710294260522797
0.850346553987639 -0.710294260522797
-0.653909667101899 -2.03438899520298
-0.653909667101899 -0.710294260522797
-0.653909667101899 -0.710294260522797
2.73587232102814 1.85221112194945
0.850346553987639 -2.03438899520298
2.73587232102814 3.15043400762813
0.850346553987639 1.85221112194945
-0.653909667101899 -2.03438899520298
-6.92940074365447 -4.57026364365122
0.850346553987639 -0.710294260522797
-0.653909667101899 1.85221112194945
7.32879358914685 -4.57026364365122
-6.92940074365447 -0.710294260522797
-2.44042142421092 0.554821796386528
0.850346553987639 -0.710294260522797
0.850346553987639 -3.36286924658108
-2.44042142421092 -0.710294260522797
-0.653909667101899 1.85221112194945
-0.653909667101899 -0.710294260522797
0.850346553987639 4.34822792211637
0.850346553987639 0.554821796386528
-0.653909667101899 0.554821796386528
-0.653909667101899 -3.36286924658108
-0.653909667101899 -2.03438899520298
2.73587232102814 0.554821796386528
0.850346553987639 1.85221112194945
-4.77868320546781 3.15043400762813
2.73587232102814 0.554821796386528
0.850346553987639 -0.710294260522797
-2.44042142421092 -0.710294260522797
-0.653909667101899 0.554821796386528
-0.653909667101899 0.554821796386528
5.16227814809476 -4.57026364365122
0.850346553987639 -3.36286924658108
2.73587232102814 -0.710294260522797
-0.653909667101899 0.554821796386528
0.850346553987639 -3.36286924658108
-0.653909667101899 0.554821796386528
0.850346553987639 0.554821796386528
-4.77868320546781 0.554821796386528
-6.92940074365447 -0.710294260522797
-0.653909667101899 0.554821796386528
7.32879358914685 0.554821796386528
-2.44042142421092 3.15043400762813
2.73587232102814 0.554821796386528
-0.653909667101899 0.554821796386528
-0.653909667101899 0.554821796386528
0.850346553987639 3.15043400762813
-0.653909667101899 4.34822792211637
-0.653909667101899 4.34822792211637
0.850346553987639 0.554821796386528
-2.44042142421092 0.554821796386528
0.850346553987639 3.15043400762813
0.850346553987639 -0.710294260522797
2.73587232102814 -2.03438899520298
5.16227814809476 -0.710294260522797
0.850346553987639 1.85221112194945
0.850346553987639 -4.57026364365122
2.73587232102814 -3.36286924658108
0.850346553987639 -0.710294260522797
-0.653909667101899 0.554821796386528
};
\addlegendentry{TX ant. 3}
\end{axis}

\end{tikzpicture}

%% file: figures/qpsk_constellation_constellation.tikz
\begin{tikzpicture}

\begin{axis}[
tick align=outside,
tick pos=left,
x grid style={white!69.0196078431373!black},
xlabel={$I$},
xmajorgrids,
xmin=-1.21255897057416, xmax=1.2090267237373,
xtick style={color=black},
y grid style={white!69.0196078431373!black},
ylabel={$Q$},
ymajorgrids,
scale=0.9,
ymin=-1.21824761683452, ymax=1.21016625273439,
ytick style={color=black}
]
\addplot [draw=black, fill=black, mark=*, only marks]
table{%
x  y
0.707 0.707
-0.707 0.707
-0.707 -0.707
0.707 -0.707
};
\draw (axis cs:0.707,0.717) node[
  scale=0.7,
  anchor=base west,
  text=black,
  rotate=0.0
]{00};
\draw (axis cs:-0.707,0.717) node[
  scale=0.7,
  anchor=base west,
  text=black,
  rotate=0.0
]{01};
\draw (axis cs:-0.707,-0.717) node[
  scale=0.7,
  anchor=base west,
  text=black,
  rotate=0.0
]{11};
\draw (axis cs:0.707,-0.717) node[
  scale=0.7,
  anchor=base west,
  text=black,
  rotate=0.0
]{10};
\end{axis}

\end{tikzpicture}

%% file: figures/constellation_noisy_symbols.tikz
\begin{tikzpicture}

\begin{axis}[
tick align=outside,
tick pos=left,
x grid style={white!69.0196078431373!black},
xlabel={$I$},
xmajorgrids,
xmin=-1.31255897057416, xmax=1.3090267237373,
xtick style={color=black},
y grid style={white!69.0196078431373!black},
ylabel={$Q$},
ymajorgrids,
ymin=-1.31824761683452, ymax=1.31016625273439,
ytick style={color=black}
]
\addplot [draw=black, fill=black, mark=*, only marks]
table{%
x  y
2.8134379975117 -0.592461300994824
0.814909939757812 -0.591923948958666
1.62226727933672 0.763427312354368
-0.800935982592886 0.617675955678153
0.591409823864595 0.770915277605754
0.629293142171292 0.764495023407202
1.60158598894752 0.82128301029095
-0.801019251264883 0.615930163247051
-0.58735989373596 -0.804226722474986
2.57050001683642 0.804183318487464
0.208738398913891 0.589827942236149
-0.612908850041133 -0.797622436320982
1.82731936014542 -0.574968372232105
0.582450781788775 0.792310900833183
0.203344701641241 0.64721262858816
2.59999196080825 0.81829659984743
0.231681979014769 0.596059016179514
1.81139753805858 -0.566270288626546
-0.777628226992265 0.604484156301098
0.392968460566446 -0.751020507922758
1.57754827693989 0.80120055000154
1.83462122743882 -0.61893097298082
0.768702952307882 -0.593833143495208
1.57571583645515 0.814931005427262
1.61327625806908 0.81146655097573
1.8058399766389 -0.657779607248491
-0.785109843850823 0.616951567624036
-0.791205587223963 0.57459594411728
0.395344531851627 -0.772285245931032
1.80028860174445 -0.577379371002442
2.60453445369773 0.790123037964519
0.81401795984802 -0.617959290243944
0.585463321202755 0.798177180861583
-0.583312882077115 -0.832597915374373
3.79634265020213 -0.619597719731649
0.792763376756063 -0.62005434354274
1.827662906464 -0.600866853295498
1.79752452539751 -0.614848213934935
1.17267855726394 0.618166203881085
1.19491311554134 0.6198089609961
0.596789749516469 0.778799070483688
1.6199277995205 0.751709462669349
-0.595569151759954 -0.773802009336031
3.1823485492009 0.641672077833175
-0.794129701329426 0.589708889644823
-0.799561889638421 0.601150544796871
1.66018609154422 0.800535522257973
0.791046350508218 -0.616676360556137
0.792692365775665 -0.601534685357418
1.79869027958999 -0.629179218771207
0.205671185507294 0.59200027954147
-0.798853107155288 0.605902802331219
2.79547773380018 -0.635552667549152
-0.808833921023555 0.632782848349318
-0.788525466626705 0.598278472982128
2.83170439075037 -0.613811168252042
-0.610904796067125 -0.792410439135431
0.811061681169493 -0.599968030544063
-0.816147667860482 0.570096961592427
-0.594611184648894 -0.852147739758661
1.75175277481913 -0.623846972352178
1.77050196503244 -0.581334448839859
1.82487280786236 -0.594722075500692
2.55306277632672 0.744290367089781
2.8117685068744 -0.614517001629615
0.584831465597315 0.781741871469423
1.72729294360267 -0.575424565101974
-0.781591518295603 0.608629021925024
1.79823674628093 -0.638743894940609
0.618057071874161 0.801951625378809
1.55957705760458 0.754339029376559
1.81373075331257 -0.656900695521759
-0.577786015543535 -0.82322973603669
2.62858362517175 0.817886794724875
1.78339054497568 -0.629309509425696
0.229081231277476 0.587272131327133
0.391415272775373 -0.767755815813487
0.625017462601029 0.787211303655272
2.75980592999067 -0.569961196962305
0.21660314253786 0.591852564670078
1.59085949833073 0.798109688682424
0.61771176437379 0.807687285022796
0.228532637002108 0.580021845232018
1.76460341010778 -0.558040347322519
0.357507874493579 -0.808946201580686
0.232949454840881 0.624988236954875
0.218306003090864 0.570770791954686
-0.610003972242759 -0.79629048257937
0.213310506326614 0.599458147432509
0.603263065911898 0.769321016128521
3.56699281596169 0.798174996838039
3.61328928756855 0.820443938136462
0.628023962588224 0.81094589366869
0.778102526495198 -0.62026522749625
1.60343389477563 0.790477449954157
0.377917026120683 -0.786102425343781
0.810782327737884 -0.638422629038636
1.76520547418545 -0.611179631112198
0.643739823788578 0.778381295254108
0.621054238344343 0.774301750598251
1.4031149417814 -0.799297367866119
0.183343260232448 0.561107191748886
-0.80894124155465 0.609810779983799
1.59381357217677 0.75645788812162
1.7786384015541 -0.583326042932827
0.609077977903927 0.776829887693341
1.42397051598199 -0.796407840360161
2.79706723765346 -0.64130317398639
1.18800321159382 0.593971787194151
1.40499610131369 -0.800396212363112
1.79814483438269 -0.628587255993086
-0.584303086692703 -0.796995714070744
3.57654311166243 0.778182127174722
1.7861426980524 -0.623340142361971
0.187035584181056 0.608418729284479
0.813883235764291 -0.619183857979423
2.20806110643371 0.60090162806608
0.815960207133647 -0.599841693240656
0.192499539413624 0.605516115551438
1.61639319723098 0.798895794694231
1.80187250750727 -0.596012642949831
1.7769943846411 -0.601856024703403
-0.779951831665395 0.600918416522244
1.21329495096545 0.643423973736498
1.80270936490781 -0.586361464139683
-0.615250820533218 -0.815313735953294
1.21095863661656 0.602933579578215
0.824577123632414 -0.606245602523914
2.39355858616807 -0.799947146944766
0.61510142445525 0.81353416426674
4.64673653560541 0.78241404769971
-0.620426968087698 -0.792888662090333
-0.612305217873013 -0.785657450054555
1.62049225563908 0.786201477288219
1.80064390441898 -0.598712251374975
2.756210067949 -0.612767266025549
0.788358765824312 -0.588303657907639
1.79803652351793 -0.62338288653971
2.38335855939626 -0.784324269623058
-0.781920867157047 0.657367138337303
2.57496771417527 0.828949006079396
3.64181426924225 0.798205481811456
0.813721432638972 -0.609840743360713
3.64615444283228 0.787632729034058
0.169780501407781 0.60866052986984
-0.765981321371352 0.563065154619552
1.78594966656071 -0.604536766925179
2.56890083529759 0.779384291724035
0.621580651454879 0.837310074766522
0.181697764558726 0.627300653795871
-0.79254164279504 0.588498730036025
-0.776013379992426 0.589071131154623
1.21751161680363 0.6138681382422
0.219817778862565 0.594053860911535
-0.827630325864159 0.587172406598686
1.82985692409999 -0.622527174766712
0.263699804790922 0.61382863543793
2.62941053160221 0.811105359641658
0.371124362156881 -0.770565344471715
2.79648583775636 -0.592043691166782
1.39959118160372 -0.787435266240464
0.788442582202532 -0.579577177716988
0.793773111101677 -0.598579976643674
0.800719999249599 -0.591507520215942
-0.808091491138855 0.580631595253722
2.63321025084195 0.734944021620064
0.79805979657868 -0.587104667768981
-0.812990971837469 0.593968401847851
0.224957270193328 0.628919095689826
-0.768257408626988 0.61498411133027
2.62449813864675 0.802741539390386
0.384284888626024 -0.787525869558548
1.19141815416758 0.596049379078046
1.77509593732318 -0.602447057750955
1.59506260044324 0.811518500010145
0.561608194051637 0.816828261686994
0.645281168846993 0.805071487887618
-0.572796455257998 -0.817978420441132
2.21994566751707 0.628308910413784
-0.629084011081928 -0.765257421466865
1.20297944297819 0.606563308739461
-0.621953273949269 -0.799335823827354
0.612422668777127 0.813814195742172
0.210988058917827 0.589082294328491
0.427317872150747 -0.777855029807959
1.77419649862702 -0.625424671596765
2.37841544182361 -0.767183663296427
-0.787181776520077 0.626116970652264
1.77592247456486 -0.613267435887095
0.601185205007536 0.758143305674014
2.60309873458272 0.815691510479645
0.384679656898355 -0.756072044840763
1.15877397919478 0.662169008567891
1.18994792808434 0.622725258996907
1.39148103355919 -0.807220537048137
1.7964743528396 -0.627572532037438
-0.632848175112631 -0.767262502005786
0.793095899419371 -0.623075362993207
1.40279490463654 -0.777956396309075
1.60047884810601 0.769976256561539
1.36885265349308 -0.810480927747434
1.81090411945306 -0.611517102259847
2.78436610683576 -0.656603553940543
0.371729915307467 -0.800665898295274
0.182395475916159 0.629318395444937
0.342059872090396 -0.777564039023047
-0.596270342007696 -0.762335702714246
0.200928124670159 0.606492521200243
-0.555765075599345 -0.815822770609272
0.241300953412062 0.593681445630432
3.77501068101401 -0.587459994179802
-0.629730812700799 -0.792061620066862
0.604350851731675 0.803892736430836
1.39630589094173 -0.854344676257
1.59620505351607 0.78663312842659
0.596875898264864 0.753926094909152
0.61054406513033 0.834224296773937
0.83484291565524 -0.590273715899848
1.81096179892169 -0.612956969219894
-0.778710667103694 0.631892067452437
-0.576545442979333 -0.7927266901443
-0.634306663517585 -0.757237207245522
1.79582126625556 -0.577369535261103
1.56066445781549 0.782036840843333
2.8206097005666 -0.597704333559725
0.228069705189248 0.597937954433349
2.79126123850525 -0.631905890647827
-0.623712872362579 -0.81783766772252
2.64121561174707 0.820477517806213
0.211243707375389 0.57350640167545
-0.591073101019744 -0.786489221226079
0.389492164417779 -0.802470181713045
0.591031522480423 0.797545259398523
-0.779560756819305 0.590860614961615
1.78885569499383 -0.63858965580202
1.40653373727452 -0.800923819625216
1.22349677701248 0.611591435791171
0.189032050186242 0.5883409236748
-0.828828492858994 0.625127144963607
0.37623075779837 -0.808040918125071
0.190696888755375 0.590766705551411
0.218488206983137 0.603921935654925
1.6229328128553 0.814161942621823
0.796828785449511 -0.577535695636835
2.62971723809678 0.799070180472732
1.62923759416517 0.805958621813172
0.199242659427075 0.597577223041672
0.63099908929946 0.766500468925088
2.81388469535871 -0.597703881514214
1.63487191230489 0.772400663194677
-0.784184093615586 0.609109770317839
0.819086458850246 -0.59757564369155
0.201517295922084 0.603245752179218
0.387846660368372 -0.777438324045363
1.8161278279504 -0.617712455974503
0.784185978011669 -0.636236974815473
0.772871409898123 -0.616426634547641
1.20665834938586 0.620092925021227
0.376567908302837 -0.805194393829821
0.801266582280661 -0.574038457391436
1.41113790006368 -0.81905225676726
0.391539567318563 -0.8146710877047
0.365051259397038 -0.79799319514702
0.603672973735774 0.801689236173979
1.8304544764929 -0.56063036592717
0.200768821910533 0.634195112910819
0.414242202604362 -0.808175877182539
1.61738570113603 0.791590173216089
0.387338020589603 -0.765976468378193
0.187405626731511 0.660638027676518
-0.775317717587954 0.594425075106566
0.221609753321634 0.586595730032176
0.610485098075839 0.778621314904602
0.411733794249184 -0.792108063877491
0.78178143641908 -0.622493252458901
1.77667338352887 -0.644842061099418
1.7858628283808 -0.558551125097819
1.19659619744124 0.572718688322433
0.579159269369841 0.77070033947316
0.791499323434074 -0.577684346210384
1.16189010564044 0.586288679915108
0.204037988325426 0.60603569493629
2.39536337669011 -0.773786653729151
1.59063844253796 0.796096288174289
0.635702586889446 0.786315636686597
-0.59525565335267 -0.767927184871748
1.80755635119985 -0.538778453487939
0.398202667732323 -0.813286338697172
1.81094545025248 -0.618202257766698
1.62563194737689 0.810476171758532
0.380518516445775 -0.805380568212826
1.6189541297708 0.807680136300577
0.203693415288811 0.603630893873672
0.610766956806071 0.783202822895678
2.80324114255198 -0.617539373865546
0.384295878384007 -0.77792097466544
0.381900483988581 -0.817059749735425
1.41561826586866 -0.767587128931378
1.6111079220623 0.821709498655019
1.79562449183168 -0.625078193213088
1.80127105866454 -0.602068351663692
0.371864540523442 -0.838322780913989
1.6039196860701 0.764988308525909
0.796921394112101 -0.599704119157897
1.58387757714296 0.796923851801731
0.779989041916396 -0.589624010879364
0.406626148661196 -0.816377890658407
0.590901970490428 0.800779210716499
0.38783650058566 -0.787479875960603
0.601972462851093 0.807444851671814
-0.622623245315578 -0.800809859611773
1.82684811235307 -0.625739277492207
1.77648493551252 -0.600824354137366
3.77128237370916 -0.627670827909588
0.589821407390924 0.827644173485623
0.608278507152869 0.817127218459463
1.59955213330746 0.852709920971319
1.39100278890383 -0.806438652925732
0.788053340704808 -0.608839276546079
1.60677160579275 0.74889208910004
0.217049103975574 0.620714784788891
0.387942299496483 -0.773153665317136
1.76745514273961 -0.582052866002916
0.212949353813022 0.614666422420478
1.60232612098512 0.792391393111957
0.427192608371664 -0.785124274667495
1.34662049207265 -0.782496107048762
0.399087096566838 -0.776709387776866
1.42662983534554 -0.820962657791633
2.79497767094083 -0.624362401494201
1.79332985537638 -0.572542355715735
1.3818133790529 -0.776377867935343
-0.775662654362528 0.607618308489808
-0.60777637745408 -0.794190495658262
-0.624353182877531 -0.801464501159906
-0.75980688071265 0.603636410425452
0.778131097491795 -0.602093954898721
3.77920125863942 -0.623416105773213
1.60046876383152 0.795562548180346
-0.584137295158326 -0.803842285611458
3.58787198361117 0.817941619350752
0.401364593344088 -0.808758159218308
-0.589754276571825 -0.8226729988607
0.637388418557786 0.831359013028189
3.59749546384903 0.794164482361278
2.6607515430211 0.789338855952236
0.38797541182951 -0.801889968053614
-0.59605161284276 -0.850205546991742
1.58651140424586 0.793248052976757
0.2002910665208 0.568136085006825
0.578900293054139 0.770517641432881
2.80868797543691 -0.611899490519496
1.7882212347522 -0.610698584253228
3.80027975056489 -0.585365852975101
0.390992019020681 -0.755429546024953
-0.608754741684569 -0.803234385935698
-0.588901106512418 -0.803541595854094
0.192638389871765 0.612584916473533
-0.810455367973304 0.615382905106718
-0.611398564877521 -0.823609486676276
1.78615150031378 -0.612200277795395
0.770561717777535 -0.57937776864427
0.619969470775864 0.802258078299473
0.191653746216189 0.587245094240643
0.797204456048795 -0.644749115180059
0.772655531839301 -0.611233867772707
0.413409611265692 -0.802423005718125
-0.584347050894697 -0.771229764999817
-0.772766497266584 0.568656407729697
1.59277673539744 0.820877829553471
0.624545988033798 0.841795260931491
0.379929808394574 -0.773417355207774
3.79992344024952 -0.627856397876148
2.19079305044129 0.614261853518638
0.382890353080171 -0.774900073708005
0.84116975894495 -0.587888723928251
2.60225984505347 0.780606005661676
1.80917705528895 -0.611778842066909
2.61161001453557 0.820256619510966
1.75834160642791 -0.603552102606279
0.586324762310718 0.780366728418696
2.58443750229096 0.817096179107789
-0.745744682505928 0.619257690282104
1.81411369540148 -0.604457933957003
0.795454074266801 -0.594642348999165
2.8070223170255 -0.633001932235322
2.80763321749232 -0.624354401984191
-0.589080029218789 -0.836817429396726
0.210174255776175 0.628886953457713
2.40706162606882 -0.750391754217984
1.80377935868611 -0.595468367126415
1.62106693302209 0.793724816616449
0.411766387183908 -0.763311256936581
2.74238160491474 -0.648821401257948
4.80649452993233 -0.613084110142923
1.58930716093077 0.820927079310446
0.384865310828941 -0.805702729995503
-0.768558546274208 0.584744484010545
1.23788999612824 0.625681143139771
-0.617995244056181 -0.836389784409807
0.573456978260008 0.805002906353916
-0.596632186425747 -0.7739754351843
3.60080212895403 0.797277377685591
1.57376053064483 0.825176508002866
1.79206651458432 -0.625693714396281
1.77553486341531 -0.601542803020987
0.817349464886209 -0.601386599397014
0.612647661661227 0.794316781265424
-0.613862241725977 -0.818155324003013
1.61961289846393 0.798474496821414
0.808637368716352 -0.605629647094615
-0.613144352984337 -0.811629115825027
1.60359348110342 0.818990470446341
0.791354819962494 -0.598451736608436
2.20297029897951 0.622228386157933
0.645157162826467 0.801603739846719
0.202388920773576 0.568637143549041
2.78821067348024 -0.589487819003315
2.58840898845477 0.801749895416109
3.77748710860042 -0.587720960144197
1.8060031348077 -0.617195192558924
-0.593044669120627 -0.782444461747314
0.399889498181054 -0.787298047930722
0.801872414050556 -0.64810070407596
2.78046205814129 -0.605192389674978
2.78793423709485 -0.606307820262699
1.80647213235074 -0.621675928276048
-0.610126772889459 -0.782470574253672
1.57053960984599 0.737888127809566
3.642978322127 0.795540569486684
2.77650303208538 -0.586783252751892
0.241847065503248 0.631336189707597
-0.774602001277733 0.589288150580765
0.785016098868753 -0.629563517609511
0.800119448215587 -0.623684596668892
-0.601961673138216 -0.820687611686201
2.20286724200512 0.614214666240075
1.7807474209092 -0.610578188612908
-0.640555668081196 -0.79477934472474
1.4258122022914 -0.803471854222432
2.79634675089145 -0.592356705949101
1.23520149963166 0.626142266334878
0.602000310513541 0.820407609217498
-0.593149871562655 -0.834866694906575
0.609298857218151 0.790330575509879
1.78643330906608 -0.577022906051439
1.18914996609364 0.579981886042076
0.366393214058183 -0.779307840443572
-0.777325428076285 0.595441669590624
1.38261457597002 -0.791265928079983
0.229547270957893 0.597360602749794
0.403273487129018 -0.786351061099225
4.80502160642183 -0.65635762000369
1.75105778978123 -0.600052428973757
2.82765599176042 -0.587560495207578
-0.758247087030429 0.590997680064746
-0.618256073654592 -0.794573796963055
0.438355239167667 -0.819206512840638
2.80093146256183 -0.576299625669642
-0.755305731331192 0.612722905305924
-0.59932613107443 -0.834818279095188
0.221933130964432 0.628046611287669
-0.799752136108244 0.617995984153324
1.79897895361853 -0.640483408580224
2.83235828413207 -0.607269517707184
0.213482052382407 0.616902226042219
1.42532515411053 -0.799689357419087
-0.793317556716815 0.6384343819462
-0.802557887667143 0.610637635862361
0.177727001193088 0.610203336627826
0.594930989937595 0.767942604275822
1.21938781440792 0.592157964517919
0.234688099897051 0.646870673326642
2.20431284202696 0.570200798562068
3.58548363232777 0.779006031802439
0.613013661750485 0.799748183889712
0.621905025341829 0.799158911792098
-0.646159912769058 -0.802246317651921
0.774508711738459 -0.629603350347457
0.34472263732983 -0.771712030178826
0.788174643881626 -0.61572500666562
-0.618296099037599 -0.772862107599574
-0.599648794839947 -0.760421214709624
1.78192682642806 -0.583034453958191
1.36066084771681 -0.765658596158007
-0.810516147115634 0.638598519934393
0.570306376783456 0.78955442318974
1.60438413795666 0.783749755501322
2.75571820076604 -0.565760817283797
0.227586876315464 0.621485456993
-0.746655450623336 0.622058480587716
2.62743795200267 0.830984385955549
-0.767768609672158 0.60693909996752
0.218275068979841 0.590051610255668
1.80253802964059 -0.609528378380543
-0.647188076865732 -0.803718662490335
-0.59243394024321 -0.79975087710852
-0.800676517589481 0.552109683809029
2.39707104277839 -0.804135345200093
0.807840114738072 -0.656582891431912
0.215896007494584 0.598287685378091
-0.580248556214349 -0.79233797607964
-0.805446491967741 0.598215700156286
2.59531883813801 0.820685898668859
0.543547799233723 0.806248455237168
-0.638025262358633 -0.764975042884264
-0.617204680036818 -0.824701891416058
1.37998078363711 -0.77139129860289
1.60876030393856 0.798970179869505
-0.586972970480117 -0.77424727585483
1.64417425606774 0.827610488721315
0.787870163942761 -0.620754733438553
0.18406808230715 0.63948984503902
0.620014841512705 0.828776724347794
-0.585404263849453 -0.790940607920653
1.76395409276208 -0.601068402232052
0.431061062573057 -0.857225154837025
1.40011591641629 -0.78213318813512
0.409792593719832 -0.774399371745046
0.424068871483969 -0.800282063223533
2.82817173550629 -0.567745268796604
-0.607606774435478 -0.798134778016269
1.36074416849483 -0.791259732845755
0.192057337958463 0.597923120871998
1.79373267372026 -0.608469872674238
0.394111076118722 -0.750617580784002
1.82914476444974 -0.591792962877406
1.62814319935977 0.805876632359293
1.60428472558169 0.790773264680546
0.371098021708289 -0.79092160487571
0.618347593294679 0.818208362804105
2.78097106463979 -0.579194337046423
0.631528300163688 0.827895579190813
-0.793461873986631 0.627428040880341
1.56553446514736 0.765938735232712
1.41418460132778 -0.778306480414186
-0.83285362166794 0.6534860457519
0.62569782265602 0.789498201474298
2.42143544483863 -0.797168324548667
1.5876307679623 0.791627490544821
0.614489926682724 0.7840392129631
0.58196344079892 0.800014838968316
-0.797889011460608 0.622548137944535
1.77703783967099 -0.601431052200178
1.22279324731196 0.576676038284959
3.62455785635445 0.773287060199126
1.58080973428418 0.845339016938811
1.62130243600005 0.804097530977385
-0.624368964569983 -0.823778428879739
0.665235300648761 0.815771203164487
1.83391873898718 -0.66580171334783
1.79663096149978 -0.569233829032136
2.38568642938841 -0.796446396634636
0.57758447047714 0.79198387890213
2.79312028395567 -0.660723729700361
3.7788503871456 -0.638752297795992
0.179021590914421 0.584105451685704
0.379273685018824 -0.799666444086678
0.409471564286395 -0.779158496904752
2.81217540799453 -0.60313228299932
1.79745088759802 -0.601993026024563
1.41737749445374 -0.774398786999685
0.622758835338389 0.770885847174594
0.231823956559551 0.613303816947602
0.397079633473541 -0.787866322953789
0.375475423690965 -0.785026141974252
0.627768084689262 0.790810522111524
-0.625782148752083 -0.800636117119696
3.79773835530718 -0.620487559370141
0.793485541371093 -0.631796075253068
0.768875347804129 -0.607829618027115
0.612811708307486 0.781566735408192
0.565473783132881 0.817521249470045
2.59298856288287 0.796754590475477
1.58752387792253 0.770240784918077
1.79786242543826 -0.633221464264797
1.81933084298675 -0.573695292366378
0.205379948402568 0.595169943679773
2.20020392532861 0.606916270387257
-0.606229353574708 -0.767183746731179
-0.58086724014907 -0.805023689644969
1.61149430516504 0.836199208072432
1.77042567949509 -0.57140123208295
-0.779506164416841 0.651029251986167
-0.658135385685232 -0.775280012970884
0.777505135126263 -0.621097909086915
0.800920112672258 -0.577407454533368
0.794903360696492 -0.601753081763659
1.16410175597125 0.640985428171984
0.782250611438801 -0.584720883898191
1.59687015141827 0.80462712912142
1.14683840103136 0.620064294745478
0.782018681710733 -0.60033664940071
1.88947871786376 -0.601347885428787
1.18555447950866 0.615226584792611
-0.78044697655097 0.618216874268226
-0.800536015403773 0.629747910653829
1.59411085635562 0.785839735415421
1.62423892196272 0.82074801040749
0.802637624750473 -0.598199368583333
1.82257924659792 -0.604149882380221
1.76561426709733 -0.577781945529348
1.20910342474599 0.604739257303601
0.589407253019563 0.814098894914115
0.809075409755641 -0.628882363339904
5.80841547501945 -0.60918010807584
1.60721336636776 0.808786873200382
2.39306306114299 -0.794446700775706
5.77766366875874 -0.596629632817252
-0.665673362185034 -0.776362765794621
1.56597899349161 0.778813445623088
1.62272489387469 0.818695805504022
-0.800728854728734 0.614621961884664
1.5750050186101 0.797604867890297
-0.786001343432377 0.632267202542071
0.39956779956967 -0.765618260115422
-0.598442802214907 -0.826610121689831
0.212237025504582 0.559985490088962
1.79153703072532 -0.626059875608635
0.371444666161085 -0.798278564582223
0.217414544954957 0.615176171392997
0.229941995342914 0.603895524943742
1.60345927795776 0.770598785211137
2.82814551990549 -0.561240733024224
0.193069625802871 0.635270476306851
1.76553853132376 -0.621862603341431
1.39684869290828 -0.772292612573855
2.79779700966334 -0.626609864523338
1.57933183183634 0.789132197343202
4.60834796105474 0.784474580855954
1.58416079023511 0.805606152023517
0.807793358730505 -0.629391066164481
0.188670478679766 0.629008152403143
0.405442785386144 -0.809421365717847
-0.645105691579832 -0.801596175144404
3.60738341473356 0.785251986825783
0.628635121741408 0.773035643002885
1.77471017792718 -0.603783780085561
4.80936582276762 -0.613900708304
-0.591783471168497 -0.758533972805525
1.19615614283136 0.636219694039402
0.783633551139942 -0.593954125472633
1.59615096741484 0.814102121984345
0.618069851828655 0.783497767410156
2.4125888927995 -0.803908971945859
2.18755270663076 0.632413215594013
2.62563087251018 0.827357314141592
-0.580815670182813 -0.796416641481933
0.209239738144948 0.623104426702115
2.78888285654593 -0.672420480959223
0.18323027342562 0.632398974513361
-0.617874758567074 -0.743214715875974
0.202588323754977 0.581050745910369
1.59358008050996 0.771183792760986
1.58707307133983 0.781650547333422
2.20557329323077 0.634460049016584
0.825375210763467 -0.636934959796954
3.21668960259089 0.622714219714236
-0.60043533585403 -0.780050443505852
-0.831337217591636 0.586361581500455
1.19854123770348 0.622262000326233
-0.631377957307565 -0.77952843041183
2.77639213372595 -0.591484238209749
0.234838243337028 0.621596857202837
-0.831337901170441 0.628578170113991
1.83506953136567 -0.584939218439103
0.196777640399325 0.639089006596476
1.80826359171555 -0.622830256982072
0.579402706579391 0.817761338933374
-0.593770154513612 -0.792532484057716
2.61771582204697 0.789403306314394
1.77679667531774 -0.599914581072526
1.82763291607508 -0.607788987783272
-0.783108326411034 0.634659281092662
0.395874377632574 -0.824611928131368
0.609083818812891 0.800525191153098
0.196553418976192 0.582168449823204
1.21611159864539 0.590703396204209
0.771837176720147 -0.620549048679001
2.59700170288149 0.823994141088746
0.828270414325395 -0.621747957948443
1.20475847693535 0.636165379646008
-0.806809513641091 0.628216766411358
4.60035948288622 0.746642206840409
2.74865249070154 -0.560737142957543
-0.834317106046447 0.566811457690899
1.25168321891439 0.615198013844032
-0.618429110366131 -0.766878909750665
1.62112526334199 0.760307569465947
1.21686870700924 0.627195933116361
0.387889811323274 -0.781996688647099
1.76888574185008 -0.628721405091689
2.3584485083655 -0.78727045936588
1.59891286832801 0.795341436309759
0.757740286805375 -0.597101929245193
2.21066620791593 0.612941627223177
0.41555267710408 -0.80789294317744
0.761937238136476 -0.620053273431799
1.60197766970125 0.776392570820541
0.794069551299573 -0.629596780036182
-0.60991998998121 -0.771092437560536
1.79452599726451 -0.60990023291892
1.59939510643173 0.76288973641974
1.60284886476767 0.765037125947376
1.57340961565264 0.772173971329745
1.61233039305608 0.748787501618329
1.3957922283273 -0.784640564217342
0.626789990912426 0.779595249329265
1.61803562627494 0.782202646176509
-0.634505973876585 -0.789468060262697
2.60318128531661 0.771919276249119
1.40987297820644 -0.780513795790161
2.35898043180566 -0.779580863043845
0.232088845338284 0.602198401904813
0.190259661359741 0.599773187396478
0.608885435627649 0.789448095659198
1.16811439399576 0.605585326196839
-0.754040843372799 0.604582747281508
1.60865524519871 0.802873171795523
1.81471432440468 -0.618302735831702
3.80453188629724 -0.612557662155792
1.40090987293848 -0.809420935976659
0.414742202367092 -0.780257458260152
2.38061116718928 -0.782384968532825
0.409172972065301 -0.787640916654364
-0.808508990802391 0.610348395799653
0.801056659404156 -0.627921540241385
0.639717699470269 0.813231048631294
-0.59720116351554 -0.804986552714128
0.653815409187815 0.814520835801138
1.62298181213706 0.773140294171644
1.5956109981202 0.806134131490062
2.63226412300151 0.777818431785598
0.822308417776202 -0.590151365168555
1.62918115142606 0.792256541580553
-0.823402943320768 0.57603987657471
-0.604696359394998 -0.746421399899252
0.586357957872533 0.794064137237932
0.817020958255296 -0.634260871317942
3.79616444259451 -0.599041251478426
1.82204481409238 -0.589366253805585
0.358171077815657 -0.801963200854731
0.763484385411231 -0.57270422032808
0.619599619800597 0.838677444433217
0.197359345070304 0.606685574729476
-0.811116691330526 0.613941073011455
0.583012882333864 0.813279685788297
0.621292037803371 0.74617466075112
1.61004205503326 0.830648921879544
2.58705649848284 0.807655248105295
2.80408124198924 -0.625147493212287
2.59225576184484 0.814432592870619
0.562312864943611 0.798134280535173
0.606709476718643 0.801638862964261
-0.597323590792259 -0.789258689222285
2.79546447507215 -0.5891730572166
1.81208472700928 -0.599862566115263
1.38142929163932 -0.787019593152362
1.36394581440481 -0.747098845680691
0.838411974163661 -0.604040006791085
0.621520155475478 0.802394946276272
0.790331625485545 -0.618694234079965
0.184203057595784 0.603207408193363
4.60632800858441 0.764330561269932
0.797481984460608 -0.628463129364411
0.590584927801689 0.770639557853452
-0.808446004298555 0.618755821830833
2.60686675078204 0.770623089030432
0.827992107104479 -0.584295551174954
1.62848950099243 0.795534652136131
2.58867593034741 0.789386148651817
0.803436921636458 -0.631382628486019
1.77573914558655 -0.611004669706276
0.220397378374788 0.640736585446964
1.8073624820581 -0.614258049020994
0.367850384624211 -0.81682887169413
0.627304725567789 0.825241913786014
1.64542066899034 0.825483419478398
1.79791722715894 -0.578019806217248
-0.631129338616567 -0.73195458200134
0.390447790446869 -0.793573806512386
0.791358639577312 -0.611281601555969
0.38402910797567 -0.810555508327408
0.408187123320915 -0.786416705905887
-0.600996640080673 -0.810998266789184
1.60789279729504 0.84167754242955
-0.602710063919223 -0.771109151644121
1.36827554427224 -0.763176080926263
1.7624371440381 -0.608158757143016
2.62160033077493 0.792090718097805
1.6362681996704 0.805512651089804
1.4371353955952 -0.78379364514786
1.61246377143849 0.803005316606522
0.563819712072609 0.756513194907063
2.83797600909167 -0.623363021987491
0.805453015363246 -0.616376809156464
0.370631988239327 -0.786568654659944
-0.804328353236142 0.605771353205557
2.37472695924917 -0.756118940714867
1.36817055161456 -0.7825330844087
0.770875086009409 -0.590260343721271
-0.618328730330861 -0.773925897763758
1.21589790786846 0.627712124481536
0.347776963749473 -0.75600173610273
1.20780199688142 0.613207790556643
-0.594068239732059 -0.800085264779567
-0.790649232847515 0.643967628139105
1.42025534618618 -0.804746522868168
1.19769928099047 0.580149498120417
1.7805663280799 -0.62926235294222
0.799790125044678 -0.623959707995847
0.184423930313244 0.596247134209533
2.61034708140683 0.759885228578952
1.80353374767378 -0.582589247621188
1.60411194735172 0.812647322601168
2.79236168182559 -0.61771323193638
1.73077494298726 -0.581506224503887
3.60578253798558 0.755080551606511
1.76978380432434 -0.604028674191213
1.37583031551311 -0.802444996268742
0.387483902400567 -0.80960833725987
-0.633829993656883 -0.774318351805911
0.411336357908452 -0.78001552179634
0.743376882891896 -0.643402003067007
0.797376275502543 -0.603653087296411
-0.759457807136738 0.631295868341795
1.81861783175884 -0.62599817331327
0.376786897885094 -0.780760895182792
0.867979468523145 -0.627018662609514
0.170417443767897 0.597241356281835
0.202818027921399 0.631573347197285
0.638611032914841 0.772704569750627
0.6132798037225 0.774369968685786
2.61449517485462 0.819897552439162
0.809361629651289 -0.623448269885466
0.759797448316606 -0.608046010000495
1.82403595854452 -0.627569128326725
0.769179920490686 -0.5794528171297
2.40898745533881 -0.793753292219163
1.38961041660206 -0.759195407452792
2.19899072118282 0.613703794949878
0.232484660309614 0.604621971833125
0.616930316996677 0.822292456201943
-0.622862546069939 -0.791869917195258
1.78516854782328 -0.571385305670924
1.78260343981872 -0.618553224727046
0.588159386905148 0.754247796542198
0.633927747144618 0.795686187206448
2.79189678744303 -0.613837447770988
2.19541880953271 0.59665774429374
0.785084085151495 -0.588794948010779
-0.783009856699105 0.632085757514181
1.63135640221738 0.801817023094142
-0.811828130600661 0.615598385118647
1.80819401034978 -0.627117364222295
0.626267102428541 0.832650529811527
2.79969113696748 -0.608032105509675
0.374166943727009 -0.77132100703953
0.415234905149426 -0.78635752186628
4.60772726026906 0.778397792105798
3.17668941622923 0.600050527930705
0.23048042187732 0.612425062583477
0.59345473552863 0.803358817254845
2.79300661322872 -0.598920055895094
2.21176187779907 0.617686879933952
1.36870910519264 -0.753608339024576
-0.776209303963836 0.627596184218443
3.81564932241237 -0.620238940998319
-0.614980386951461 -0.824986664373189
0.799495972480835 -0.587342280135317
1.81174021072946 -0.596944637214462
0.623799239482642 0.780550466501317
-0.797671211979646 0.613398654863734
1.6417046293193 0.846781056479943
0.24650869739491 0.588138151416252
0.609020907287514 0.742457129388993
1.66546428409493 0.779162964710262
0.791965198494788 -0.675617717237616
-0.769001084642139 0.616711661771731
0.622307352101175 0.847965754856984
1.40817039477821 -0.794499211953096
2.20926977618153 0.617708388798846
-0.624440784625953 -0.799035421956006
1.38512197751468 -0.781883410021879
0.792931658523442 -0.598775182556027
2.78620814683996 -0.623404266838553
2.76826590125038 -0.590824842839493
0.441168057032293 -0.764403028352779
1.81436856867975 -0.5945141889126
-0.578950405844294 -0.82389731598073
1.2105568796578 0.598941466383389
-0.578629804997997 -0.757843791998418
0.179995398054868 0.579344373776059
0.387106974970427 -0.749757885583757
1.58416437858283 0.80249753786699
0.741534433238955 -0.604186105381725
1.58100701934048 0.823148991547201
1.5981426326857 0.77249236178135
1.39767663221519 -0.763160665337995
0.380839436586738 -0.811848691592416
-0.580667640679221 -0.768577994017121
0.591388226519621 0.75432243840048
0.604516690460162 0.792712686184192
0.18383487356112 0.60534982579518
-0.59930784211077 -0.787982074092509
3.82112256126558 -0.581910767437735
0.379470271736157 -0.76411353154508
0.368102787990483 -0.804181186616643
1.16706257666064 0.569898484837465
0.802103063427498 -0.601038789178586
0.437510508297656 -0.80706885187728
-0.595833791535096 -0.80420530890163
1.78653741206728 -0.612580983881894
1.60683861963369 0.816807699628123
1.78858311069023 -0.582368262416646
0.624598104946121 0.827307142130856
1.40110648376926 -0.764086282100425
2.61641653799179 0.775346457259288
2.58967969590141 0.795561354662869
0.767867881613999 -0.624674419635099
0.574057713563775 0.761452973614136
1.19169081435086 0.606180239306692
0.55231714181785 0.793632591442917
0.420871169951572 -0.785296087399344
0.792808072248532 -0.568269791751488
0.607617703538509 0.771531268008763
0.16358316024674 0.593312434346095
0.785617319099559 -0.586445512626988
1.79742154910851 -0.584635346747953
2.57037358249658 0.821063427467652
0.223477864157734 0.599711530408313
0.773187324957259 -0.616752823967548
-0.779328706986026 0.581299604565877
0.215169320647444 0.593871250046731
1.78211019549101 -0.637186542586806
1.36158100291775 -0.792640432120412
2.73980461062 -0.612195453560578
0.407375168916364 -0.800091213563921
-0.604906368075793 -0.802775841039646
-0.610818026702603 -0.821348500164275
0.593667595840616 0.783184392283126
1.56839495623435 0.792375970337149
-0.616134615291416 -0.789674908294604
1.78481798644793 -0.598472104114235
0.417277864204971 -0.814150361913471
1.59777441608334 0.817949191870124
-0.587699373378668 -0.785454718854107
1.79253919240659 -0.609009232956719
2.83329611018759 -0.58390097192154
0.640512289422079 0.808916544641843
-0.620892796241209 -0.785195480523268
4.80163427998102 -0.588964192198088
1.18436240533231 0.632997408708265
2.17481726412339 0.597129790453252
0.605818647749708 0.800782056005806
2.61433687254511 0.771860454531416
0.793804751165645 -0.611309046507577
0.401991129343855 -0.783607352416077
0.173931336157616 0.584921406234872
0.579254207368285 0.765292929129741
};
\end{axis}

\end{tikzpicture}

%% file: figures/pdf_empirical_sinr.tikz
\begin{tikzpicture}

\begin{axis}[
width=4in,
height=2in,
tick align=outside,
tick pos=left,
x grid style={white!69.0196078431373!black},
xlabel={$X$},
xmajorgrids,
xmin=25, xmax=29,
xtick style={color=black},
y grid style={white!69.0196078431373!black},
ylabel={$p_X(x)$},
ymajorgrids,
ymin=0, ymax=0.44,
ytick style={color=black},
ytick={0, 0.1, ..., 0.5},
legend cell align={left},
legend style={at={(.82,0.72)}, fill opacity=0.6, draw opacity=1, text opacity=1, draw=white!80!black, nodes={scale=.8}},
]
\addplot [line width=1.75, blue]
table {%
24.6214829057238 0.00333333333333332
25.0230982789307 0.0166666666666668
25.4247136521375 0.0266666666666666
25.8263290253444 0.0699999999999998
26.2279443985513 0.130000000000001
26.6295597717581 0.13
27.031175144965 0.13
27.4327905181719 0.183333333333334
27.8344058913787 0.15
28.2360212645856 0.0799999999999998
28.6376366377925 0.0533333333333332
29.0392520109993 0.0266666666666668
};
\addlegendentry{SINR at the receiver (empirical)}

\addplot [line width=1.75, red]
table {%
21.8101752932757 6.47106308136862e-20
21.8198237106501 8.82773135646878e-20
21.8294721280244 1.20276689552933e-19
21.8391205453987 1.63671366050299e-19
21.8487689627731 2.22445093632827e-19
21.8584173801474 3.01947778116063e-19
21.8680657975218 4.09354650247034e-19
21.8777142148961 5.54276545432252e-19
21.8873626322704 7.495699333501e-19
21.8970110496448 1.01241063203691e-18
21.9066594670191 1.36571512760963e-18
21.9163078843935 1.84001950105974e-18
21.9259563017678 2.47595994659639e-18
21.9356047191421 3.32754308287873e-18
21.9452531365165 4.46645170823202e-18
21.9549015538908 5.98770618622015e-18
21.9645499712651 8.01709821033933e-18
21.9741983886395 1.07209386148083e-17
21.9838468060138 1.43188222650292e-17
21.9934952233882 1.91003212086733e-17
22.0031436407625 2.54467853681263e-17
22.0127920581368 3.38597748720643e-17
22.0224404755112 4.49980909643808e-17
22.0320888928855 5.97259403263345e-17
22.0417373102599 7.91754948524675e-17
22.0513857276342 1.04828038754622e-16
22.0610341450085 1.38619082044131e-16
22.0706825623829 1.83074332949067e-16
22.0803309797572 2.41485348514624e-16
22.0899793971316 3.1813613168026e-16
22.0996278145059 4.18595051530074e-16
22.1092762318802 5.50090402828824e-16
22.1189246492546 7.2199290611931e-16
22.1285730666289 9.46434748750927e-16
22.1382214840033 1.23910271463692e-15
22.1478699013776 1.62025296260034e-15
22.1575183187519 2.11600760726573e-15
22.1671667361263 2.76000907459771e-15
22.1768151535006 3.59552804172581e-15
22.1864635708749 4.67814561124644e-15
22.1961119882493 6.07916142831663e-15
22.2057604056236 7.8899182216299e-15
22.215408822998 1.02272815791916e-14
22.2250572403723 1.32405739201513e-14
22.2347056577466 1.71203363708561e-14
22.244354075121 2.21093849958833e-14
22.2540024924953 2.85167427278327e-14
22.2636509098697 3.67351704529135e-14
22.273299327244 4.72631962185406e-14
22.2829477446183 6.07327579440987e-14
22.2925961619927 7.79438414530055e-14
22.302244579367 9.99078231503162e-14
22.3118929967414 1.27901628608199e-13
22.3215414141157 1.63535310828314e-13
22.33118983149 2.0883625447484e-13
22.3408382488644 2.6635394831648e-13
22.3504866662387 3.39290165587604e-13
22.3601350836131 4.31660484670516e-13
22.3697835009874 5.48494410735922e-13
22.3794319183617 6.96082970913335e-13
22.3890803357361 8.8228460861546e-13
22.3987287531104 1.11690256513407e-12
22.4083771704847 1.41214978969898e-12
22.4180255878591 1.78322085905399e-12
22.4276740052334 2.24899452804945e-12
22.4373224226078 2.83289550818666e-12
22.4469708399821 3.56395003985291e-12
22.4566192573564 4.47807697244871e-12
22.4662676747308 5.61966461339795e-12
22.4759160921051 7.04349380830394e-12
22.4855645094795 8.8170798697169e-12
22.4952129268538 1.10235204304505e-11
22.5048613442281 1.37649534658055e-11
22.5145097616025 1.71667500798118e-11
22.5241581789768 2.13825907313334e-11
22.5338065963512 2.66006020198836e-11
22.5434550137255 3.30507646891201e-11
22.5531034310998 4.10138429799159e-11
22.5627518484742 5.08321318410953e-11
22.5724002658485 6.2922372892627e-11
22.5820486832229 7.77912536996258e-11
22.5916971005972 9.60539793009314e-11
22.6013455179715 1.18456491687024e-10
22.6109939353459 1.45902013903562e-10
22.6206423527202 1.79482712780679e-10
22.6302907700946 2.20517410337721e-10
22.6399391874689 2.70596431371212e-10
22.6495876048432 3.31634856613616e-10
22.6592360222176 4.05935660523675e-10
22.6688844395919 4.96264453999729e-10
22.6785328569662 6.05937829283229e-10
22.6881812743406 7.38927621685927e-10
22.6978296917149 8.99983765665737e-10
22.7074781090893 1.09477883685165e-09
22.7171265264636 1.33007784308547e-09
22.7267749438379 1.61393736320544e-09
22.7364233612123 1.95593873953927e-09
22.7460717785866 2.36746071693036e-09
22.755720195961 2.86199769627083e-09
22.7653686133353 3.45553064326714e-09
22.7750170307096 4.16695867351284e-09
22.784665448084 5.01860043347836e-09
22.7943138654583 6.03677562499315e-09
22.8039622828327 7.2524783899942e-09
22.813610700207 8.70215579502475e-09
22.8232591175813 1.04286063440187e-08
22.8329075349557 1.2482015316297e-08
22.84255595233 1.49211457878116e-08
22.8522043697043 1.78147064608189e-08
22.8618527870787 2.12429199135431e-08
22.871501204453 2.52993175999792e-08
22.8811496218274 3.00927908928165e-08
22.8907980392017 3.5749930681123e-08
22.900446456576 4.24176915187511e-08
22.9100948739504 5.02664200779591e-08
22.9197432913247 5.94932917016495e-08
22.9293917086991 7.03262031708284e-08
22.9390401260734 8.30281744117778e-08
22.9486885434477 9.79023167569082e-08
22.9583369608221 1.15297430537163e-07
22.9679853781964 1.35614300209101e-07
22.9776337955708 1.5931276088897e-07
22.9872822129451 1.86919616053544e-07
22.9969306303194 2.19037492241947e-07
23.0065790476938 2.56354722813902e-07
23.0162274650681 2.99656359137678e-07
23.0258758824425 3.49836413938982e-07
23.0355242998168 4.0791144786789e-07
23.0451727171911 4.75035616556427e-07
23.0548211345655 5.52517301450465e-07
23.0644695519398 6.41837453401468e-07
23.0741179693142 7.44669783277458e-07
23.0837663866885 8.62902938566513e-07
23.0934148040628 9.98664808953866e-07
23.1030632214372 1.15434910699549e-06
23.1127116388115 1.3326443721137e-06
23.1223600561858 1.53656554701461e-06
23.1320084735602 1.76948827507777e-06
23.1416568909345 2.03518606507951e-06
23.1513053083089 2.33787046556123e-06
23.1609537256832 2.68223438501924e-06
23.1706021430575 3.07349868562426e-06
23.1802505604319 3.51746216715176e-06
23.1898989778062 4.02055504396106e-06
23.1995473951806 4.58989600097534e-06
23.2091958125549 5.23335289445553e-06
23.2188442299292 5.95960713968558e-06
23.2284926473036 6.77822180033567e-06
23.2381410646779 7.69971336300607e-06
23.2477894820523 8.73562714516928e-06
23.2574378994266 9.89861624527282e-06
23.2670863168009 1.12025239000726e-05
23.2767347341753 1.2662469066297e-05
23.2863831515496 1.42949349915212e-05
23.296031568924 1.61178604827386e-05
23.3056799862983 1.81507335207126e-05
23.3153284036726 2.04146868039745e-05
23.324976821047 2.29325947386801e-05
23.3346252384213 2.57291713197406e-05
23.3442736557957 2.88310682753263e-05
23.35392207317 3.22669727714479e-05
23.3635704905443 3.60677038967093e-05
23.3732189079187 4.02663070701133e-05
23.382867325293 4.4898145438098e-05
23.3925157426673 5.00009872515836e-05
23.4021641600417 5.56150881410006e-05
23.411812577416 6.17832671382377e-05
23.4214609947904 6.8550975230301e-05
23.4311094121647 7.59663551720287e-05
23.440757829539 8.4080291235402e-05
23.4504062469134 9.29464475327912e-05
23.4600546642877 0.000102621293522197
23.4697030816621 0.00011316411528587
23.4793514990364 0.000124637011171197
23.4889999164107 0.000137104870396008
23.4986483337851 0.000150635333250984
23.5082967511594 0.000165298731581128
23.5179451685338 0.00018116800829765
23.5275935859081 0.00019831861476207
23.5372420032824 0.00021682838499765
23.5468904206568 0.000236777385818919
23.5565388380311 0.000258247742130239
23.5661872554054 0.00028132343682814
23.5758356727798 0.000306090084950003
23.5854840901541 0.000332634681943676
23.5951325075285 0.00036104532618714
23.6047809249028 0.000391410916164154
23.6144293422771 0.000423820822999427
23.6240777596515 0.000458364539371974
23.6337261770258 0.000495131306158283
23.6433745944002 0.000534209718501664
23.6530230117745 0.00057568731336013
23.6626714291488 0.000619650140947099
23.6723198465232 0.000666182322843641
23.6819682638975 0.000715365599923194
23.6916166812719 0.000767278873584826
23.7012650986462 0.000821997744134023
23.7109135160205 0.000879594050475274
23.7205619333949 0.000940135415582049
23.7302103507692 0.00100368480248286
23.7398587681436 0.00107030008573927
23.7495071855179 0.00114003364358748
23.7591556028922 0.0012129319760659
23.7688040202666 0.00128903535454689
23.7784524376409 0.00136837750813189
23.7881008550153 0.00145098535234526
23.7977492723896 0.001536878765473
23.8073976897639 0.00162607041773203
23.8170461071383 0.00171856565822283
23.8266945245126 0.00181436246430706
23.8363429418869 0.00191345145766749
23.8459913592613 0.00201581599084232
23.8556397766356 0.0021214323074862
23.86528819401 0.00223026977899587
23.8749366113843 0.00234229121945159
23.8845850287586 0.0024574532800725
23.894233446133 0.00257570692356918
23.9038818635073 0.00269699797790673
23.9135302808817 0.00282126776807549
23.923178698256 0.00294845382351241
23.9328271156303 0.0030784906578337
23.9424755330047 0.00321131061654392
23.952123950379 0.00334684478738234
23.9617723677534 0.00348502396697703
23.9714207851277 0.00362577967650663
23.981069202502 0.00376904521813529
23.9907176198764 0.00391475676310641
24.0003660372507 0.00406285446156078
24.0100144546251 0.00421328356341025
24.0196628719994 0.00436599553895165
24.0293112893737 0.00452094918736963
24.0389597067481 0.00467811172085995
24.0486081241224 0.00483745981181598
24.0582565414967 0.0049989805903761
24.0679049588711 0.00516267257963277
24.0775533762454 0.00532854655596389
24.0872017936198 0.00549662632226919
24.0968502109941 0.00566694938238069
24.1064986283684 0.0058395675055689
24.1161470457428 0.00601454717088388
24.1257954631171 0.00619196988204626
24.1354438804915 0.00637193234473804
24.1450922978658 0.00655454649941993
24.1547407152401 0.00673993940421468
24.1643891326145 0.00692825296393384
24.1740375499888 0.00711964350296382
24.1836859673632 0.00731428118146138
24.1933343847375 0.00751234925610592
24.2029828021118 0.00771404318850527
24.2126312194862 0.00791956960622799
24.2222796368605 0.00812914512330724
24.2319280542349 0.00834299502892108
24.2415764716092 0.00856135185475749
24.2512248889835 0.00878445383330929
24.2608733063579 0.00901254326098269
24.2705217237322 0.00924586478142053
24.2801701411065 0.00948466360581522
24.2898185584809 0.00972918368819458
24.2994669758552 0.00997966587468601
24.3091153932296 0.0102363460465837
24.3187638106039 0.0104994532776442
24.3284122279782 0.0107692080264026
24.3380606453526 0.0110458203844299
24.3477090627269 0.0113294884013326
24.3573574801013 0.0116203965069168
24.3670058974756 0.011918714050322
24.3766543148499 0.0122245939750538
24.3863027322243 0.0125381716477309
24.3959511495986 0.0128595638570207
24.405599566973 0.0131888679976767
24.4152479843473 0.0135261614528314
24.4248964017216 0.0138715011857591
24.434544819096 0.0142249235502262
24.4441932364703 0.0145864443263216
24.4538416538447 0.0149560589863237
24.463490071219 0.015333743192754
24.4731384885933 0.0157194535283117
24.4827869059677 0.0161131284549132
24.492435323342 0.0165146894966075
24.5020837407164 0.0169240426387274
24.5117321580907 0.0173410799333079
24.521380575465 0.0177656812985772
24.5310289928394 0.0181977164982291
24.5406774102137 0.0186370472842601
24.550325827588 0.0190835296853991
24.5599742449624 0.0195370164216075
24.5696226623367 0.0199973594238085
24.5792710797111 0.0204644124369038
24.5889194970854 0.020938033683298
24.5985679144597 0.0214180885635532
24.6082163318341 0.021904452370465
24.6178647492084 0.0223970129927756
24.6275131665828 0.0228956735849081
24.6371615839571 0.0234003551795453
24.6468100013314 0.0239109992205218
24.6564584187058 0.0244275699943959
24.6661068360801 0.0249500569401518
24.6757552534545 0.0254784768177657
24.6854036708288 0.0260128757178264
24.6950520882031 0.026553330895982
24.7047005055775 0.0270999524177148
24.7143489229518 0.0276528846007515
24.7239973403262 0.028212307244309
24.7336457577005 0.0287784366363008
24.7432941750748 0.0293515263315918
24.7529425924492 0.02993186769634
24.7625910098235 0.0305197902153868
24.7722394271978 0.0311156615615494
24.7818878445722 0.0317198874274751
24.7915362619465 0.0323329111224628
24.8011846793209 0.0329552129382829
24.8108330966952 0.0335873092895592
24.8204815140695 0.0342297516356824
24.8301299314439 0.0348831251924976
24.8397783488182 0.0355480474431562
24.8494267661926 0.0362251664585372
24.8590751835669 0.0369151590385098
24.8687236009412 0.0376187286860589
24.8783720183156 0.0383366034269087
24.8880204356899 0.0390695334877703
24.8976688530643 0.0398182888467232
24.9073172704386 0.0405836566695132
24.9169656878129 0.0413664386457275
24.9266141051873 0.0421674482389115
24.9362625225616 0.0429875078647019
24.945910939936 0.043827446011026
24.9555593573103 0.0446880943143206
24.9652077746846 0.0455702846055861
24.974856192059 0.0464748459399411
24.9845046094333 0.0474026016231424
24.9941530268076 0.0483543662483409
25.003801444182 0.0493309427561224
25.0134498615563 0.0503331195306581
25.0230982789307 0.0513616675445738
25.032746696305 0.0524173375648912
25.0423951136793 0.0535008574321866
25.0520435310537 0.054612929424848
25.061691948428 0.0557542277200651
25.0713403658024 0.05692539596292
25.0809887831767 0.0581270449546496
25.090637200551 0.0593597504708373
25.1002856179254 0.0606240512199342
25.1099340352997 0.0619204469521101
25.1195824526741 0.0632493967279907
25.1292308700484 0.064611317356313
25.1388792874227 0.0660065820089725
25.1485277047971 0.0674355190212594
25.1581761221714 0.0688984108843685
25.1678245395458 0.0703954934364222
25.1774729569201 0.0719269552573558
25.1871213742944 0.0734929372719835
25.1967697916688 0.0750935325644879
25.2064182090431 0.0767287864063568
25.2160666264174 0.0783986964985134
25.2257150437918 0.080103213427024
25.2353634611661 0.0818422413302754
25.2450118785405 0.0836156387740316
25.2546602959148 0.0854232198291529
25.2643087132891 0.0872647553451361
25.2739571306635 0.0891399744109474
25.2836055480378 0.0910485659929272
25.2932539654122 0.0929901807378347
25.3029023827865 0.0949644329274103
25.3125508001608 0.0969709025691766
25.3221992175352 0.0990091376065931
25.3318476349095 0.101078656230149
25.3414960522839 0.103178949269536
25.3511444696582 0.105309482645732
25.3607928870325 0.107469699860658
25.3704413044069 0.109659024501005
25.3800897217812 0.111876862732063
25.3897381391556 0.114122605756653
25.3993865565299 0.116395632213934
25.4090349739042 0.118695310492601
25.4186833912786 0.121021000933096
25.4283318086529 0.123372057893769
25.4379802260272 0.125747831656527
25.4476286434016 0.128147670148405
25.4572770607759 0.130570920456677
25.4669254781503 0.133016930116561
25.4765738955246 0.135485048152364
25.4862223128989 0.137974625854934
25.4958707302733 0.140485017280599
25.5055191476476 0.143015579459381
25.515167565022 0.145565672303069
25.5248159823963 0.148134658206832
25.5344643997706 0.150721901341289
25.544112817145 0.153326766635408
25.5537612345193 0.15594861845421
25.5634096518937 0.158586818978929
25.573058069268 0.161240726301072
25.5827064866423 0.163909692245621
25.5923549040167 0.166593059942434
25.602003321391 0.169290161168608
25.6116517387654 0.172000313488234
25.6213001561397 0.174722817219409
25.630948573514 0.177456952261722
25.6405969908884 0.180201974820335
25.6502454082627 0.182957114065628
25.6598938256371 0.185721568769653
25.6695422430114 0.188494503962678
25.6791906603857 0.191275047654608
25.6888390777601 0.194062287667193
25.6984874951344 0.196855268623443
25.7081359125087 0.199652989140771
25.7177843298831 0.202454399273837
25.7274327472574 0.205258398251994
25.7370811646318 0.208063832554603
25.7467295820061 0.210869494365228
25.7563779993804 0.213674120442998
25.7660264167548 0.216476391446016
25.7756748341291 0.219274931737887
25.7853232515035 0.222068309704125
25.7949716688778 0.224855038600325
25.8046200862521 0.227633577948932
25.8142685036265 0.230402335495819
25.8239169210008 0.233159669732156
25.8335653383752 0.235903892981074
25.8432137557495 0.238633275042447
25.8528621731238 0.241346047383033
25.8625105904982 0.24404040785303
25.8721590078725 0.246714525904091
25.8818074252469 0.249366548278059
25.8914558426212 0.25199460513011
25.9011042599955 0.254596816544854
25.9107526773699 0.257171299399197
25.9204010947442 0.259716174521544
25.9300495121185 0.262229574093293
25.9396979294929 0.264709649235519
25.9493463468672 0.267154577721455
25.9589947642416 0.269562571753714
25.9686431816159 0.27193188574434
25.9782915989902 0.274260824035652
25.9879400163646 0.276547748500489
25.9975884337389 0.278791085961868
26.0072368511133 0.280989335374222
26.0168852684876 0.28314107471118
26.0265336858619 0.285244967508445
26.0361821032363 0.28729976901432
26.0458305206106 0.289304331905223
26.055478937985 0.291257611528564
26.0651273553593 0.293158670640915
26.0747757727336 0.295006683615247
26.084424190108 0.296800940097013
26.0940726074823 0.298540848095048
26.1037210248567 0.300225936499399
26.113369442231 0.301855857024335
26.1230178596053 0.303430385580693
26.1326662769797 0.304949423087399
26.142314694354 0.306412995737325
26.1519631117283 0.30782125473753
26.1616115291027 0.309174475548356
26.171259946477 0.310473056649645
26.1809083638514 0.31171751786563
26.1905567812257 0.312908498282583
26.2002051986 0.314046753795226
26.2098536159744 0.315133154319127
26.2195020333487 0.316168680706813
26.2291504507231 0.317154421405153
26.2387988680974 0.318091568890772
26.2484472854717 0.318981415918755
26.2580957028461 0.319825351617936
26.2677441202204 0.320624857463445
26.2773925375948 0.321381503154246
26.2870409549691 0.322096942419977
26.2966893723434 0.322772908777698
26.3063377897178 0.323411211255299
26.3159862070921 0.324013730094225
26.3256346244665 0.324582412440145
26.3352830418408 0.325119268026111
26.3449314592151 0.325626364848914
26.3545798765895 0.326105824835617
26.3642282939638 0.326559819493883
26.3738767113381 0.326990565536679
26.3835251287125 0.327400320469366
26.3931735460868 0.327791378125047
26.4028219634612 0.328166064132478
26.4124703808355 0.32852673129978
26.4221187982098 0.328875754896792
26.4317672155842 0.329215527818977
26.4414156329585 0.329548455616557
26.4510640503329 0.329876951373807
26.4607124677072 0.330203430425317
26.4703608850815 0.330530304898336
26.4800093024559 0.330859978073151
26.4896577198302 0.331194838556689
26.4993061372046 0.331537254268079
26.5089545545789 0.331889566238816
26.5186029719532 0.332254082234196
26.5282513893276 0.332633070206935
26.5378998067019 0.333028751598129
26.5475482240763 0.333443294504949
26.5571966414506 0.333878806738612
26.5668450588249 0.334337328800103
26.5764934761993 0.334820826804848
26.5861418935736 0.335331185390914
26.595790310948 0.335870200648299
26.6054387283223 0.336439573109471
26.6150871456966 0.337040900843364
26.624735563071 0.337675672696636
26.6343839804453 0.338345261726973
26.6440323978196 0.339050918873704
26.653680815194 0.339793766910829
26.6633292325683 0.34057479472687
26.6729776499427 0.341394851974697
26.682626067317 0.342254644132613
26.6922744846913 0.3431547280157
26.7019229020657 0.344095507773567
26.71157131944 0.345077231407398
26.7212197368144 0.346099987835553
26.7308681541887 0.347163704532985
26.740516571563 0.348268145765496
26.7501649889374 0.349412911435377
26.7598134063117 0.350597436550393
26.7694618236861 0.351820991323352
26.7791102410604 0.353082681904827
26.7887586584347 0.354381451746933
26.7984070758091 0.3557160835915
26.8080554931834 0.357085202071666
26.8177039105578 0.358487276911686
26.8273523279321 0.359920626705926
26.8370007453064 0.361383423254439
26.8466491626808 0.362873696429241
26.8562975800551 0.364389339542637
26.8659459974294 0.36592811518643
26.8755944148038 0.367487661508844
26.8852428321781 0.369065498894356
26.8948912495525 0.370659037010416
26.9045396669268 0.372265582184281
26.9141880843011 0.37388234507271
26.9238365016755 0.375506448587326
26.9334849190498 0.3771349360387
26.9431333364242 0.378764779462884
26.9527817537985 0.380392888095043
26.9624301711728 0.382016116955949
26.9720785885472 0.383631275518543
26.9817270059215 0.385235136423212
26.9913754232959 0.38682444421212
27.0010238406702 0.388395924054651
27.0106722580445 0.389946290437729
27.0203206754189 0.391472255796584
27.0299690927932 0.392970539063189
27.0396175101676 0.394437874111249
27.0492659275419 0.395871018078144
27.0589143449162 0.39726675954561
27.0685627622906 0.39862192656222
27.0782111796649 0.399933394491799
27.0878595970392 0.401198093672817
27.0975080144136 0.402413016874585
27.1071564317879 0.403575226536611
27.1168048491623 0.404681861777928
27.1264532665366 0.40573014516344
27.1361016839109 0.406717389214488
27.1457501012853 0.407641002650828
27.1553985186596 0.40849849635119
27.165046936034 0.40928748901942
27.1746953534083 0.410005712543078
27.1843437707826 0.410651017031222
27.193992188157 0.411221375517968
27.2036406055313 0.411714888318395
27.2132890229057 0.412129787023389
27.22293744028 0.412464438120217
27.2325858576543 0.412717346225921
27.2422342750287 0.412887156921151
27.251882692403 0.412972659172732
27.2615311097774 0.412972787334163
27.2711795271517 0.412886622714386
27.280827944526 0.41271339470649
27.2904763619004 0.412452481469599
27.3001247792747 0.412103410158958
27.3097731966491 0.41166585670127
27.3194216140234 0.411139645114455
27.3290700313977 0.41052474637342
27.3387184487721 0.409821276825871
27.3483668661464 0.409029496164825
27.3580152835207 0.408149804967177
27.3676637008951 0.407182741810361
27.3773121182694 0.406128979981905
27.3869605356438 0.40498932379935
27.3966089530181 0.403764704560623
27.4062573703924 0.402456176147408
27.4159057877668 0.401064910306426
27.4255542051411 0.399592191635643
27.4352026225155 0.398039412304302
27.4448510398898 0.396408066537333
27.4544994572641 0.394699744896007
27.4641478746385 0.392916128387732
27.4737962920128 0.391058982438555
27.4834447093872 0.389130150762292
27.4930931267615 0.387131549160166
27.5027415441358 0.385065159284479
27.5123899615102 0.382933022399121
27.5220383788845 0.380737233168649
27.5316867962589 0.378479933506302
27.5413352136332 0.376163306509626
27.5509836310075 0.373789570510464
27.5606320483819 0.371360973263849
27.5702804657562 0.368879786297947
27.5799288831305 0.366348299444683
27.5895773005049 0.363768815567954
27.5992257178792 0.361143645503613
27.6088741352536 0.358475103222598
27.6185225526279 0.355765501225792
27.6281709700022 0.35301714617647
27.6378193873766 0.350232334773542
27.6474678047509 0.347413349866247
27.6571162221253 0.344562456808687
27.6667646394996 0.341681900050307
27.6764130568739 0.338773899956599
27.6860614742483 0.33584064985255
27.6957098916226 0.332884313279963
27.705358308997 0.32990702145859
27.7150067263713 0.326910870940179
27.7246551437456 0.323897921443919
27.73430356112 0.320870193861452
27.7439519784943 0.317829668419587
27.7536003958687 0.314778282989059
27.763248813243 0.311717931528082
27.7728972306173 0.308650462650124
27.7825456479917 0.305577678306175
27.792194065366 0.302501332572714
27.8018424827403 0.299423130537761
27.8114909001147 0.296344727278531
27.821139317489 0.293267726925506
27.8307877348634 0.290193681809017
27.8404361522377 0.287124091685633
27.850084569612 0.284060403042976
27.8597329869864 0.281004008482609
27.8693814043607 0.27795624618178
27.8790298217351 0.274918399435663
27.8886782391094 0.271891696282513
27.8983266564837 0.268877309214801
27.9079750738581 0.265876354979711
27.9176234912324 0.262889894472751
27.9272719086068 0.259918932728155
27.9369203259811 0.256964419009681
27.9465687433554 0.254027247005041
27.9562171607298 0.251108255126691
27.9658655781041 0.248208226921067
27.9755139954785 0.245327891587456
27.9851624128528 0.242467924606822
27.9948108302271 0.239628948479797
28.0044592476015 0.236811533571921
28.0141076649758 0.234016199063041
28.0237560823502 0.23124341399654
28.0334044997245 0.228493598422876
28.0430529170988 0.225767124630704
28.0527013344732 0.223064318457779
28.0623497518475 0.220385460672767
28.0719981692218 0.217730788418183
28.0816465865962 0.215100496703901
28.0912950039705 0.212494739940057
28.1009434213449 0.209913633497696
28.1105918387192 0.207357255285264
28.1202402560935 0.204825647328983
28.1298886734679 0.202318817345262
28.1395370908422 0.199836740293676
28.1491855082166 0.197379359899545
28.1588339255909 0.194946590135952
28.1684823429652 0.192538316655907
28.1781307603396 0.190154398166518
28.1877791777139 0.187794667738291
28.1974275950883 0.185458934044046
28.2070760124626 0.183146982523511
28.2167244298369 0.180858576471177
28.2263728472113 0.178593458046738
28.2360212645856 0.176351349209056
28.2456696819599 0.174131952576309
28.2553180993343 0.171934952216649
28.2649665167086 0.169760014375215
28.274614934083 0.167606788144934
28.2842633514573 0.165474906089824
28.2939117688316 0.163363984830799
28.303560186206 0.161273625605004
28.3132086035803 0.15920341481052
28.3228570209547 0.157152924549012
28.332505438329 0.155121713179205
28.3421538557033 0.153109325894295
28.3518022730777 0.151115295336341
28.361450690452 0.149139142260311
28.3710991078264 0.147180376259936
28.3807475252007 0.145238496566685
28.390395942575 0.14331299293214
28.4000443599494 0.141403346602805
28.4096927773237 0.139509031394913
28.4193411946981 0.13762951487519
28.4289896120724 0.135764259651722
28.4386380294467 0.133912724777214
28.4482864468211 0.1320743672649
28.4579348641954 0.130248643715345
28.4675832815698 0.128435012050284
28.4772316989441 0.126632933347572
28.4868801163184 0.124841873769313
28.4965285336928 0.123061306573262
28.5061769510671 0.121290714195774
28.5158253684414 0.119529590392849
28.5254737858158 0.117777442424293
28.5351222031901 0.116033793264692
28.5447706205645 0.114298183823717
28.5544190379388 0.112570175157452
28.5640674553131 0.11084935065173
28.5737158726875 0.109135318158127
28.5833642900618 0.10742771206312
28.5930127074362 0.105726195271082
28.6026611248105 0.104030461082216
28.6123095421848 0.102340234947207
28.6219579595592 0.1006552760813
28.6316063769335 0.0989753789217118
28.6412547943079 0.0973003744136443
28.6509032116822 0.095630131111771
28.6605516290565 0.0939645560858394
28.6702000464309 0.0923035956208998
28.6798484638052 0.0906472357047296
28.6894968811796 0.0889955022971015
28.6991452985539 0.0873484613777073
28.7087937159282 0.0857062187717393
28.7184421333026 0.0840689197542658
28.7280905506769 0.0824367484367095
28.7377389680512 0.0808099269407505
28.7473873854256 0.0791887143669513
28.7570358027999 0.0775734055672366
28.7666842201743 0.0759643297320083
28.7763326375486 0.0743618488042445
28.7859810549229 0.0727663557341923
28.7956294722973 0.0711782725894464
28.8052778896716 0.069598048536099
28.814926307046 0.0680261577073317
28.8245747244203 0.0664630969763523
28.8342231417946 0.0649093836507906
28.843871559169 0.0633655531057668
28.8535199765433 0.0618321563726624
28.8631683939177 0.0603097577003032
28.872816811292 0.0587989321047315
28.8824652286663 0.0573002629230455
28.8921136460407 0.0558143393859713
28.901762063415 0.0543417542228569
28.9114104807894 0.052883101311722
28.9210588981637 0.0514389733858495
28.930707315538 0.0500099598071953
28.9403557329124 0.0485966444156527
28.9500041502867 0.0471996034619429
28.959652567661 0.0458194036306441
28.9693009850354 0.0444566001586525
28.9789494024097 0.0431117350531576
28.9885978197841 0.0417853354120972
28.9982462371584 0.0404779118490034
29.0078946545327 0.0391899570231581
29.0175430719071 0.0379219442751258
29.0271914892814 0.0366743263669273
29.0368399066558 0.0354475343254722
29.0464883240301 0.0342419763873013
29.0561367414044 0.0330580370422308
29.0657851587788 0.0318960761731757
29.0754335761531 0.0307564282891599
29.0850819935275 0.0296394018484053
29.0947304109018 0.0285452786683316
29.1043788282761 0.0274743134193303
29.1140272456505 0.0264267331992777
29.1236756630248 0.0254027371859129
29.1333240803992 0.0244024963644137
29.1429724977735 0.0234261533277451
29.1526209151478 0.0224738221476233
29.1622693325222 0.0215455883142192
29.1719177498965 0.020641508742992
29.1815661672709 0.019761611847328
29.1912145846452 0.0189058976758886
29.2008630020195 0.018074338113804
29.2105114193939 0.017266877147028
29.2201598367682 0.0164834311892912
29.2298082541425 0.0157238894712129
29.2394566715169 0.0149881144911391
29.2491050888912 0.0142759425272833
29.2587535062656 0.0135871842106744
29.2684019236399 0.012921625158291
29.2780503410142 0.0122790266656093
29.2876987583886 0.0116591264575582
29.2973471757629 0.0110616394966317
29.3069955931373 0.010486258846609
29.3166440105116 0.00993265659000795
29.3262924278859 0.00940048479705281
29.3359408452603 0.00888937654356784
29.3455892626346 0.00839894697483881
29.355237680009 0.00792879441210782
29.3648860973833 0.00747850149799777
29.3745345147576 0.00704763637681367
29.384182932132 0.00663575390532462
29.3938313495063 0.00624239688933742
29.4034797668807 0.00586709734109059
29.413128184255 0.00550937775226918
29.4227766016293 0.00516875237725282
29.4324250190037 0.00484472852106214
29.442073436378 0.00453680782638242
29.4517218537523 0.00424448755400074
29.4613702711267 0.00396726185100394
29.471018688501 0.0037046230011577
29.4806671058754 0.00345606265199749
29.4903155232497 0.00322107301333849
29.499963940624 0.00299914802212665
29.5096123579984 0.00278978446881789
29.5192607753727 0.00259248308077788
29.5289091927471 0.00240674955853855
29.5385576101214 0.00223209556112354
29.5482060274957 0.00206803963705987
29.5578544448701 0.00191410809811948
29.5675028622444 0.00176983583328093
29.5771512796188 0.00163476706085569
29.5867996969931 0.00150845601718844
29.5964481143674 0.00139046758080524
29.6060965317418 0.0012803778313422
29.6157449491161 0.00117777454304086
29.6253933664905 0.00108225761303219
29.6350417838648 0.000993439425053404
29.6446902012391 0.000910945149639343
29.6543386186135 0.00083441298220456
29.6639870359878 0.000763494320778695
29.6736354533621 0.000697853885473068
29.6832838707365 0.000637169782042285
29.6929322881108 0.000581133512153514
29.7025807054852 0.00052944993319419
29.7122291228595 0.000481837170630604
29.7218775402338 0.000438026486076182
29.7315259576082 0.000397762104342788
29.7411743749825 0.000360801002826801
29.7508227923569 0.000326912666629784
29.7604712097312 0.000295878812830008
29.7701196271055 0.000267493087308719
29.7797680444799 0.00024156073749596
29.7894164618542 0.000217898264335866
29.7990648792286 0.000196333056685333
29.8087132966029 0.000176703011252351
29.8183617139772 0.000158856141055826
29.8280101313516 0.000142650175248886
29.8376585487259 0.000127952152994724
29.8473069661003 0.000114638013921321
29.8569553834746 0.000102592187510209
29.8666038008489 9.17071835979768e-05
29.8762522182233 8.18831859889428e-05
29.8859006355976 7.30276509956544e-05
29.895549052972 6.50549125426193e-05
29.9051974703463 5.78857952892201e-05
29.9148458877206 5.14472370520217e-05
29.924494305095 4.56719216358181e-05
29.9341427224693 4.04979230180277e-05
29.9437911398436 3.5868361673437e-05
29.953439557218 3.17310736766522e-05
29.9630879745923 2.80382930786164e-05
29.9727363919667 2.47463479217677e-05
29.982384809341 2.181537013604e-05
29.9920332267153 1.92090194455013e-05
30.0016816440897 1.68942213126893e-05
30.011330061464 1.48409188549552e-05
30.0209784788384 1.30218385840865e-05
30.0306268962127 1.14122697469058e-05
30.040275313587 9.98985698032712e-06
30.0499237309614 8.73440593886827e-06
30.0595721483357 7.62770150568815e-06
30.0692205657101 6.65333815926472e-06
30.0788689830844 5.79656203630893e-06
30.0885174004587 5.04412420705846e-06
30.0981658178331 4.38414466097487e-06
30.1078142352074 3.80598648869862e-06
30.1174626525818 3.30013973925076e-06
30.1271110699561 2.85811442938225e-06
30.1367594873304 2.47234218418921e-06
30.1464079047048 2.13608599401939e-06
30.1560563220791 1.84335758190282e-06
30.1657047394534 1.58884188771118e-06
30.1753531568278 1.36782818956631e-06
30.1850015742021 1.17614739928363e-06
30.1946499915765 1.01011508642954e-06
30.2042984089508 8.66479804587317e-07
30.2139468263251 7.42376313307526e-07
30.2235952436995 6.35283309696028e-07
30.2332436610738 5.42985304408857e-07
30.2428920784482 4.63538297735117e-07
30.2525404958225 3.95238932274381e-07
30.2621889131968 3.36596819267671e-07
30.2718373305712 2.86309755779202e-07
30.2814857479455 2.43241569526605e-07
30.2911341653199 2.0640234711375e-07
30.3007825826942 1.74930819657601e-07
30.3104310000685 1.48078697250416e-07
30.3200794174429 1.25196760317749e-07
30.3297278348172 1.05722531688539e-07
30.3393762521916 8.91693680679819e-08
30.3490246695659 7.51168235900399e-08
30.3586730869402 6.32021512274672e-08
30.3683215043146 5.31128200646924e-08
30.3779699216889 4.45799378117505e-08
30.3876183390632 3.73724784799181e-08
30.3972667564376 3.12922248816762e-08
30.4069151738119 2.61693445918534e-08
30.4165635911863 2.18585262491222e-08
30.4262120085606 1.82356106256136e-08
30.4358604259349 1.51946577858509e-08
30.4455088433093 1.26453979347283e-08
30.4551572606836 1.05110192569939e-08
30.464805678058 8.72625121752363e-09
30.4744540954323 7.23570646087734e-09
30.4841025128066 5.99244865767437e-09
30.493750930181 4.95675743061393e-09
30.5033993475553 4.09506488878847e-09
30.5130477649297 3.37904133852383e-09
30.522696182304 2.78481045333556e-09
30.5323445996783 2.29227660409918e-09
30.5419930170527 1.88454920073788e-09
30.551641434427 1.5474508041232e-09
30.5612898518014 1.26909745536845e-09
30.5709382691757 1.03954116115969e-09
30.58058668655 8.50465788480613e-10
30.5902351039244 6.94928778620582e-10
30.5998835212987 5.67142105638013e-10
30.6095319386731 4.62286793981738e-10
30.6191803560474 3.76356087749673e-10
30.6288287734217 3.06023042804627e-10
30.6384771907961 2.48528904131058e-10
30.6481256081704 2.01589144703229e-10
30.6577740255447 1.63314488006176e-10
30.6674224429191 1.32144622469897e-10
30.6770708602934 1.06792649829551e-10
30.6867192776678 8.61985973773073e-11
30.6963676950421 6.94905720504122e-11
30.7060161124164 5.59523474675883e-11
30.7156645297908 4.49963579349117e-11
30.7253129471651 3.6141230106671e-11
30.7349613645395 2.8993116933375e-11
30.7446097819138 2.32302128482881e-11
30.7542581992881 1.85899265410767e-11
30.7639066166625 1.48582705016444e-11
30.7735550340368 1.18610968442706e-11
30.7832034514112 9.45686852727568e-12
30.7928518687855 7.53070551450328e-12
30.8025002861598 5.98948802131008e-12
30.8121487035342 4.75783490554763e-12
30.8217971209085 3.7748054992406e-12
30.8314455382828 2.99119858574413e-12
30.8410939556572 2.36734354471651e-12
30.8507423730315 1.8712965424965e-12
30.8603907904059 1.47736957599144e-12
30.8700392077802 1.16493264253442e-12
30.8796876251545 9.17439696682465e-13
30.8893360425289 7.21637699170857e-13
30.8989844599032 5.66925244461132e-13
30.9086328772776 4.44833209444612e-13
30.9182812946519 3.4860479798049e-13
30.9279297120262 2.72856433539801e-13
30.9375781294006 2.13304317976172e-13
30.9472265467749 1.66544248202285e-13
30.9568749641493 1.29874564814481e-13
30.9665233815236 1.011539816097e-13
30.9761717988979 7.8687582761336e-14
30.9858202162723 6.11355337899572e-14
30.9954686336466 4.74400821910669e-14
31.005117051021 3.67672642258431e-14
31.0147654683953 2.84604196005446e-14
31.0244138857696 2.20031734251423e-14
31.034062303144 1.69899979975533e-14
31.0437107205183 1.31028346258436e-14
31.0533591378926 1.00925535468705e-14
31.063007555267 7.76427091683327e-15
31.0726559726413 5.96573642059981e-15
31.0823043900157 4.57816194941473e-15
31.09195280739 3.50898813137672e-15
31.1016012247643 2.68618706722596e-15
31.1112496421387 2.05378115773213e-15
31.120898059513 1.56832326152097e-15
31.1305464768874 1.19613572626371e-15
31.1401948942617 9.11147634415211e-16
31.149843311636 6.93202957376535e-16
31.1594917290104 5.26738885933526e-16
31.1691401463847 3.99754746306858e-16
31.1787885637591 3.03008705745455e-16
31.1884369811334 2.29392793327069e-16
31.1980853985077 1.73447313001269e-16
31.2077338158821 1.3098407073953e-16
31.2173822332564 9.8794427659577e-17
31.2270306506308 7.44233875559168e-17
31.2366790680051 5.5995019025557e-17
31.2463274853794 4.20777295015144e-17
31.2559759027538 3.15804342629293e-17
31.2656243201281 2.36726435652828e-17
31.2752727375025 1.77230410632411e-17
31.2849211548768 1.32523383205984e-17
31.2945695722511 9.89713592564862e-18
31.3042179896255 7.38225884185561e-18
31.3138664069998 5.49960730930439e-18
31.3235148243742 4.09201046199693e-18
31.3331632417485 3.04091582678096e-18
31.3428116591228 2.25701599701562e-18
31.3524600764972 1.67312125662519e-18
31.3621084938715 1.23874725444038e-18
31.3717569112458 9.1601053606617e-19
31.3814053286202 6.76520057574327e-19
31.3910537459945 4.99026142816879e-19
31.4007021633689 3.67644692102838e-19
31.4103505807432 2.70517679549842e-19
31.4199989981175 1.98804095804965e-19
31.4296474154919 1.45920830517773e-19
31.4392958328662 1.06972345868479e-19
31.4489442502406 7.83227594094816e-20
};
\addlegendentry{SINR at the receiver (KDE)};

\end{axis}

\end{tikzpicture}

%% file: figures/history_keras_lstm.tikz
\begin{tikzpicture}

\begin{axis}[
legend cell align={left},
legend style={fill opacity=0.8, draw opacity=1, text opacity=1, nodes={scale=.8}, draw=white!80!black},
tick align=outside,
tick pos=left,
x grid style={white!69.0196078431373!black},
xlabel={Epoch},
xmajorgrids,
xmin=0, xmax=50,
xminorgrids,
xtick style={color=black},
y grid style={white!69.0196078431373!black},
ylabel={Loss},
ymajorgrids,
scale=0.9,
ymin=1.48684486150742, ymax=2.1117729306221,
yminorgrids,
ytick style={color=black}
]
\addplot [line width=1.75, red]
table {%
0 1.82190132141113
1 1.72662389278412
2 1.65988218784332
3 1.63471472263336
4 1.62005019187927
5 1.60897815227509
6 1.60590887069702
7 1.61250925064087
8 1.61659741401672
9 1.61570310592651
10 1.61467361450195
11 1.61078667640686
12 1.60619056224823
13 1.60560321807861
14 1.60685777664185
15 1.60969495773315
16 1.61388647556305
17 1.61912250518799
18 1.62237763404846
19 1.62257134914398
20 1.62124121189117
21 1.62051796913147
22 1.61946523189545
23 1.61748397350311
24 1.61579382419586
25 1.61495387554169
26 1.61641824245453
27 1.61968994140625
28 1.62230372428894
29 1.62016916275024
30 1.61595892906189
31 1.61263024806976
32 1.61189877986908
33 1.61161196231842
34 1.61322057247162
35 1.61403012275696
36 1.61163735389709
37 1.60965394973755
38 1.61141800880432
39 1.61365818977356
40 1.61550915241241
41 1.61392784118652
42 1.61079466342926
43 1.61030733585358
44 1.60826146602631
45 1.60637533664703
46 1.60455012321472
47 1.60610592365265
48 1.61041831970215
49 1.61495995521545
};
\addlegendentry{Validation loss}
\addplot [line width=1.75, blue]
table {%
0 2.08336710929871
1 1.95774900913239
2 1.80362546443939
3 1.72501003742218
4 1.69896984100342
5 1.65659701824188
6 1.61825346946716
7 1.6139007806778
8 1.61577117443085
9 1.58999049663544
10 1.58822858333588
11 1.59079432487488
12 1.58639359474182
13 1.57251179218292
14 1.58418393135071
15 1.58217835426331
16 1.58212065696716
17 1.57701992988586
18 1.5767126083374
19 1.57042694091797
20 1.54971587657928
21 1.56009364128113
22 1.56184387207031
23 1.55634760856628
24 1.56857264041901
25 1.55338847637177
26 1.55976271629333
27 1.54168581962585
28 1.5621532201767
29 1.54475688934326
30 1.56283390522003
31 1.55256974697113
32 1.54428434371948
33 1.55606389045715
34 1.55265355110168
35 1.55148553848267
36 1.54597961902618
37 1.52553009986877
38 1.52632319927216
39 1.54406177997589
40 1.53077256679535
41 1.5314918756485
42 1.55224621295929
43 1.51525068283081
44 1.54460799694061
45 1.53807842731476
46 1.56412887573242
47 1.53632342815399
48 1.52937936782837
49 1.52474367618561
};
\addlegendentry{Loss}
\end{axis}

\end{tikzpicture}

%% file: figures/history_keras_CNN_equalization_learning.tikz
\begin{tikzpicture}

\begin{axis}[
tick align=outside,
tick pos=left,
x grid style={white!69.0196078431373!black},
xlabel={Epoch},
xmajorgrids,
xmin=-4.75, xmax=90,
xminorgrids,
xtick style={color=black},
y grid style={white!69.0196078431373!black},
ylabel={Loss},
ymajorgrids,
ymin=-0.00657773418355632, ymax=0.13813732938309,
yminorgrids,
scale=0.9,
ytick style={color=black},
yticklabels={$0$, $0.05$, $0.1$, $0.15$},
legend cell align={left},
legend style={at={(.94,0.96)}, fill opacity=0.6, draw opacity=1, text opacity=1, draw=white!80!black, nodes={scale=.8}},
]
\addplot [line width=1.75, red]
table {%
0 0.131559371948242
1 0.0103642651811242
2 0.00473615806549788
3 0.000346826069289818
4 0.00801152735948563
5 0.0106243127956986
6 0.00129236828070134
7 0.000862309825606644
8 0.000115886665298603
9 0.000110175438749138
10 1.49676952787559e-05
11 5.43674468644895e-05
12 9.50727189774625e-06
13 0.000613795185927302
14 0.00319079007022083
15 0.00470801116898656
16 0.000182461299118586
17 0.00103749369736761
18 0.00445706024765968
19 0.0185614340007305
20 0.00707923760637641
21 0.00117396365385503
22 0.000415046524722129
23 0.000341429637046531
24 0.000175237990333699
25 2.80440781352809e-05
26 1.47311766340863e-05
27 4.90232159791049e-05
28 0.00040633010212332
29 0.00125422200653702
30 0.000865281093865633
31 0.000725402496755123
32 0.00163883715867996
33 3.68464970961213e-05
34 0.0123683242127299
35 0.000142759308801033
36 2.71356129815103e-05
37 5.56194800083176e-06
38 3.37710880558006e-05
39 2.13511202673544e-06
40 5.09422443428775e-06
41 0.000572202901821584
42 2.73167697741883e-05
43 0.000221997615881264
44 0.00046751502668485
45 0.00183648557867855
46 8.42111185193062e-05
47 0.000217776789213531
48 0.000879252038430423
49 0.00198941538110375
50 0.000126078928587958
51 0.000259086868027225
52 0.00102926336694509
53 3.22402120218612e-05
54 0.00162720039952546
55 5.79052320972551e-05
56 6.17217447143048e-05
57 0.000262055487837642
58 2.22909075091593e-05
59 0.000129602150991559
60 0.000580900406930596
61 0.000198950117919594
62 1.52091661220766e-05
63 6.80253142490983e-05
64 0.000382940255803987
65 0.0113222328945994
66 0.000150488674989901
67 0.000126001294120215
68 2.39785586018115e-05
69 0.000218627290450968
70 0.000831175420898944
71 3.15833422064316e-05
72 7.59497052058578e-05
73 2.31437006732449e-05
74 8.22972651803866e-06
75 0.00597420195117593
76 0.00876775570213795
77 0.000135979309561662
78 0.00235034245997667
79 6.28254383627791e-06
80 2.92544409603579e-05
81 0.00120731547940522
82 0.000393192283809185
83 3.42583261954132e-05
84 2.23251291231463e-07
85 0.00017857042257674
86 4.8409958708362e-07
87 0.00624543614685535
88 0.000798192748334259
89 0.00456478074193001
90 0.00308038224466145
91 9.77250238065608e-05
92 2.56614839599933e-05
93 5.39393513463438e-05
94 0.000192244071513414
95 0.000194127482245676
};
\addlegendentry{Validation loss}
\addplot [line width=1.75, blue]
table {%
0 0.0248571932315826
1 0.00488382810726762
2 0.00660106725990772
3 0.00449458276852965
4 0.00185177195817232
5 0.00925357080996037
6 0.00414204923436046
7 0.00624064402654767
8 0.00281039322726429
9 0.00271045067347586
10 0.000356456643203273
11 0.000321258761687204
12 0.000320641207508743
13 0.000276801263680682
14 0.0063744829967618
15 0.00930777564644814
16 0.00273123593069613
17 0.00265928567387164
18 0.000825589057058096
19 0.0191736873239279
20 0.00481484876945615
21 0.00540979718789458
22 0.00128893589135259
23 0.000876725709531456
24 0.000688796339090914
25 0.000350923946825787
26 0.000307902315398678
27 0.000973700720351189
28 0.000434447400039062
29 0.00176436745095998
30 0.00373746710829437
31 0.00486696511507034
32 0.00400182791054249
33 0.00151557277422398
34 0.000668452063109726
35 0.00319376401603222
36 0.00056422280613333
37 0.000420698634115979
38 0.000478434056276456
39 0.00035051119630225
40 0.000344895233865827
41 0.000992657849565148
42 0.000392399320844561
43 0.000354892428731546
44 0.00219113240018487
45 0.00264979340136051
46 0.00253677531145513
47 0.000565644877497107
48 0.00391788501292467
49 0.00249787792563438
50 0.00179491052404046
51 0.00122920039575547
52 0.0034234409686178
53 0.00248655350878835
54 0.00257475394755602
55 0.000690672546625137
56 0.00215888046659529
57 0.00183474633377045
58 0.00160804984625429
59 0.00120901071932167
60 0.00708889169618487
61 0.000867067195940763
62 0.000572399643715471
63 0.000381392484996468
64 0.00511619867756963
65 0.00734839355573058
66 0.00490128388628364
67 0.000827427254989743
68 0.000853782752528787
69 0.00124321796465665
70 0.00401601288467646
71 0.00387142226099968
72 0.00344615266658366
73 0.0010510120773688
74 0.000299720501061529
75 0.00172364385798573
76 0.0066316588781774
77 0.00346369645558298
78 0.00348321185447276
79 0.00150844873860478
80 0.000463793083326891
81 0.00347529607824981
82 0.00372046953998506
83 0.000529072945937514
84 0.000447965547209606
85 0.000396029936382547
86 0.000320859951898456
87 0.00224202638491988
88 0.00948767457157373
89 0.004039007704705
90 0.00939210597425699
91 0.00749642727896571
92 0.00103157933335751
93 0.000418502080719918
94 0.000509979785420001
95 0.000352284056134522
};
\addlegendentry{Loss}
\end{axis}

\end{tikzpicture}

%% file: figures/history_keras_autoencoder_compression_0.25.tikz
\begin{tikzpicture}

\begin{axis}[
legend cell align={left},
legend style={fill opacity=0.8, draw opacity=1, text opacity=1, nodes={scale=.8}, draw=white!80!black},
tick align=outside,
tick pos=left,
x grid style={white!69.0196078431373!black},
xlabel={Epoch},
xmajorgrids,
xmin=0, xmax=200, 
xminorgrids,
xtick style={color=black},
xtick={-50,0,50,100,150,200,250},
xticklabels={
  \(\displaystyle {\ensuremath{-}50}\),
  \(\displaystyle {0}\),
  \(\displaystyle {50}\),
  \(\displaystyle {100}\),
  \(\displaystyle {150}\),
  \(\displaystyle {200}\),
  \(\displaystyle {250}\)
},
y grid style={white!69.0196078431373!black},
ylabel={Loss},
ymajorgrids,
scale=0.9,
ymin=0, ymax=0.007,
yminorgrids,
ytick style={color=black},
]
\addplot [line width=1.75, red]
table {%
1 0.00638733571395278
0 0.00638978509232402
2 0.00638506561517715
3 0.00638292031362653
4 0.00638091657310724
5 0.00637896312400699
6 0.00637704227119684
7 0.00637517962604761
8 0.00637330953031778
9 0.00637144222855568
10 0.00636955350637436
11 0.00636766664683819
12 0.00636574253439903
13 0.0063637369312346
14 0.00636164844036102
15 0.00635944167152047
16 0.00635713106021285
17 0.00635472033172846
18 0.00635221973061562
19 0.00634953798726201
20 0.00634668534621596
21 0.00634362455457449
22 0.0063403220847249
23 0.00633675185963511
24 0.00633288314566016
25 0.00632867310196161
26 0.00632407981902361
27 0.00631907815113664
28 0.00631361128762364
29 0.00630762102082372
30 0.00630106357857585
31 0.00629388308152556
32 0.00628600968047976
33 0.00627736933529377
34 0.00626788940280676
35 0.00625750701874495
36 0.00624615140259266
37 0.00623371219262481
38 0.00622011907398701
39 0.00620531942695379
40 0.00618924992159009
41 0.00617183418944478
42 0.00615300098434091
43 0.00613270094618201
44 0.00611085537821054
45 0.00608744006603956
46 0.00606243731454015
47 0.00603586109355092
48 0.00600766856223345
49 0.00597788626328111
50 0.00594654306769371
51 0.00591369811445475
52 0.00587941845878959
53 0.00584376789629459
54 0.00580686656758189
55 0.00576863717287779
56 0.00572923058643937
57 0.00568865472450852
58 0.00564688071608543
59 0.00560388341546059
60 0.00555962510406971
61 0.00551406433805823
62 0.00546717178076506
63 0.00541894510388374
64 0.00536941038444638
65 0.00531862350180745
66 0.00526665756478906
67 0.00521361455321312
68 0.0051595619879663
69 0.00510454922914505
70 0.00504860933870077
71 0.00499176094308496
72 0.00493402173742652
73 0.00487541221082211
74 0.00481596915051341
75 0.00475574703887105
76 0.00469481572508812
77 0.00463325204327703
78 0.00457113375887275
79 0.00450852606445551
80 0.00444548623636365
81 0.00438205106183887
82 0.00431824894621968
83 0.00425410876050591
84 0.00418966170400381
85 0.00412495387718081
86 0.00406003929674625
87 0.00399498082697392
88 0.00392983946949244
89 0.00386467040516436
90 0.00379952089861035
91 0.00373442890122533
92 0.00366942957043648
93 0.00360455689951777
94 0.00353984790854156
95 0.00347534217871726
96 0.00341108022257686
97 0.00334710208699107
98 0.0032834461890161
99 0.00322014442645013
100 0.00315722823143005
101 0.00309472647495568
102 0.00303266663104296
103 0.00297107943333685
104 0.00290999328717589
105 0.00284943636506796
106 0.00278943544253707
107 0.0027300154324621
108 0.00267119845375419
109 0.0026130061596632
110 0.00255547277629375
111 0.00249858340248466
112 0.00244239019230008
113 0.00238689943216741
114 0.00233212765306234
115 0.002278090454638
116 0.00222480203956366
117 0.00217227428220212
118 0.00212051812559366
119 0.00206954288296402
120 0.00201935810036957
121 0.00196997122839093
122 0.00192139041610062
123 0.00187362241558731
124 0.00182667269837111
125 0.0017805447569117
126 0.0017352404538542
127 0.0016907611861825
128 0.00164710788521916
129 0.00160428136587143
130 0.00156228104606271
131 0.00152110552880913
132 0.00148075097240508
133 0.00144121388439089
134 0.00140248995739967
135 0.0013645754661411
136 0.00132746493909508
137 0.00129115267191082
138 0.00125563202891499
139 0.0012208956759423
140 0.00118693674448878
141 0.00115374766755849
142 0.00112132076174021
143 0.00108964741230011
144 0.0010587191209197
145 0.00102852680720389
146 0.000999062089249492
147 0.000970315712038428
148 0.000942278711590916
149 0.000914941250812262
150 0.00088829395826906
151 0.000862327113281935
152 0.00083703052951023
153 0.000812394311651587
154 0.000788408156950027
155 0.000765061529818922
156 0.000742343894671649
157 0.000720244424883276
158 0.000698751886375248
159 0.000677855510730296
160 0.000657543772831559
161 0.000637805438600481
162 0.000618628866504878
163 0.000600002880673856
164 0.000581915839575231
165 0.000564356334507465
166 0.000547312840353698
167 0.000530773831997067
168 0.000514728017151356
169 0.000499163987115026
170 0.000484070478705689
171 0.000469436054117978
172 0.000455249712103978
173 0.000441500247688964
174 0.000428176776040345
175 0.000415268295910209
176 0.000402764038881287
177 0.00039065329474397
178 0.000378925615223125
179 0.000367570610251278
180 0.000356577918864787
181 0.000345937442034483
182 0.000335639313561842
183 0.00032567442394793
184 0.000316031248075888
185 0.000306702277157456
186 0.00029767764499411
187 0.000288948300294578
188 0.000280505424598232
189 0.000272340374067426
190 0.000264444737695158
191 0.000256810191785917
192 0.000249428674578667
193 0.000242292269831523
194 0.000235393337788992
195 0.000228724253247492
196 0.000222277696593665
197 0.000216046450077556
198 0.000210023601539433
199 0.000204202282475308
};
\addlegendentry{Validation loss}
\addplot [thick, blue]
table {%
0 0.00639251852408051
1 0.00638978509232402
2 0.00638733571395278
3 0.00638506561517715
4 0.00638292031362653
5 0.00638091657310724
6 0.00637896312400699
7 0.00637704227119684
8 0.00637517962604761
9 0.00637330953031778
10 0.00637144222855568
11 0.00636955350637436
12 0.00636766664683819
13 0.00636574253439903
14 0.0063637369312346
15 0.00636164844036102
16 0.00635944167152047
17 0.00635713106021285
18 0.00635472033172846
19 0.00635221973061562
20 0.00634953798726201
21 0.00634668534621596
22 0.00634362455457449
23 0.0063403220847249
24 0.00633675185963511
25 0.00633288314566016
26 0.00632867310196161
27 0.00632407981902361
28 0.00631907815113664
29 0.00631361128762364
30 0.00630762102082372
31 0.00630106357857585
32 0.00629388308152556
33 0.00628600968047976
34 0.00627736933529377
35 0.00626788940280676
36 0.00625750701874495
37 0.00624615140259266
38 0.00623371219262481
39 0.00622011907398701
40 0.00620531942695379
41 0.00618924992159009
42 0.00617183418944478
43 0.00615300098434091
44 0.00613270094618201
45 0.00611085537821054
46 0.00608744006603956
47 0.00606243731454015
48 0.00603586109355092
49 0.00600766856223345
50 0.00597788626328111
51 0.00594654306769371
52 0.00591369811445475
53 0.00587941845878959
54 0.00584376789629459
55 0.00580686656758189
56 0.00576863717287779
57 0.00572923058643937
58 0.00568865472450852
59 0.00564688071608543
60 0.00560388341546059
61 0.00555962510406971
62 0.00551406433805823
63 0.00546717178076506
64 0.00541894510388374
65 0.00536941038444638
66 0.00531862350180745
67 0.00526665756478906
68 0.00521361455321312
69 0.0051595619879663
70 0.00510454922914505
71 0.00504860933870077
72 0.00499176094308496
73 0.00493402173742652
74 0.00487541221082211
75 0.00481596915051341
76 0.00475574703887105
77 0.00469481572508812
78 0.00463325204327703
79 0.00457113375887275
80 0.00450852606445551
81 0.00444548623636365
82 0.00438205106183887
83 0.00431824894621968
84 0.00425410876050591
85 0.00418966170400381
86 0.00412495387718081
87 0.00406003929674625
88 0.00399498082697392
89 0.00392983946949244
90 0.00386467040516436
91 0.00379952089861035
92 0.00373442890122533
93 0.00366942957043648
94 0.00360455689951777
95 0.00353984790854156
96 0.00347534217871726
97 0.00341108022257686
98 0.00334710208699107
99 0.0032834461890161
100 0.00322014442645013
101 0.00315722823143005
102 0.00309472647495568
103 0.00303266663104296
104 0.00297107943333685
105 0.00290999328717589
106 0.00284943636506796
107 0.00278943544253707
108 0.0027300154324621
109 0.00267119845375419
110 0.0026130061596632
111 0.00255547277629375
112 0.00249858340248466
113 0.00244239019230008
114 0.00238689943216741
115 0.00233212765306234
116 0.002278090454638
117 0.00222480203956366
118 0.00217227428220212
119 0.00212051812559366
120 0.00206954288296402
121 0.00201935810036957
122 0.00196997122839093
123 0.00192139041610062
124 0.00187362241558731
125 0.00182667269837111
126 0.0017805447569117
127 0.0017352404538542
128 0.0016907611861825
129 0.00164710788521916
130 0.00160428136587143
131 0.00156228104606271
132 0.00152110552880913
133 0.00148075097240508
134 0.00144121388439089
135 0.00140248995739967
136 0.0013645754661411
137 0.00132746493909508
138 0.00129115267191082
139 0.00125563202891499
140 0.0012208956759423
141 0.00118693674448878
142 0.00115374766755849
143 0.00112132076174021
144 0.00108964741230011
145 0.0010587191209197
146 0.00102852680720389
147 0.000999062089249492
148 0.000970315712038428
149 0.000942278711590916
150 0.000914941250812262
151 0.00088829395826906
152 0.000862327113281935
153 0.00083703052951023
154 0.000812394311651587
155 0.000788408156950027
156 0.000765061529818922
157 0.000742343894671649
158 0.000720244424883276
159 0.000698751886375248
160 0.000677855510730296
161 0.000657543772831559
162 0.000637805438600481
163 0.000618628866504878
164 0.000600002880673856
165 0.000581915839575231
166 0.000564356334507465
167 0.000547312840353698
168 0.000530773831997067
169 0.000514728017151356
170 0.000499163987115026
171 0.000484070478705689
172 0.000469436054117978
173 0.000455249712103978
174 0.000441500247688964
175 0.000428176776040345
176 0.000415268295910209
177 0.000402764038881287
178 0.00039065329474397
179 0.000378925615223125
180 0.000367570610251278
181 0.000356577918864787
182 0.000345937442034483
183 0.000335639313561842
184 0.00032567442394793
185 0.000316031248075888
186 0.000306702277157456
187 0.00029767764499411
188 0.000288948300294578
189 0.000280505424598232
190 0.000272340374067426
191 0.000264444737695158
192 0.000256810191785917
193 0.000249428674578667
194 0.000242292269831523
195 0.000235393337788992
196 0.000228724253247492
197 0.000222277696593665
198 0.000216046450077556
199 0.000210023601539433
};
\addlegendentry{Loss}
\end{axis}

\end{tikzpicture}

%% file: figures/Qlearning_perf_tabular.tikz
\begin{tikzpicture}

\begin{axis}[
tick align=outside,
tick pos=left,
x grid style={white!69.0196078431373!black},
xlabel={Episode},
xmajorgrids,
xmin=-0.1, xmax=100.2,
xtick style={color=black},
y grid style={white!69.0196078431373!black},
ylabel={Expected Action-Value $Q$},
ymajorgrids,
ymin=-0.0151285788868104, ymax=0.35, 
ytick style={color=black}
]
\addplot [line width=2, blue]
table {%
1 0.000657318426666667
2 0.00378896379333333
3 0.00509777815762083
4 0.0051564354730265
5 0.00612048877297586
6 0.00855346808529447
7 0.0103483043307166
8 0.0124394037050199
9 0.012942572898966
10 0.0156048553582391
11 0.0183176913770463
12 0.0196562187165186
13 0.0202906387446779
14 0.0208092039710185
15 0.0234535784000339
16 0.0275157383032773
17 0.0293726204874749
18 0.0323327490243121
19 0.0345782937561701
20 0.0366737335936683
21 0.0396304427995692
22 0.0412612826608809
23 0.045288757172996
24 0.0482743074898101
25 0.0510495547063234
26 0.0534211172696053
27 0.0561583336009971
28 0.0612824594627198
29 0.06726208694134
30 0.0717752896475987
31 0.0742102373682333
32 0.0792836213747586
33 0.0842288900915684
34 0.0869666874349323
35 0.0899566685181003
36 0.0940193018071762
37 0.0988185835057609
38 0.102174212863048
39 0.106186074377718
40 0.10965416920332
41 0.117019404675078
42 0.12484957998593
43 0.126800384919684
44 0.129370757548281
45 0.133098800115447
46 0.135975624914337
47 0.138323099502926
48 0.140382217227811
49 0.142575258975604
50 0.144865004784022
51 0.148605778946738
52 0.153735767464284
53 0.157351204530269
54 0.161295798955668
55 0.163072469775728
56 0.164492585367963
57 0.165213681730543
58 0.16887084116325
59 0.172700095701873
60 0.175084514533532
61 0.179993563706647
62 0.18701784466513
63 0.18932967925842
64 0.191816360019184
65 0.19572441707727
66 0.201837800350758
67 0.207147066843032
68 0.209483360270031
69 0.214779659190776
70 0.220737201751375
71 0.222687117155648
72 0.22451224862521
73 0.226554012475005
74 0.231219222114725
75 0.235028603893712
76 0.237038205402403
77 0.243078476255558
78 0.250432925828413
79 0.25675459723013
80 0.260886603516084
81 0.263085251127981
82 0.267126799432357
83 0.271164459553094
84 0.275654155423716
85 0.277838558751628
86 0.279901666051721
87 0.281347829117224
88 0.283277178978721
89 0.284762444338924
90 0.28707383811984
91 0.289860219927964
92 0.296571095305195
93 0.299216857883143
94 0.300425781956726
95 0.303209725926171
96 0.305897221462655
97 0.309103392805666
98 0.312325692944568
99 0.316786373271792
100 0.31951946432208
};
\end{axis}

\end{tikzpicture}

%% file: figures/Qlearning_perf_dqn.tikz
\begin{tikzpicture}

\begin{axis}[
tick align=outside,
tick pos=left,
x grid style={white!69.0196078431373!black},
xlabel={Episode},
xmajorgrids,
xmin=0, xmax=200,
xtick style={color=black},
y grid style={white!69.0196078431373!black},
ylabel={Expected Action-Value $Q_\pi(s,a)$},
ymajorgrids,
ymin=-0.1, ymax=2.2,
ytick style={color=black}
]
\addplot [line width=1.5, blue]
table {%
1 0.223487936060556
2 0.0685357728076876
3 -0.0599273784057004
4 0.015516184542624
5 -0.0551829760261171
6 0.00637863745006533
7 0.176877457060896
8 0.0576054721939727
9 0.0714408381536487
10 0.0317002769194611
11 0.0830268767969553
12 0.197049811313627
13 0.10833177195794
14 0.171461026122525
15 0.0993429472857517
16 0.0767021654076719
17 0.00665135524468497
18 0.0664609382719001
19 0.209851075801646
20 0.251168953259131
21 0.0974072447279468
22 0.223804682260379
23 0.263030600065549
24 0.202826153995314
25 0.164085840263094
26 0.195514723868109
27 0.289299063545297
28 0.258525505784847
29 0.192792913436683
30 0.299229508809124
31 0.157374882143423
32 0.244848901813384
33 0.201612742966972
34 0.278618463644913
35 0.301233567969272
36 0.248637737997342
37 0.203162279891937
38 0.359653601597529
39 0.412057031702716
40 0.234793792275013
41 0.319209925233736
42 0.257370938553423
43 0.337967593620609
44 0.22931796569416
45 0.412003659420103
46 0.276914223873367
47 0.360740676527106
48 0.363247264098997
49 0.234953343573337
50 0.298717792301937
51 0.362218622064878
52 0.155575340342087
53 0.305760654962311
54 0.512512796131584
55 0.418241127200114
56 0.323313622588844
57 0.316314870598338
58 0.176935628703278
59 0.347915388415762
60 0.2185856938983
61 0.55443346289394
62 0.325059909711681
63 0.475868144926305
64 0.420974653576895
65 0.489529412905944
66 0.391862397314981
67 0.35577898998506
68 0.562767534031688
69 0.32649900833591
70 0.293288763728924
71 0.34900364170446
72 0.453304167544573
73 0.355078036615547
74 0.613704529396879
75 0.443727192047259
76 0.484694755173479
77 0.645301955415764
78 0.466053589746459
79 0.42715963255614
80 0.484174668265041
81 0.540243402899553
82 0.302113810166096
83 0.563935178894705
84 0.490185784428225
85 0.392197026114445
86 0.419830268253227
87 0.444573177276955
88 0.705084081040695
89 0.455723641960261
90 0.544490891633797
91 0.585570092933873
92 0.379428069999752
93 0.519246232012261
94 0.466249661822157
95 0.631336719816318
96 0.360435615316318
97 0.584181988068546
98 0.668147305702968
99 0.517060389380074
100 1.02194044824379
101 0.67832203702225
102 0.719959404785186
103 0.691849503577881
104 0.12308379712825
105 0.676578231466313
106 0.703806798093461
107 0.433727028027837
108 0.663127278011346
109 0.700272085542122
110 0.404424526955053
111 0.677534460399849
112 0.799504218163201
113 0.721677299511308
114 0.767772067726279
115 0.762185758807593
116 0.701175217866613
117 0.750073052025982
118 0.666849706797317
119 0.707886812315943
120 0.799353029966975
121 0.707859625039245
122 0.818924407513502
123 0.91317203687989
124 0.723531296917437
125 0.870913143329866
126 0.816019785978521
127 0.622192392251842
128 0.902068293085463
129 0.894392462118935
130 1.03096434328472
131 0.908358343598795
132 0.816334949069035
133 0.874454206745092
134 0.774004007369513
135 0.780557307763956
136 0.954321099120231
137 0.921087732159858
138 0.705986865262578
139 0.859222277987283
140 1.01063852990462
141 0.844699692226843
142 0.783016490245548
143 0.946336188916272
144 1.31708821184778
145 0.917278402014068
146 1.01973799910241
147 1.00247793149902
148 1.28048599416312
149 1.09558714934004
150 0.963136528469383
151 1.14700561237987
152 1.05931865501229
153 1.05138867080132
154 1.03985730902683
155 0.991282104080967
156 1.16185574150378
157 1.0817247376622
158 1.11392760598877
159 1.0498735730588
160 1.20040228690557
161 1.516194122533
162 1.16413459371943
163 1.18599925961462
164 1.30137908903675
165 1.40934355644276
166 1.19994075216043
167 1.34167047434797
168 1.57325013845305
169 1.49974058005804
170 1.54224596719723
171 1.45530328771306
172 1.64922307744079
173 1.66776112230016
174 1.38469433491118
175 2.06500199964891
176 1.51047906618203
177 1.26250423469658
178 1.5669013774265
179 1.60501009925113
180 1.53091453638238
181 1.34025306933284
182 1.78557183605584
183 1.78594382072632
184 1.45871404306187
185 1.66879617818114
186 2.11900002004889
187 1.67169399100821
188 1.4240500240121
189 1.47638280385096
190 1.7709926298509
191 1.62806124277025
192 1.60018553555322
193 1.8933456291755
194 1.73732033476958
195 1.36797812115401
196 1.97013027896173
197 1.69254429722205
198 1.22114392877039
199 1.8326126228009
200 1.95337868555604
};
\end{axis}

\end{tikzpicture}

%% file: figures/environment_SINR_dB_tabular.tikz
\begin{tikzpicture}
\tikzstyle{every node}=[font=\Large]
\begin{axis}[
tick align=outside,
tick pos=left,
x grid style={white!69.0196078431373!black},
xlabel={Time step $t$},
xmajorgrids,
xmin=0.25, xmax=16.75,
xtick style={color=black},
y grid style={white!69.0196078431373!black},
ylabel={SINR [dB]},
ymajorgrids,
ymin=-3.5, ymax=15,
ytick style={color=black}
]
\addplot [line width=2, black]
table {%
1 0
2 3
3 4
4 3
5 0
6 -3
7 0
8 1
9 4
10 1
11 4
12 5
13 6
14 7
15 10
16 13
};

\addplot [line width=1, dash pattern=on 8pt off 4pt, gray]
table {%
0 12
17 12
};
\end{axis}

\end{tikzpicture}

%% file: figures/environment_SINR_dB_dqn.tikz
\begin{tikzpicture}
\tikzstyle{every node}=[font=\Large]
\begin{axis}[
tick align=outside,
tick pos=left,
x grid style={white!69.0196078431373!black},
xlabel={Time step $t$},
xmajorgrids,
xmin=0.25, xmax=16.75,
xtick style={color=black},
y grid style={white!69.0196078431373!black},
ylabel={SINR [dB]},
ymajorgrids,
ymin=-3.5, ymax=15,
ytick style={color=black}
]
\addplot [line width=2, black]
table {%
1 0
2 3
3 4
4 5
5 4
6 3
7 0
8 3
9 0
10 3
11 6
12 9
13 10
14 9
15 10
16 13
};

\addplot [line width=1, dash pattern=on 8pt off 4pt, gray]
table {%
0 12
17 12
};
\end{axis}

\end{tikzpicture}

%% file: figures/actions_tabular.tikz
\begin{tikzpicture}
\tikzstyle{every node}=[font=\Large]
\begin{axis}[
tick align=outside,
tick pos=left,
x grid style={white!69.0196078431373!black},
xlabel={Time step $t$},
xmajorgrids,
xmin=0, xmax=15,
xtick style={color=black},
y grid style={white!69.0196078431373!black},
ylabel={Action $a_t$},
ymajorgrids,
ymin=-0.1, ymax=2.1,
ytick style={color=black},
ytick={0,1,2}
]
\addplot [line width=2, black, const plot mark right]
table {%
1 2
2 1
3 1
4 1
5 2
6 1
7 1
8 0
9 0
10 1
11 1
12 2
13 2
14 0
15 2
};
\end{axis}

\end{tikzpicture}

%% file: figures/actions_dqn.tikz
\begin{tikzpicture}
\tikzstyle{every node}=[font=\Large]
\begin{axis}[
tick align=outside,
tick pos=left,
x grid style={white!69.0196078431373!black},
xlabel={Time step $t$},
xmajorgrids,
xmin=000, xmax=15,
xtick style={color=black},
y grid style={white!69.0196078431373!black},
ylabel={Action $a_t$},
ymajorgrids,
ymin=-0.1, ymax=2.1,
ytick style={color=black},
ytick={0,1,2}
]
\addplot [line width=2, black, const plot mark right]
table {%
1 2
2 1
3 2
4 1
5 2
6 1
7 1
8 0
9 1
10 2
11 2
12 1
13 1
14 0
15 0
};
\end{axis}

\end{tikzpicture}